\documentclass[a4paper,twocolumn,11pt,accepted=2025-11-26]{quantumarticle}
\pdfoutput=1
\usepackage[utf8]{inputenc}
\usepackage[english]{babel}
\usepackage[T1]{fontenc}
\usepackage{amsmath}
\usepackage{hyperref}

\usepackage{tikz}
\usepackage{lipsum}

\usepackage[caption=false]{subfig}
\usepackage{cleveref}
\usepackage{physics}
\usepackage{amssymb}
\usepackage{ulem}
\usepackage[numbers,sort&compress]{natbib}

\begin{document}

\title{Optically Hyperpolarized Materials for Levitated Optomechanics}

\author{Marit O. E. Steiner}
\affiliation{Institut für Theoretische Physik, Albert-Einstein-Allee 11, Universität Ulm, D-89081 Ulm, Germany}
\affiliation{Center for Integrated Quantum Science and Technology (IQST), Ulm University, Albert-Einstein-Allee 11, 89081 Ulm, Germany}
\email{marit.steiner@uni-ulm.de}
\author{Julen S. Pedernales}
\email{julen.pedernales@uni-ulm.de}
\affiliation{Institut für Theoretische Physik, Albert-Einstein-Allee 11, Universität Ulm, D-89081 Ulm, Germany}
\affiliation{Center for Integrated Quantum Science and Technology (IQST), Ulm University, Albert-Einstein-Allee 11, 89081 Ulm, Germany}
\author{Martin B. Plenio}
\affiliation{Institut für Theoretische Physik, Albert-Einstein-Allee 11, Universität Ulm, D-89081 Ulm, Germany}
\affiliation{Center for Integrated Quantum Science and Technology (IQST), Ulm University, Albert-Einstein-Allee 11, 89081 Ulm, Germany}
\email{martin.plenio@uni-ulm.de}
\maketitle

\begin{abstract}
We explore the potential of levitating solids embedded with non-permanent, optically controllable electron spins, which can be used to hyperpolarize their nuclear spin environment 
with exceptionally long lifetimes.
For example, pentacene-doped naphthalene, which will also serve as our prime example, can achieve bulk polarization 
exceeding 80\,\% at cryogenic temperatures with polarization lifetimes extending over weeks. These materials make a compelling case for applications such as matter-wave interferometry and novel uses of established NMR techniques. In that spirit, we design a multi-spin Stern-Gerlach-type interferometry protocol which, thanks to the homogeneous spin distribution and the absence of a preferential nuclear-spin quantization axis in such materials, avoids many of the limitations associated with solid state crystals
hosting electronic spin defects, such as nanodiamonds containing NV centers. We assess the potential of our interferometer to enhance existing bounds on the free parameters of objective collapse models.  Beyond matter-wave interferometry, we analyze the prospects for implementing magic angle spinning at frequencies surpassing the current standard in NMR, capitalizing on the exceptional rotational capabilities offered by levitation. Additionally, we outline a novel protocol for measuring spin ensemble polarization via the position of the nanoparticle and conduct an analysis of dominant noise sources, benchmarking the required isolation levels for various applications.
\end{abstract}

\section{Introduction}
Levitated optomechanics has opened a new window into the quantum mechanics of mesoscopic systems~\cite{Black2004,GonzalezBallestero2021}. The levitation of nano- and microscale solids in a vacuum,  pioneered by Ashkin more than 50 years ago~\cite{Ashkin1970}, started its incursion into the quantum regime barely 10 years ago, with the first proposals for cooling the mechanical degrees of freedom of levitated particles~\cite{Cirac2010, Zoller2010,Schneider2010}. Today, with the recent demonstration of motional ground state cooling~\cite{Delic2020, Aspelmeyer2021, Novotny2021}, it is generally acknowledged that the first stage of this incursion has been completed, establishing levitated optomechanics as a quantum platform that promises a plethora of scientific as well as technological applications. Among others, the quantum control of microscopic objects may open the door to testing the linearity of quantum mechanics in unprecedented regimes, and offer a platform for ultrasensitive force measurements with both commercial and fundamental-science applications.

In comparison to clamped optomechanical platforms~\cite{Marquardt2014}, levitated systems distinguish themselves by their inherent lack of direct material contact with their surroundings. This unique characteristic affords exceptional isolation conditions, rendering them mechanical quantum systems with ultra-low dissipation. Furthermore, from a mechanical point of view, levitated systems possess the capability to undergo large displacements of their center of mass, including displacements larger than their size. They also exhibit controllable rotational degrees of freedom~\cite{Millen2017,Kim2021, Plenio2020a}, with the ability to spin at unprecedented frequencies, only limited by the tensile strength of the levitated material~\cite{Novotny2018,Li2018,Li2020}. Additionally, the capacity to dynamically adjust trapping fields over time or deactivate them entirely offers avenues for exploring dynamics in complex potential landscapes as well as free dynamics in the absence of external fields. All these capabilities render levitated systems as unique quantum platforms, offering unparalleled opportunities to explore the realm of quantum mechanics at the macroscopic scale.

In terms of levitation methods, these have evolved from the original optical tweezer technique to encompass diamagnetic levitation as well as electric levitation in Paul traps, each presenting distinct advantages and drawbacks. The catalog of commonly levitated materials includes dielectric materials such as silica, diamond with color centers~\cite{Vamivakas2015,Li2016,Quidant2018}, magnets~\cite{Budker2019, Ulbricht2020,Lukin2020}, or even superfluid helium~\cite{Harris2023}. A notable subset comprises materials containing accessible spin degrees of freedom, as they offer a pathway for harnessing the motion of the levitated particle through spin-dependent forces~\cite{Hetet2020, Hetet2021,Hetet2022, Plenio2021}. For instance, diamond nanoparticles hosting color centers hold promise for implementing matter-wave interferometry, enabling advancements in force sensing and fundamental physics applications, based on Stern-Gerlach-effect \cite{Bose2013, Kim2016,Plenio2020} or photon recoil \cite{Albrecht2014}.

However, the presence of color centers such as the nitrogen-vacancy (NV) center in these setups comes with important challenges. In particular, they are anticipated to introduce significant channels of decoherence, potentially rendering the most advanced matter-wave interferometry experiments extraordinarily challenging and, potentially, unattainable~\cite{Plenio2020,Pedernales2022}. For instance, the negatively charged NV center induces an electric dipole moment in the neutrally charged diamond nanoparticle, making it susceptible to interaction with stray electric fields. Moreover, the preferential axis of quantization of the NV center leads to undesired torques as it seeks alignment with the background magnetic field. Additional torques arise in the presence of magnetic field gradients, as these exert forces on the NV center which will, typically, be offset from the center of mass of the nanoparticle. Besides these drawbacks directly associated with the presence of the NV center, despite recent advances \cite{march2023long}, the achievable purity of diamond crystals remains a significant concern. Even the purest diamond samples harbor a substantial number of electronic impurities both in the bulk of the nanodiamond, e.g. P1 centers, as well as in its surface, e.g. dangling bonds. These can result in a significant source of decoherence, for which the possibility of mitigation is uncertain. These challenges underscore the need for innovative approaches in the use of spin-dependent forces including the exploration of alternative materials.

Here, we examine the potential of a material hitherto unexplored in the realm of levitated optomechanics, one that offers solutions to several of the aforementioned challenges: diamagnetically levitated, pentacene-doped naphthalene. Remarkably, even in macroscopic samples, the nuclear spins of the hydrogen atoms in naphthalene crystals can be polarized to levels 
exceeding $80\,\%$, with ultralong decay times surpassing $900$~h at $25$~K~\cite{Henstra1990, Eichhorn2013,Eichhorn2013a, Can2015, Quan2019}. This polarization is achieved through the short-lived photo-excited triplet state of embedded pentacene molecules. The transient nature of this state allows for the elimination of associated noise channels from the experiment following the polarization of the nuclear spin ensemble. Additionally, naphthalene crystals demonstrate experimentally the near-absence of magnetically active impurities, as evidenced by the extended polarization times. Beyond naphthalene, these advantageous aspects point towards the possibility of exploring much broader classes of materials with embedded photo-excitable electron spins for applications in high-mass matter-wave interferometry. For instance, biological materials such as p-dibromobenzene doped with p-dichlorobenzene~\cite{Deimling1980} and p-terphenyl doped with pentacene~\cite{Iinuma2000} can also be optically hyperpolarized. Recent advances in chemically modifying pentacene-doped naphthalene to enhance its robustness during the hyperpolarization process~\cite{Sakamoto2023} shows the potential for engineered materials tailored specifically for optomechanical applications. Following the strategies in quantum information processing, where molecules are chemically designed to meet specific needs~\cite{Wasielewski2020}, we can envision a new class of materials optimized for the challenges of levitated optomechanics, further expanding the possibilities in this field.

As an exemplary application, we design a multi-spin matter-wave interferometry protocol that leverages the high polarization rates of the nuclear spin ensemble. Benefiting from a homogeneous spin distribution and the absence of a preferential axis of quantization, we eliminate undesired torques, enhancing the visibility of the interferometer. Beyond matter-wave interferometry, levitating naphthalene suggests exciting prospects for pushing the boundaries of NMR applications. The ability to rotate naphthalene nanoparticles, or any other material, at unprecedented frequencies opens avenues for attaining unmatched nuclear spin coherence times through the technique of magic-angle spinning thus opening novel avenues for solid-state nuclear magnetic resonance. Furthermore, we design a novel technique for measuring nanoparticle magnetization by observing displacements induced by the interaction with a background magnetic field gradient.

The manuscript proceeds as follows: Section~\ref{sec:LevitatedNaphthalene} provides a comprehensive analysis of levitated naphthalene crystals and relevant parameter regimes, including procedures for hyperpolarization of the nuclear spin ensemble and for magic angle spinning. Section~\ref{sec:Interferometer} presents a novel multi-spin matter-wave interferometry protocol and its applicability in testing the linearity of quantum mechanics and constraining free parameters of collapse models. Furthermore, a novel measurement scheme for detecting the polarization rate of the nuclear spin ensemble is introduced. Section~\ref{sec:Noise} explores sources of noise and benchmarks the isolation required for a successful implementation of the ideas presented in the previous sections. Finally, Section~\ref{sec:Conclusions} concludes with a summary and outlook for future lines of research.

%------------------------------------------------

\section{Naphthalene - a material for levitation}
\label{sec:LevitatedNaphthalene}

\begin{figure}
\centering

\subfloat[\label{fig:Naphthalene}]{%
  \includegraphics[width=0.45\columnwidth]{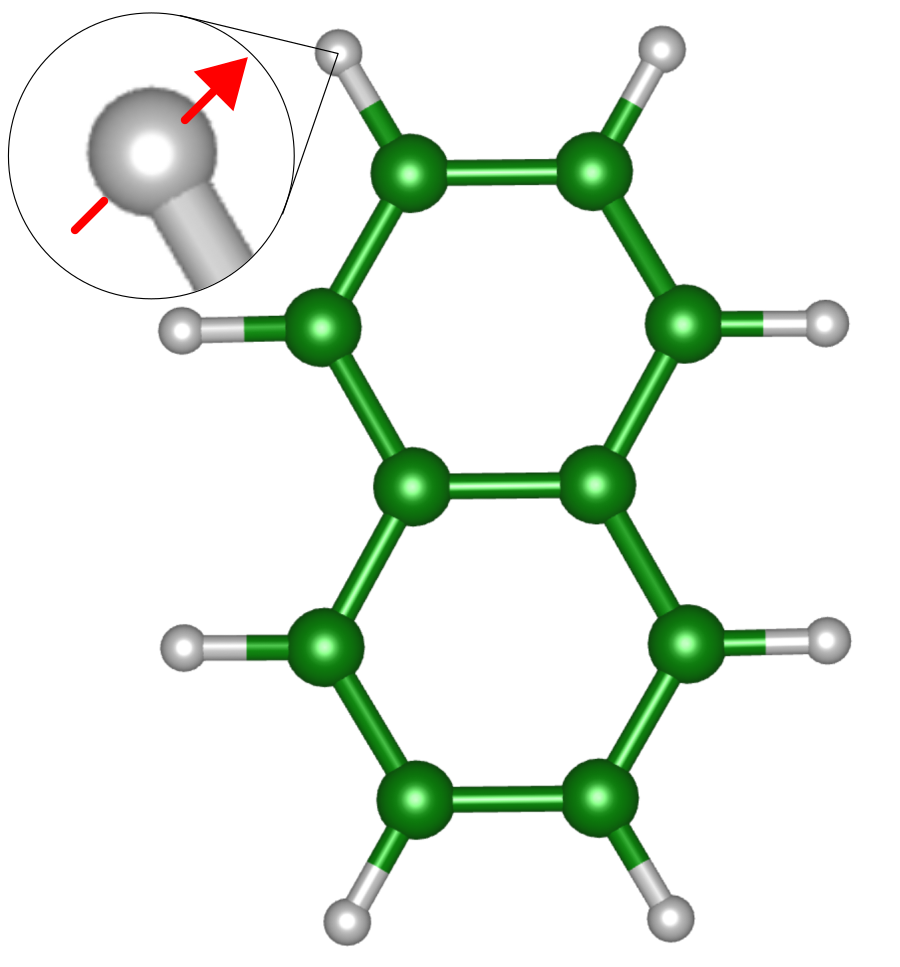}%
}%
\subfloat[\label{fig:NaphthaleneCrystal}]{%
\centering
  \includegraphics[width=0.45\columnwidth]{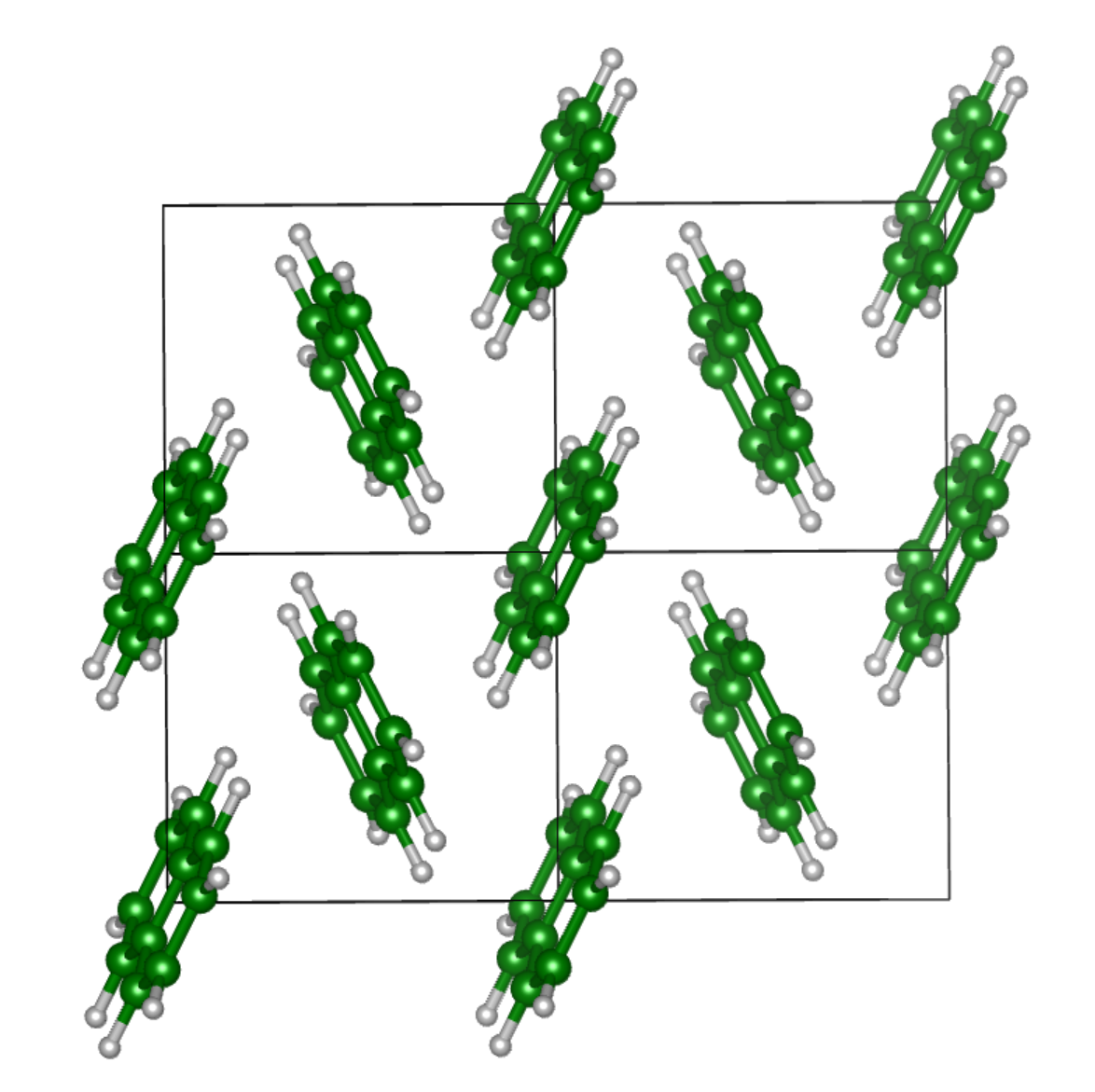}%
}

\subfloat[\label{fig:Pentacene}]{%
  \includegraphics[width=0.7\columnwidth]{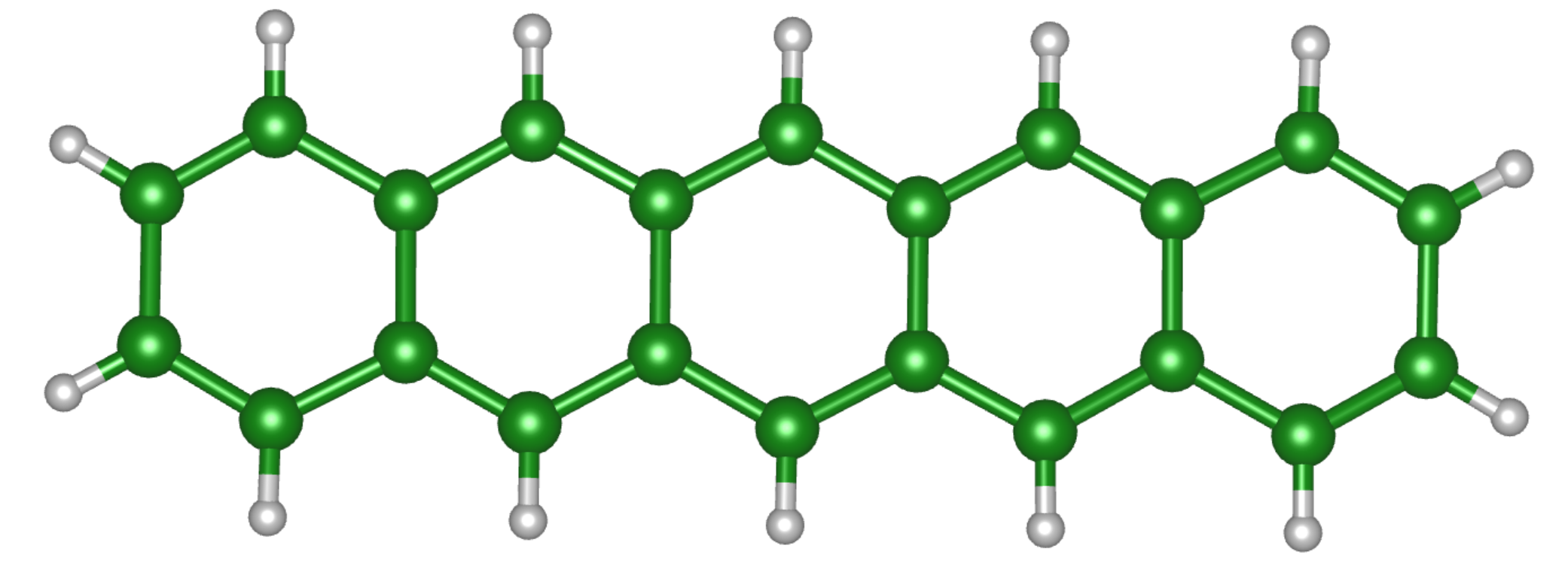}%
}
\caption{(a) and (b): The structure of an individual naphthalene molecule and naphthalene in its crystalline form. (c) Structure of pentacene. The carbon atoms are colored green while the hydrogen atoms are colored gray. 
The hydrogen atoms carry a nuclear spin of $1/2$. Naturally, the $^{13}$C isotope of carbon, which carries a spin, constitutes only $1.1\,\%$ of all carbon atoms, whereas the $^{12}$C isotope does not possess a spin. To ensure that only the hydrogen atoms carry a spin, $^{12}$C enriched naphthalene can be used.} 
\label{fig:Structures}
\end{figure}

Naphthalene is the simplest of the polycyclic aromatic hydrocarbons, consisting of two fused benzene rings as shown in \cref{fig:Structures}. At room temperature, it forms a white, crystalline solid held together by Van der Waals forces. It has a distinctive, pungent odor and, while it is best known as the main ingredient of mothballs, its usage is wide ranging~\cite{Komatsu1993, Wang2010, Dash2015}. Naphthalene can be doped with pentacene, another polycyclic aromatic hydrocarbon, where each pentacene molecule substitutes two naphthalene molecules without perturbing the lattice. Pentacene-doped naphthalene has gained attention in the nuclear magnetic resonance community due to its large nuclear spin polarizability. When doped with pentacene, the spins of the hydrogen atoms in naphthalene can be polarized~\cite{Henstra1990, Eichhorn2013,Eichhorn2013a, Can2015, Quan2019} to over $80\,\%$ even in a macroscopic crystal. Hyperpolarized naphthalene is used as a spin filter in neutron scattering~\cite{Haag2012} and as a means to enhance sensitivity in nuclear magnetic resonance~\cite{Eichhorn2022}.

In this section, we discuss the particularities of a hyperpolarized naphthalene nanoparticle for levitation. Specifically, we explore the crucial aspects of nuclear spin hyperpolarization protocols, the potential enabled by levitation for conducting magic angle spinning at remarkable rotation frequencies, and the constraints imposed by sublimation rates on the permissible bulk temperature of naphthalene.

\subsection{Hyperpolarization}

The nuclear spins of the hydrogen atoms in naphthalene can be hyperpolarized by using embedded pentacene molecules as a source of polarization. Pentacene, in its electronic ground state, is a spin-0 system while in its optically excited electronic state is a short-lived triplet state, where selection rules strongly favor the occupation of one of its three triplet states. This offers a mechanism for the preparation of a highly polarized spin state of the electron, whose polarization can then be transferred to the surrounding nuclear spins via dipole-dipole magnetic interactions. 

It is noteworthy to mention that the optical excitation of the electron spin in pentacene requires a good alignment of the pentacene molecule with the external magnetic field. Nevertheless, a few degrees of deviation against perfect alignment is admissible, as this results in only a few percent loss of the polarization of the pentacene electron spin. However, for larger deviations, the achievable polarization drops quickly. For a more detailed analysis, we refer the reader to Ref.~\cite{VanStrien1980}. The alignment might be achieved by suitable trapping conditions. Alternatively, it has been shown that instead of orientating the crystal in the magnetic field, the pentacene molecule can be modified with polarizing agents such as thiophene. Thiophene-modified pentacene exhibits a sharper and stronger electron spin resonance spectrum compared to unmodified pentacene. Moreover, the triplet electron distribution is more isotropic without loss of triplet polarization, making it more robust to deviations from the ideal orientation in the external magnetic field~\cite{Sakamoto2023}. 

The transfer of electron spin polarization from the pentacene molecule to the nuclear spin ensemble can be accomplished by the application of a suitable microwave drive (or, a sequence of pulses) that brings the electron spin in resonance with the nuclear spin lattice~\cite{Eichhorn2013}. However, the eigenfrequency of the electron spin depends on the spatial orientation of the pentacene molecule with respect to the external magnetic field and thus this will also require a good alignment between the magnetic field and the pentacene molecule. In this case, the effect of imperfect alignments might be mitigated by using pulse sequences that exhibit robustness against detuning errors such as sweep schemes \cite{Eichhorn2013} or robust pulsed sequences such as PulsePol \cite{Schwartz2018}.

The triplet state of pentacene has a lifetime in the order of tens of microseconds, which is sufficiently long to allow for the transfer of polarization to the nuclear spin-lattice. After the transfer, the electron quickly decays back to form a singlet ground state, which does not interact magnetically with its environment. This differentiates pentacene-doped naphthalene apart from persistent embedded spins, such as color centers in diamond, where the color center can also be used as a source of polarization of the nuclear spin-lattice. However, in contrast to pentacene-doped naphthalene, persistent electronic spins used for polarization remain magnetically active, typically in the form of a ground state triplet. This leads to spurious interactions with the spin-lattice and environmental fields, even after the polarization transfer is complete, resulting in unwanted decoherence and dissipation.
In contrast, the short lifetime of the photo-excited triplet states in pentacene allows the opening of a channel for polarization transfer that is subsequently closed, thus removing an important source of dissipation, and resulting in unmatched nuclear polarization lifetimes. In bulk, up to 80\,\% nuclear polarization was achieved with an ultralong $T_1$ time of up to 920\,h when operating at 25\, K and 500\, mT~\cite{Quan2019}. In a nanoparticle of 70\,nm radius, with just one or two embedded pentacenes, even higher polarization levels
can be expected.

As will be discussed in the next chapter, the ability to polarize the nuclear spins in naphthalene offers a promising alternative to create macroscopic superpositions, diverging from the conventional use of single electron spins like NV centers in diamond. Despite the electron spin's higher gyromagnetic ratio, the collective strength of naphthalene's hydrogen spins, numbering around $6\cdot10^7$ in a particle with $70$~nm radius, surpasses that of an individual electron spin by more than 4 orders of magnitude. 

\subsection{Magic Angle Spinning}\label{sec:Magic Angle Spinning}

While the longitudinal relaxation time $T_1$ is very long, the coherence time of each nuclear spin is constrained by the, generally much shorter, $T_2$ time. For each nuclear spin, the main source of decoherence is provided by its interaction with the rest of the nuclear spin lattice, which constitutes its immediate environment. Thus, to achieve extended coherence times, it would be desirable to suppress the inter-nuclear spin interactions. A technique, well known in the field of nuclear magnetic resonance spectroscopy, that can achieve this is that of magic angle spinning (MAS), which is routinely used in solid-state NMR to increase the $T_2$ time of spins in a crystal by suppressing their dipole-dipole interaction. The technique works by rotating the probe sufficiently fast around an axis that is tilted with respect to the direction of the magnetic field by a specific angle of $\arccos(1/\sqrt{3})\approx54.74$°, referred to as the magic angle~\cite{Hennel2004}.
The achievable $T_2$ time is approximately determined by the rotation frequency $\nu_\mathrm{MAS}$ and is given by~\cite{Zorin2006}
\begin{equation}
    T_2 \approx \frac{\nu_\mathrm{MAS}}{\pi G \left(\frac{d_\mathrm{rss}}{2 \pi}\right)^2},
\end{equation}
where $d_\mathrm{rss}$ is the effective dipolar coupling. The effective dipolar coupling experienced by a given proton $j$ can be written as $d_\mathrm{rss,j}=(\sum_{k\neq j} d_{j,k}^2)^{1/2}$, where $d_{j,k}$ is the dipolar coupling between the $j^\mathrm{th}$ and $k^\mathrm{th}$ spin. $G$ is a geometric factor given by the structure of the lattice, with a value that can range between $0.04$ and $0.11$~\cite{Zorin2006}. While the exact value for the naphthalene crystal is not known, for our calculations we will use the maximum value of $G=0.11$, leading to the most conservative estimations of the $T_2$ time. To estimate $d_\mathrm{rss}$, we first compute the inter-nuclear dipolar coupling strength $d_{j,k} = \frac{\mu_0 \gamma_H^2\hbar}{4 \pi \abs{\vec r_{jk}}^3}(1 - 3 \cos^2 \theta_{jk})$ between a nuclear spin $j$ and all the spins $k$ in its lattice cell as well as in the first-neighbor cells, and use it to compute $d_{{\rm rss},j}$. We then average this value over all the spins $j$ within a lattice cell. Here, $\gamma_H \approx (2\pi) 42.576 \, \mathrm{MHz/T}$
is the gyromagnetic ratio of the hydrogen nucleus, $\vec r_{jk}$ is the vector joining the spins $j$ and $k$, and $\theta_{jk}$ is the angle formed by the magnetic field and the vector $\vec r_{jk}$. We take the positions of the hydrogen atoms in naphthalene from X-ray scattering data~\cite{Oddershede2004}. The computation yields $d_\mathrm{rss} \approx (2\pi)8.1\,$kHz, resulting in the following relation between the achievable $T_2$ time and the rotation frequency of the crystal
\begin{equation}
T_2 \approx 4.4\,\nu_\mathrm{MAS}\cdot10^{-8}\,{\rm s}. 
\end{equation}

Thanks to their inherent rotational freedom, nanoparticles levitated in a vacuum offer an unprecedented opportunity for MAS with rotational frequencies well beyond the state of the art~\cite{Barnes2024} in solid-state NMR. The most advanced rotation methods in this field typically operate at frequencies on the order of $100$~kHz, with the largest demonstrated rotation frequencies being on the order of $200$~kHz~\cite{Ramamoorthy2023}. This is achieved by pneumatic techniques, which employ compressed air to spin an air-bearing rotor containing the sample of interest. In contrast, levitated nano- and microparticles, exploit the transfer of angular momentum from circularly polarized light to the rotational degrees of freedom of the particle. This transfer can occur through photon absorption, or due to the birefringence or geometric anisotropy of the particles~\cite{Rudolph2021, Barnes2024}. Exploiting this mechanism, microparticles levitated in a vacuum can attain rotational frequencies up to the MHz range~\cite{Arita2013,Moore2018} while a nanoparticle can achieve frequencies as high as GHz~\cite{Novotny2018, Zhang2021}. Notably, naphthalene crystals exhibit strong birefringence with three optical axes, with refractive indices along these axes measured at $1.442, 1.775$, and $1.932$ respectively~\cite{Bhagavantam1929}, surpassing even the birefringence of vaterite~\cite{Haynes2014}. This suggests that under similar conditions, naphthalene could achieve rotational frequencies as high as those observed in the rotation of vaterite~\cite{Arita2013} or potentially higher.

Nevertheless, the ultimate rotation frequencies achievable with any material are constrained by the yield strength of the material. The maximal rotation frequency a sphere of naphthalene can survive before breaking is given by $\nu_\mathrm{max}=\sqrt{\sigma_\mathrm{UTS}/(r^2 \rho)}/(2 \pi)$ where $\sigma_\mathrm{UTS}=49\,$MPa is its ultimate tensile strength~\cite{Schuck2018,Bassi1983}. For example, for a $70$\, nm naphthalene particle with a density of $\rho=1145$\, kg/m$^3$ we find $\nu_\mathrm{max}=470\,$MHz, which suggests that nuclear spins in naphthalene can reach coherence times of $T_2 \approx 20$~s. As will be discussed in the next section, this exceeds by several orders of magnitude the required coherence times to perform matter-wave interferometry using nuclear spins.

An important aspect that could limit the achievable rotation frequencies is the absorption of photons coming from the laser used to induce the rotation. An excessive degree of heating of the internal temperature of the particle will have implications for the sublimation rate, as discussed in the next section, and for the intensity
of the emitted black body radiation, which can result in excessive decoherence and could be detrimental for interferometric applications. It is, however, noteworthy to mention that in contrast to the optically levitated setups where rotation has been demonstrated, here we are proposing a protocol where the nanoparticle is magnetically levitated and the laser is used only for a short time to induce the desired rotations. The short duration of the laser pulse together with the high degree of purity attainable in the production of naphthalene crystals, which results in low absorption rates, leads us to expect heating rates below those observed in optical traps with other materials. In the following, we present an estimation of the expected heating and conclude that, while the increase in internal temperature may not be negligible, it does not impede achieving the required rotation frequencies.

The rotation frequency $\nu_{\rm MAS}$ obeys the differential equation
\begin{equation}
    I \dot{\nu}_\mathrm{MAS}=\tau_\mathrm{opt}-\tau_\mathrm{drag},
\end{equation}
where $I$ is the moment of inertia, $\tau_\mathrm{opt}$ is the light-induced torque, and $\tau_\mathrm{drag}$ is the damping torque due to viscous interaction of the rotating nanoparticle with the surrounding gas molecules. $\tau_\mathrm{drag}$ is proportional to $\nu_\mathrm{MAS} P$, with $P$ the pressure of the gas, and the maximal rotation frequency is reached, as soon as $\tau_\mathrm{opt}=\tau_\mathrm{drag}$\cite{Novotny2018}. At frequencies much lower than this maximum, one can safely assume that $\tau_{\rm opt} \gg \tau_{\rm drag}$, and thus that the rotation frequency increases linearly in time as $\nu_{\rm MAS} = (\tau_{\rm opt}/I) t$.

For a nanoparticle that is significantly smaller than the wavelength of the incident light, the optical torque due to the birefringence of the particle is given by~\cite{Rudolph2021}
\begin{align}
   \mathbf{\tau_{\rm opt}}&=\frac{\epsilon_0 V}{2}\Re[(\chi\mathbf{E}^*)\cross\mathbf{E}] \nonumber \\
    &+\frac{\epsilon_0 k^3 V^2}{12 \pi} \Im[(\chi^2\mathbf{E}^*)\cross\mathbf{E}-(\chi \mathbf{E}^*)\cross(\chi\mathbf{E})],
\end{align}
where $V$ is the volume of the particle, and $k=2\pi/\lambda_L$ is the wave number of the laser. $\chi$ is the orientation-dependent polarizability tensor, which for an anisotropic, ellipsoidal particle is diagonal in a frame coinciding with its principal axes. In that frame, it is determined by the diagonal elements~\cite{Bohren1998}
\begin{equation}
    \chi_j=3\frac{\epsilon_j-1}{\epsilon_j+2}.
\end{equation}
Here, $\epsilon_j = n_j^2$ is the relative permittivity along the $j^\mathrm{th}$ principal axis, and $\mathbf{E}$ is the possibly complex amplitude of the electric field $\mathbf{E}(t)=\mathbf{E}\exp(-i\omega t)$. Notably, for an isotropic material, the torque is zero. To compute the torque, the complex contribution to $\chi$ will be neglected, as the absorption of naphthalene in the visible range is minimal. If the naphthalene is orientated such that the principal axes of $n$ align with the lab frame, $n$ is given by $n=\mathrm{diag}(1.442, 1.932, 1.775)$. The largest torque is achieved with circularly polarized light traveling along the $z$ axis, namely $\mathbf{E}=E_0/\sqrt{2}(1, \exp(i \pi/2), 0)$. For a spherical naphthalene particle of radius $r$, with moment of inertia $I=2/5\,m r^2$ and a laser with wavelength $\lambda_L$, we find
\begin{equation}
    \nu_\mathrm{MAS} = a \frac{r}{\lambda_L^3} \mathcal{I} t\, ,
    \label{eq:Heating_nuMAS}
\end{equation}
where $a=8.15\cdot10^{-11}\,\mathrm{m^4/(W s^2)}$ is a constant, $\mathcal{I}=c\epsilon_0|E_0|^2/2$ is the intensity of the laser and $t$ the time the particle is exposed to the laser light. 

We now proceed to estimate the amount of heating expected as a function of the rotation frequency. If heat losses due to other mechanisms like black body radiation are neglected, the heating rate is given by~\cite{Bateman2014}
\begin{equation}
    m c_m(T) \dot{T} = 8 \pi^2 \mathcal{I}  \frac{r^3}{\lambda_L} \Im\left[\frac{\epsilon-1}{\epsilon+2}\right],
    \label{eq:Heating_HeatingLaser}
\end{equation}
where $c_m(T)$ is the temperature-dependent specific heat capacity in units of $\mathrm{J/(kg K)}$.  
This differential equation can be solved by the technique of separation of variables. Plugging \cref{eq:Heating_nuMAS} in we get
\begin{equation} 
    \int_{T_i}^{T_f} c_m(T)\,\mathrm{d}T = 6 \pi \nu_\mathrm{MAS} \frac{\lambda_L^2 }{a \, r\,\rho} \Im\left[\frac{\epsilon-1}{\epsilon+2}\right].
\end{equation}

Remarkably, the expression is independent of the employed laser intensity, as the heating rate is proportional to the laser intensity while the duration of the laser pulse is inversely proportional to it. We compute specific values for the expected temperature increase, using data from Ref.~\cite{Chirico2002} to model the temperature dependence of the heat capacity of naphthalene.
To find $\epsilon=n^2$, we assume that the real part of $n$ is given by the average over the two axes involved in the rotation, therefore $\Re(n)=1.687$. On the other hand, we calculate an upper bound to the imaginary part of $n$ from one of the main absorption peaks of naphthalene in the optical range, which is found at $470$~nm and shows an absorbance of $\alpha=5\cdot10^{-4}\,\mathrm{cm}^{-1}$~\cite{Schwoerer2006a}. $\alpha$ is related to the imaginary part of the refraction index via $\Im(n)=\alpha c/(2 \omega)$, where $\omega$ is the frequency of the light. We therefore find $\Im(n_\mathrm{naph})=1.175\cdot10^{-8}$. In the neighborhood of this peak, the absorption is expected to be even lower~\cite{Port1978}. A pentacene concentration of $c_\mathrm{pen}=10^{-7}\,\mathrm{mol/mol}$ would allow for multiple pentacene molecules within a spherical naphthalene crystal of $70\,$nm.  If the effective absorption index is estimated via linear mixing, it is given by $\mathrm{Im}(n_\mathrm{eff})=n_\mathrm{naph}(1-c_\mathrm{pen})+n_\mathrm{pen}c_\mathrm{pen}$. For the individual pentacene molecules to have a significant effect, the absorption coefficient would need to be on the order of $10^{-1}$.While solid-state pentacene exhibits strong absorption near its electronic transitions at around 600 nm, the absorption of single pentacene molecules decreases rapidly. As there are no quantitative measurements of pentacene’s absorption coefficient in the blue spectral range, we refer to the work of Faltermeier et al.~\cite{Faltermeier2006}, where the extinction coefficient of pentacene is not resolvable for wavelengths shorter than 400\,nm. Based on this, we conclude that absorption of the incident laser light by the pentacene molecules can be neglected. Naturally, in experiments, the frequency of the employed laser should be suitably placed outside any absorption line width of both naphthalene and pentacene, to minimize the internal heating of the particle.  With these considerations we use $n=1.687+1.175\cdot10^{-8} i$ as an upper bound.  While this estimate is based on rough calculations, the true absorption coefficient of this hybrid system would need to be determined experimentally.

Putting everything together, we find that, for example, to achieve a $T_2$ time of $1\,$s, a spinning frequency of $23\,$MHz is required. For a particle with $70\,$nm radius that was initially prepared at an internal temperature of 5\,K this would result in its temperature increasing up to not more than 11\,K when being accelerated to this frequency using $500\,$nm light. 

\subsection{Sublimation}

Because naphthalene is a van der Waals crystal, the question arises as to whether, at the low pressures necessary to conduct matter-wave interferometry experiments, sublimation events may occur that are detrimental to the coherence of the experiment. Each time a molecule of naphthalene evaporates from the crystal, departing with some momentum $\vec p$, it induces a recoil on the levitated particle, that constitutes a form of noise on its mechanical degrees of freedom. Moreover, each evaporated naphthalene molecule carries into the environment information on the position and the velocity of the naphthalene crystal, which may also destroy the desired interference pattern. In particular, for situations where the de Broglie wavelength of the sublimated molecule is shorter than the spatial superposition of the interferometer, a single molecule loss could be enough to resolve the position of the naphthalene particle and in turn destroy the interference. Thus, we would like to estimate the rate at which naphthalene molecules detach from the bulk of the crystal and restrict each of our experimental runs to times for which the probability of a single naphthalene molecule evaporating is negligible. We assume that {the rate at which particles detach from the crystal depends only on the internal temperature of the crystal, and follow the argument presented by Tennakone and Peiris to find this rate~\cite{Tennakone1978}. 

Let us consider a bulk sphere of naphthalene with radius $r$, levitated in a completely evacuated container. Over time, some of the naphthalene will evaporate until the pressure in the chamber is the same as the vapor pressure $p_n$ of naphthalene, at which point, an equilibrium is reached between the rate of particles that leave the crystal and those that are deposited onto it. Following statistical physics considerations, one can estimate the rate at which, at a pressure of $p_n$, naphthalene molecules in the gas collide with the bulk sphere of radius $r$. This is given by
\begin{equation}
    \dot{N}=\frac{4\pi r^2 p_\mathrm{n}}{ \sqrt{2 \pi m_\mathrm{N} k_B T} },
    \label{eq:SublimationRate}
\end{equation}
where $m_\mathrm{N}=128\,\mathrm{a.m.u.}=2.1\cdot10^{-25}\,$kg is the mass of a single naphthalene molecule and $T$ is its temperature, which is also the temperature of the bulk, as the situation is described at equilibrium. If all incident particles were to get deposited onto the surface of the naphthalene nanoparticle, \cref{eq:SublimationRate} would also constitute the rate at which naphthalene molecules sublimate from the crystal. While this is not true in general, \cref{eq:SublimationRate} provides an upper bound to the real rate of sublimation and allows us to make a conservative estimation of the available time before a single molecule escapes (or gets deposited onto) the crystal, as a function of its temperature and size. 

The relation between the vapor pressure of naphthalene and its temperature is found to be well described by the equation ~\cite{Ruzicka2005} 
\begin{equation}
\ln \frac{p_n}{p_0} = \left(1 - \frac{T_0 / K}{ T / K}\right) \exp (\sum_{i=0}^n A_i (T / K)^i),
\label{eq:vaporPressure}
\end{equation}
where the parameters $A_i$ need to be fitted by experiment,
and $p_0$ is the vapor pressure at an arbitrarily chosen temperature $T_0$. As a reference, we use the values at room temperature, $p_0 = 13.08$~Pa and $T_0 = 300$~K. As for the parameters $A_i$, the values $\{A_0=3.272310, A_1=- 2.663498 \cdot 10^{-4}, A_2= - 2.929123 \cdot 10^{-9} \}$ for the first three orders provide a good fit in the temperature range from $150$~K to the triple point at $353.37$~K~\cite{Ruzicka2005}.

With this, we can compute that, for example, at room temperature the emission rate of naphthalene molecules from a particle of $70$~nm is $\dot N \approx 10^{10}$~s$^{-1}$ which would limit the duration of a single experimental run to $10^{-10}$~s. This naturally rules out any option of performing interferometry experiments without cooling the internal temperature of the naphthalene nanoparticle. However, this rate exhibits a dramatic dependence on temperature. At $150$~K, which is the lowest temperature for which \cref{eq:vaporPressure} provides a reliable value,} the emission frequency is $\dot{N}=40$~s$^{-1}$, extending the available time for each experimental run to several milliseconds. 

Moreover, for matter-wave interferometry experiments, internal temperatures on the order of a few Kelvin will be required to reduce noise coming from the emission of black-body radiation, as will be explained in the next section. At these temperatures, sublimation rates are expected to be even lower by orders of magnitude. Thus, we do not expect sublimation to be a factor limiting the coherence of matter-wave interferometry experiments with naphthalene.

On the other hand, any experiment will consist of a statistically significant number of runs that allow the reconstruction of the observables of interest. Thus, while one could guarantee that on a single experimental run, no naphthalene molecules are sublimated, it is to be expected that the total experimental time will extend over periods where the emission rate of naphthalene particles is not negligible. This could lead to a drift of the mass of the particle as the experiment progresses. To understand the extent of this mass variation, we compute the rate of change of the total mass of a naphthalene particle of radius $r$ relative to its initial mass $M_0$. From \cref{eq:SublimationRate} we see that, at pressures below the vaporization pressure, the rate of mass loss is upper bounded by
\begin{equation}
\frac{d}{dt} \left(\frac{M}{M_0}\right) = - \sqrt{\frac{m_N}{2 \pi k_B T}} \frac{3 p_N}{\rho \, r},
\end{equation}
where $\rho= 1.145 \cdot 10^{3}$~kg/m$^3$ is the density of naphthalene. For a particle of $70$~nm, at a temperature of $150$~K, this results in a relative mass that varies in time as $M/M_0 = 1 - 1.8 \cdot 10^{-10} t$, which suggests that even on a time scale of hours the mass of the nanoparticle would not change beyond its 7th significant digit. We, thus, do not expect this to affect the experiment. 

%------------------------------------------------

\section{Naphthalene in a magnetic field  gradient}
\label{sec:Interferometer}
This section describes the dynamics of a diamagnetic particle containing N embedded nuclear spins in the presence of a magnetic field gradient. The magnetic field gradient induces a harmonic confinement of the center of mass of the diamagnetic particle and, at the same time, its interaction with the spin system leads to forces whose strength and direction depend on the state of the spin ensemble. These forces can be used to manipulate the motion of the particle and in particular to expand its position variance and enhance its sensitivity to external perturbations \cite{pedernales2023origin,Cosco2021,Weiss2021}.  We introduce a protocol based on microwave pulses that manipulates precisely the state of the spin ensemble, allowing us to engineer dynamics that involve a rapid expansion of the center of mass wave packet and its subsequent reversal. Probing the coherence of such expand-and-recombine dynamics, which can be done through the measurement of the spin state at the end of the protocol, allows us to test the linearity of quantum mechanics at the scales of the employed particle mass and the reached size of the wave function. This constitutes an instance of a matter-wave interferometry protocol, that is, an interferometer that harnesses the wave character of matter. The novelty with respect to other matter-wave interferometry protocols based on spin-dependent forces is the presence of multiple spins, which boosts the spread of the wave function, resulting in shorter experimental times and therefore increased robustness to several sources of noise. We discuss the scaling of the different figures of merit with the number of spins N. Moreover, we explore the intricacies of the protocol, propose a way to measure the spin state through the position of the nanoparticle, and analyze the resilience of the protocol against various sources of noise.

\subsection{Advantages of Naphthalene for Stern-Gerlach type interferometry}

Nuclear spins in naphthalene present a promising pathway for creating macroscopic superpositions of the nanoparticle in position space. While current proposals for Stern-Gerlach-type macroscopic interferometers often rely on single electron spins, such as NV centers in diamond~\cite{Yin2013, Wood2021}, here we argue that the use of nuclear spin ensembles presents an intriguing alternative. Although the gyromagnetic ratio of an electron spin is roughly three orders of magnitude higher than that of a nuclear spin, the collective strength of approximately $6\cdot10^7$ polarized hydrogen spins in a 70\,nm naphthalene particle exceeds that of an individual electron spin in the same magnetic field gradient.

Furthermore, the homogeneous distribution of nuclear spins within naphthalene presents a distinct advantage over systems reliant on single-electron spins. In traditional setups, precise positioning of the spin on the nanoparticle's center of mass is necessary to prevent undesired rotations of the entire particle. However, in naphthalene, the evenly dispersed nuclear spins exert a collective force that acts uniformly on the nanoparticle, avoiding undesired rotational effects. 

On the other hand, particles containing embedded spins with zero-field splitting, such as the NV center in diamond, are prone to additional unwanted rotations. For particles with a zero-field splitting the axis of quantization is not determined by the direction of the external magnetic field but rather by the spin direction that diagonalizes the zero-field Hamiltonian. When this quantization axis deviates from the direction of the magnetic field, it tries to realign, exerting a force that induces a torque on the particle. This presents a challenge for experiments leveraging spin-dependent forces for the generation of spatial superposition. Matter-wave interferometry experiments will generally require preparing the spin in a superposition of states $\ket{\uparrow}$ and $\ket{\downarrow}$. Unless the NV center starts exactly at a position of vanishing magnetic field, one of the superposed states will be aligned parallel to the magnetic field, while the other will be in an anti-parallel orientation. Note that generally the nanodiamond will be located at a position with non-vanishing magnetic field, due to the gravitational pull which shifts the equilibrium position of the oscillator (see next section). In this configuration, any small deviations from this perfect alignment between the NV axis and the magnetic field will induce torques dependent on the spin state, which will eventually lead to entanglement between spin and rotational degrees of freedom. In contrast, nuclear spin-$\frac{1}{2}$ systems lack such splitting. Any potential zero-field splitting observed in the nuclear spins of naphthalene arises from dipole-dipole interactions among nuclear spins, which in our setup are suppressed by the implementation of magic-angle spinning. Moreover, the rapid rotation of the nanoparticle contributes to the stabilization of its rotational degrees of freedom, akin to magnetic tops, increasing the robustness of the setup against unwanted torques and rotations~\cite{Berry1996, Gov1999}.

\subsection{Time evolution in a magnetic trap}
\label{subsec:minimalProtocol}
\iffalse

\begin{table*}[t]
  \centering
  \begin{tabular}{| c | c | c | c | c |  c |}
   \hline
    constants & $\chi_V$ & $\rho$ in kg/m$^3$ & $\sigma_\mathrm{UTS}$ in MPa & H density in 1/m$^3$\\
    \hline
     & $7.33\cdot10^{-7}$ & 1145 & 49 & $4.3\cdot 10^{28}$ \\
    \hline
    \hline
    parameters & $r$ in nm & $B^\prime$ in T/m & $\Omega$ in s$^{-1}$ & $T_\mathrm{tot}$ in ms \\
    \hline
    a) & $70$ & $5\cdot10^4$ & $1.1\cdot10^3$ & 11 \\
    \hline
    b) & $45$ & $1.88\cdot10^5$ & $4.2\cdot10^3$ & 3 \\
    \hline
  \end{tabular}
  \caption{Material constants of naphthalene and parameters used.}
  \label{tab:parameters}
\end{table*}

\fi

\begin{table*}[t]
  \centering
  \begin{tabular}{| c | c | c | c |  c |}
   \hline
    $\chi_V$ (CGS) & $\chi_V$ (SI)  &  $\rho$ in kg/m$^3$ & $\sigma_\mathrm{UTS}$ in MPa & H density in 1/m$^3$\\
    \hline
     -$7.33\cdot10^{-7}$ & -$9.21\cdot10^{-6}$  & 1145 & 49 & $4.3\cdot 10^{28}$ \\
    \hline
  \end{tabular}
  \caption{Material constants of naphthalene.}
  \label{tab:parameters}
\end{table*}

In this section, we analyze the dynamics of a hyperpolarized naphthalene sphere confined within a diamagnetic trap. Diamagnetic levitation exploits the fact that diamagnetic materials are repelled by magnetic fields, naturally tending toward regions of minimal magnetic field strength. The potential energy density $E(\vec{r})$ of a particle in a magnetic field $\vec{B}(\vec{r})$ and subject to earth's gravity, at position $\vec{r} = (x, y, z)$, is given by
\begin{equation}
    E(\vec{r}) = -\frac{\chi_V}{2 \mu_0} \vec{B}(\vec{r})^2 + \rho g x,
\end{equation}
where $\mu_0 = 4 \pi \cdot 10^{-7}~\mathrm{H/m}$ is the vacuum permeability, $\chi_V$ is the volume magnetic susceptibility, $\rho$ is the mass density of the particle, and $g$ is the gravitational acceleration. Here, and throughout the rest of the manuscript, the coordinate frame is chosen such that gravity points in the $x$-direction. The equilibrium position for levitation is determined by the condition $\vec{F}(\vec{r}) = -\nabla E(\vec{r}) = 0$, where $\vec{F}(\vec{r})$ is the force density acting on the particle. For the $x$-component, this leads to the standard levitation condition for diamagnetic trapping:
\begin{equation}
    B_0 \frac{\partial B}{\partial x} = \mu_0 g \frac{\rho}{\chi_V}.
\end{equation}
For naphthalene, with $\rho = 1145~\mathrm{kg/m^3}$ and $\chi_V = -9.21 \cdot 10^{-6}$~\cite{Haynes2014}, the condition requires $B (\partial B / \partial x)$ to be on the order of $1500~\mathrm{T^2/m}$, which is within experimental feasibility~\cite{Naito2024}.
Furthermore we would like to point out that the static contribution of the magnetic field is determining the equilibrium position of the levitated particle, while the gradient determines the trap frequency.
In principle, other levitation methods for naphthalene could be considered. These include charging the particle and trapping it in a Paul trap, utilizing its spin angular momentum to stably levitate it in a static magnetic field~\cite{Rusconi2017}, or potentially, hyperpolarizing the nuclear spin lattice and using the Meissner effect to levitate it above a superconductor. In this work, we focus solely on diamagnetic levitation and leave the exploration of alternative methods for future investigations.

The Hamiltonian describing the center-of-mass motion and spin dynamics of a naphthalene nanoparticle of mass $m$ containing $N$ nuclear spins, each with gyromagnetic ratio $\gamma_p$, and embedded in a magnetic field gradient of the form $\vec B(x)= (B' x + B_0) \hat k$, with $\hat k$ a unit vector in the $z$ direction, can be expressed as
\begin{align}
    \hat{H}=&\frac{\hat{p}^2}{2 m}-\frac{\chi_V V (B_0+B'\hat{x})^{2}}{2 \mu_0}\nonumber \\
    &+\frac{\hbar}{2}\gamma_p (B_0+B^\prime \hat{x}) \sum_{n=1}^N \hat{\sigma}_z^{(n)}.
    \label{eq:Hamiltonian0}
\end{align}
Under the coordinate transformation $x \, \rightarrow
\, x - \frac{B_0}{B'}$, and a suitable rearranging of the terms, the Hamiltonian can be rewritten as
\begin{align}
    \hat{H}=&\frac{\hat{p}^2}{2 m}+\frac{m \Omega^2}{2}\left(\hat{x}+ \chi \sum_{n=1}^N \hat{\sigma}_z^{(n)} \right)^2 \nonumber \\
    &-\eta \left(\sum_{n=1}^N \hat{\sigma}_z^{(n)}\right)^2
    \label{eq:Hamiltonian},
\end{align}
where we have introduced the parameters
\begin{align}
&\Omega \equiv \sqrt{\frac{|\chi_V|}{\rho \mu_0}}B^\prime, \qquad \chi \equiv \frac{\hbar \gamma _p \mu_0}{2 |\chi_V| V B^\prime} \quad \mathrm{and} \quad \nonumber \\ 
&\eta \equiv \frac{\hbar^2\gamma_p^2\mu_0}{8|\chi_V| V}.
\label{eq:HOparameters}
\end{align}
In this form, it becomes clear that the center of mass oscillates with frequency $\Omega$ around an equilibrium position that is displaced from the origin by an amount $x_{\rm eq} = - \chi \sum_{n=1}^N \hat{\sigma}_z^{(n)}$. This equilibrium position depends on the state of the spin ensemble. We would like to point out that the trap frequency $\Omega$ solely depends on material constants and the magnetic field gradient. With the values provided in \cref{tab:parameters}, we find $\Omega\approx 0.08\,\mathrm{m/(Ts)}\cdot B'$ and therefore trap frequencies on the order of tens to hundreds of Hertz for gradients in the range of $10^3-10^4\, \mathrm{T/m}$. The magnetic field assumed here is, of course, not a physical field as it does not satisfy with Maxwell's equations, which require that $\vec{\nabla}\cdot\vec{B}=0$. Therefore, a physical fields needs at least another contribution along the $y$ axis. In \cref{App:Spin dynamics in physical magnetic field} we discuss why this contribution can be neglected.

For clarity, we will presume that the spin ensemble is initially fully polarized---any implications of imperfect polarization are discussed in \cref{sec:Measurement schemes}. Hence we can assume that all spins are initially arranged in their lowest energy state, aligning uniformly in a single direction. Subsequently, we apply a $\pi/2$-pulse, mapping each spin into a superposition of spin up and spin down. The precision with which such a pulse can be delivered to all the spins in the ensemble is discussed in \cref{sec:Pulse}. Assuming that the center-of-mass motion is prepared in a thermal state $\hat \rho_{\rm th}$, the resulting state of the combined system is given by  
\begin{equation}
    \hat{\rho}=\hat{\rho}_\mathrm{th}\otimes\frac{1}{2^{N}}\sum_{\kappa, \kappa'=0}^N \sqrt{\binom{N}{\kappa}\binom{N}{\kappa'}}\ket{\kappa}\bra{\kappa'},
    \label{eq:InitialState}
\end{equation}
where $\ket{\kappa}$ is the $\kappa^\mathrm{th}$ Dicke state, defined as the linear combination of all spin states with $\kappa$ spins up
\begin{equation}
\sqrt{\binom{N}{\kappa}} \ket{\kappa}=\ket{\uparrow^{\otimes \kappa} \downarrow^{\otimes N-\kappa}}+\mathrm{permutations}.
\end{equation}
Hamiltonian (\ref{eq:Hamiltonian}) preserves the Dicke state, as this is an eigenstate of the operator $\sum_{n=1}^N \sigma_z^{(n)}$, and thus, of the Hamiltonian. Therefore, provided that the initial state is contained in the subspace spanned by the Dicke states, the state of the system at any time $t$ can be expressed in terms of Dicke states as
\begin{equation}
    \hat{\rho}(t)=\frac{1}{2^{N}}\sum_{\kappa, \kappa'=0}^N \sqrt{\binom{N}{\kappa}\binom{N}{\kappa'}}\hat{\rho}_{\kappa,\kappa'}(t)\otimes\ket{\kappa}\bra{\kappa'}.
    \label{eq:AnsatzRho}
\end{equation}

In the following, we want to find a differential equation describing the time evolution of the spatial components of the density matrix $\hat{\rho}_{\kappa,\kappa'}(t)$. The time evolution of $\hat{\rho}(t)$ is given by von Neumann's equation $\dot{\hat{\rho}}=-\frac{i}{\hbar} [\hat{H},\hat{\rho}]$ and with Ansatz \ref{eq:AnsatzRho}, we find
\begin{align}
     \dot{\hat{\rho}}_{\kappa,\kappa^\prime}&=-\frac{i}{\hbar}\left\{ \hat{H}_{\kappa} \hat{\rho}_{\kappa,\kappa^\prime} -  \hat{\rho}_{\kappa,\kappa^\prime} \hat{H}_{{\kappa^\prime}}  + \eta_{\kappa, \kappa'} \hat{\rho}_{\kappa,\kappa^\prime}\right\}
     \label{eq:rhoKappaKappapWithoutNoise}
\end{align}
with
\begin{align}
    \hat{H}_{\kappa}&\equiv\frac{\hat{p}^2}{2 m}+\frac{m \Omega^2}{2} [ \hat{x} + (2\kappa-N)\chi ]^2
\end{align}
and
\begin{align}
    \eta_{\kappa, \kappa'} &\equiv \eta \left[(2\kappa'-N)^2-(2\kappa-N)^2\right],
\end{align}
and which has the solution
\begin{equation}
\hat \rho_{\kappa, \kappa'} (t) = e^{- \frac{i}{\hbar} \hat H_\kappa t} \hat \rho_{\kappa, \kappa'}(0) e^{\frac{i}{\hbar} \hat H_{\kappa'} t} e^{-\frac{i }{\hbar} \eta_{\kappa, \kappa'} t}.
\end{equation}

The evolution of the initial state in \cref{eq:InitialState} can be described as follows. At first, the mechanical and spin degrees of freedom are in a product state, however, as the evolution progresses the initial center-of-mass wave packet splits into $N+1$ trajectories, each correlated with one of the Dicke states, which are parameterized by $\kappa\in [0,N]$. All trajectories follow a harmonic motion, albeit each oscillating around a different equilibrium position. Thus to each Dicke state, we can associate an equilibrium position $x_{\rm eq, \kappa} = (2 \kappa-N) \chi$.  It is not difficult to show that at any given time, one will find the $N+1$ wave packets evenly distributed along dimension $x$ with the nearest-neighbor spacing given by $2\chi \sin^2(\Omega t/2)$. 

\begin{figure}[ht]
    \centering
    \includegraphics[width=\linewidth]{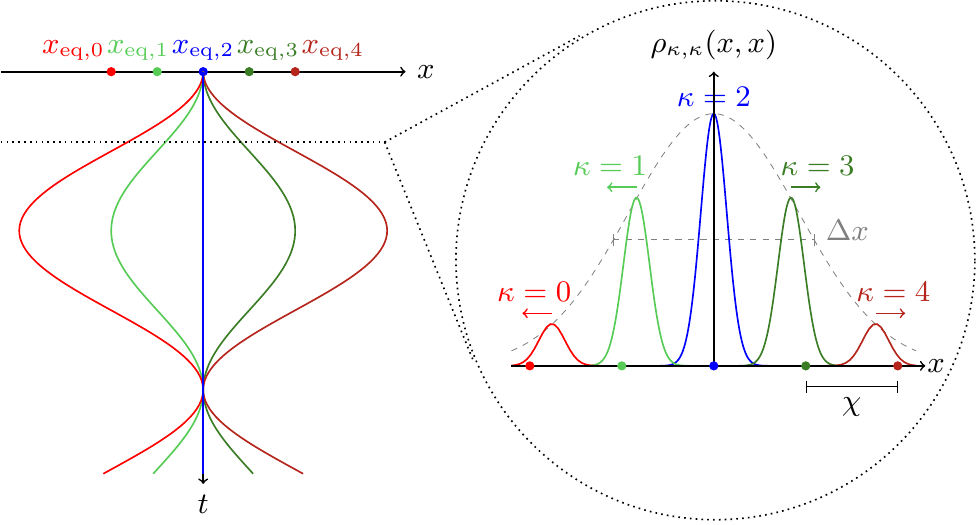}
    \caption{Here, the time evolution of the $\kappa$ wave packets in a magnetic field gradient is displayed. The $\kappa^\mathrm{th}$ wave packet oscillates around the equilibrium position $x_\mathrm{eq,\kappa}$, as depicted on the left. After a full period of $2\pi/\Omega$, all trajectories meet again in the center of the trap. The equilibrium positions $x_\mathrm{eq,\kappa}=(2\kappa-N)\chi$ are spaced by $\chi$. If the initial spin state is given by \cref{eq:InitialState}, the wave packets corresponding to $\kappa$ are in a superposition, as depicted on the right. The Gaussians are displaced by $x_\mathrm{eq,\kappa}[1-\cos(\Omega t)]$ and enclosed in an envelope that follows a binomial distribution. The width of the envelope is given by \cref{eq:PositionVariance}.}
    \label{fig:Dynamics}
\end{figure}

The specific shape of each trajectory associated with a Dicke state $\kappa$ depends on the initial position of the particle. If the particle is cooled while the spins are fully polarized, that is, when the spin ensemble is in the Dicke state $\kappa = N$—then the particle will be prepared in the ground state (or a low temperature thermal state) at the corresponding equilibrium position, $x_{\rm eq, N}$. As a result, after the first $\pi/2$ pulse, the trajectory corresponding to $\kappa=N$ will remain centered around its equilibrium position, while the other trajectories, associated with $\kappa<N$, will begin to oscillate with amplitudes that grow with the distance to their equilibrium positions. Not all trajectories are equally likely, however. According to state~(\ref{eq:InitialState}) the probability of the spin ensemble being in Dicke state $\kappa$ is given by $p_{\kappa}=\binom{N}{\kappa}2^{-N}$. Therefore, it is desirable for the trajectory corresponding to $\kappa=N/2$ which is the most probable to remain near equilibrium, while the less probable trajectories fan out symmetrically to the left and right, as depicted in \cref{fig:Dynamics}. This would ease the requirement for maintaining a linear magnetic field gradient across the region where the particle is most likely to be found. 

To realize this configuration, we propose continuously driving the nuclear spins with a Rabi frequency $\Omega_{\mathrm{M}}$ that is higher than the oscillation frequency of the mechanical oscillator~$\Omega$. 
The effective potential experienced by the particle can then be described by a harmonic oscillator whose center is given by $w(t) = N \chi \cos(\Omega_{\mathrm{M}} t)$, such that the differential equation governing the center-of-mass position~$q(t)$ of the particle reads
$m \ddot{q}(t) + m \Omega^2 \big[q(t) - w(t)\big] = 0$.
We assume that the particle has already settled into its long-time behavior, described by the particular solution
$q(t) = N \chi \, \Omega^2/(\Omega^2 - \Omega_{\mathrm{M}}^2) \cos(\Omega_{\mathrm{M}} t)$,
which is referred to as micromotion. The amplitude of the micromotion is given by
$\left| N \chi \, \Omega^2/(\Omega^2 - \Omega_{\mathrm{M}}^2) \right| \approx \left| N \chi \, \Omega^2/\Omega_{\mathrm{M}}^2 \right|$
for $\Omega^2 \ll \Omega_{\mathrm{M}}^2$. Using the parameters listed in \cref{tab:parameters}, we find
$N \chi \, \Omega^2/\Omega_{\mathrm{M}}^2 \approx (0.53\,\mathrm{m^2s^{-2}/T}) \cdot B'/\Omega_{\mathrm{M}}^2.$
For a gradient of $10^4\,\mathrm{T/m}$ and a Rabi frequency of 100\,kHz we find a resulting oscillation amplitude of approximately 13\,nm respectively. To start the interferometry protocol, the driving field is first turned off, and the sequence of pulses described above is applied. If this is done at an arbitrary time, the particle may already possess some initial momentum or displacement within the trap. In that case, its resulting motion will be the sum of two contributions: the motion induced by the pulse sequence for an initially stationary and centered particle, and the free motion corresponding to its initial state in the absence of any pulses. Let us consider the case when the particle is maximally displaced at the moment of starting the protocol, and hence, has no momentum. In this case, it oscillates in the mechanical trap with an amplitude as discussed above and frequency $\Omega$. This implies that the trajectories spread over a spatial extent that is larger than the ideal case by $13\,\mathrm{nm}$, depending on the applied Rabi frequency. For a gradient of $10^4\,\mathrm{T/m}$, the variation of the magnetic field across this space is now larger by $0.26\,\mathrm{mT}$. Alternatively, if the drive stops when the particle is at the trap center, it has a maximum momentum 
$p(0) = m N \chi \Omega^2 / \Omega_\mathrm{M}$, corresponding to an oscillation amplitude of $p(0)/(m \Omega)=6.6/\Omega_\mathrm{M}$. For $\Omega_\mathrm{M}=100\,\mathrm{kHz}$, this adds to each trajectory oscillations with amplitude $10\,\mu\mathrm{m}$, which results in an additional field variation of $0.2\,\mathrm{T}$. This requirement may be demanding, but timing the drive to stop at maximum displacement and using the largest possible Rabi frequency should, in principle, suffice to keep the particle within the linear gradient region.

In this regime, the nuclear spins become effectively decoupled from the oscillation mode, and the particle will settle near the equilibrium position corresponding to $\ket{\kappa = N/2}$, with a deviation of about $10\,\mathrm{nm}$.
While continuous spin driving may induce polarization loss through $T_2$ relaxation, the associated timescale is expected to be on the order of seconds under magic-angle spinning. The thermalization time is highly dependent on the specific realization of the experiment, but should remain shorter than this timescale to avoid significant losses in polarization.

 Shortly after stopping the drive, the initial $\pi/2$ pulse can be applied, or equivalently, the drive could be stopped when the spins have rotated by a suitable angle corresponding to a $\pi/2$ rotation with respect to their initial state. With this modification, the trajectories would evolve as depicted in~\cref{fig:Dynamics}. The drive can either be applied while hyperpolarizing the spin ensemble during polarization within the trap, or, if the particle is hyperpolarized outside the trap, the drive can be applied immediately after placing the particle in the trap.

The variance in the position of the center of mass at a given time $t$ can be computed to be
\begin{equation}
\langle \Delta x \rangle^2 = \langle \Delta x_{\rm th} \rangle ^2 + \left(2 \chi \sin^2(\Omega t/2)\sqrt{N}\right)^2,
\label{eq:PositionVariance}
\end{equation}
where $\langle \Delta x_{\rm th} \rangle$ is the position variance of the initial thermal state. Equation~(\ref{eq:PositionVariance})  shows the enhancement that can be achieved in the expansion of the center-of-mass wave function from a larger number of spins. In comparison to matter-wave interferometry based on a single electron spin, such as an NV-center in a diamond nanoparticle, we observe an enhancement factor of $\sqrt{(\expval{\Delta x}^2-\langle \Delta x_{\rm th} \rangle ^2)/(\expval{\Delta x}^{(NV)}-\langle \Delta x_{\rm th} \rangle ^2)}\approx 1.5\cdot10^{-3} \sqrt{N}$. Here, the factor of $1.5\cdot 10^{-3}$ originates from the ratio between the gyromagnetic ratios of the hydrogen and electronic spins. A naphthalene nanoparticle with a radius $100$~nm contains approximately $10^8$ nuclear spins, resulting in a 15-fold enhancement of the achievable variance at any given time over that achievable with a single electron spin. For the parameters in \cref{tab:Overview} a), one can achieve an expansion of the wave packet of $\langle \Delta x \rangle^2-\langle \Delta x_\mathrm{th}\rangle ^2 = (167\,\mathrm{nm})^2$ after just 0.5\,ms, a position variance that is larger than its radius. 

For matter-wave interferometry with spin-dependent forces, it is often desirable to revert the expansion of the wavepacket and return the system to its initial state where spin and mechanical degrees of freedom are decoupled from each other. Then the coherence of the process can be tested through a measurement of the spin system, certifying also the spatial coherence of the superposition achieved during the protocol. For the setup described in this section, this decoupling occurs without additional intervention at times that are natural multiples of the trap period, a time at which all the trajectories return simultaneously to their initial position, see Fig.~\ref{fig:Dynamics}. Nevertheless, with a suitable manipulation of the spin state, this decoupling can be achieved at times shorter than a single period of the trap, in this way reducing the influence of detrimental sources of noise \cite{Martinetz2020}. While this also reduces the size of the prepared superposition, it allows us to adjust the protocol to the available coherence time.

\begin{table*}[t]
  \centering
  \begin{tabular}{| c | c | c | c | c | c | c | c | c | c | c | c |}
   \hline
     & $r$ [nm] & $N$ &  $B' [\frac{\mathrm{T}}{\mathrm{m}}]$ & $T_\mathrm{tot}$ [ms] & $\chi$[m] &  $\frac{\langle \hat{M} \rangle (T_\mathrm{tot}) }{ \langle \hat{M} \rangle (0)}$ & $\frac{\Omega}{4 \lambda}$ & $\Delta x_\mathrm{dis}$ [m] & $P_\mathrm{flip}$ & $T_\mathrm{i, max}$ [K] \\    
    \hline
    a) &  44 & $1.5\cdot 10^7$ & $10^4$ & 7.8 & $5.4\cdot10^{-10}$ &   0.956 & 0.4 & $6.6\cdot10^{-4}$ & 0.04 & 4\\
    \hline
    b) &  44 & $1.5\cdot 10^7$ & $10^3$ & 78 & $5.4\cdot10^{-9}$ &  0.65 & 0.1 & $56\cdot10^{-3}$ & 0.3 & 4\\
     \hline
    c) &  105 & $2.1\cdot10^8$ & $10^5$ & 0.77 & $4\cdot10^{-12}$ &  0.61 &  4.3 & $6.5\cdot10^{-4}$ & 0.05 & 9\\
    \hline   
  \end{tabular}
  \caption{Overview of some critical specifications of the protocol for different sets of parameters. The parameters are chosen in such a way, that $\phi=0$ and an interference fringe can be resolved. It was assumed that the only noise acting on the state where spontaneous collapses with $\lambda_\mathrm{CSL}=10^{-16}$ and $r_\mathrm{CSL}=10^{-7}$. While the proposed field gradients are challenging, they are in principle within experimental feasibility \cite{Naito2024, Rugar2004}.}
  \label{tab:Overview}
\end{table*}

\begin{figure}
\centering
\includegraphics{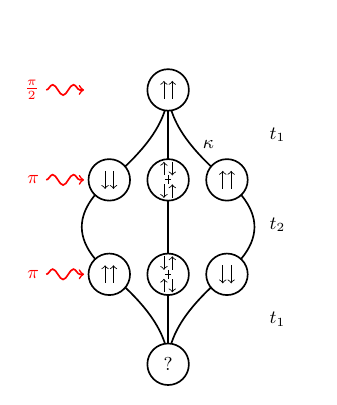}
\caption{Protocol for rapid expansion and recombination of the wave function of the center-of-mass motion in naphthalene. A hyperpolarized particle is placed in a magnetic gradient, forming a harmonic trap. Then a $\pi/2$ pulse is applied. The spin state is now a superposition of the $N+1$ possible Dicke state. Depending on the spin state, the particle travels along another trajectory and the wave function expands for a time $t_1$. By applying two well-timed $\pi$ pulses spaced by $t_2$, all trajectories recombine after another time $t_1$. The trajectories are denoted by $\kappa\in[0,N]$. To better illustrate the intrinsic dynamics of the protocol, we consider an initial particle position at $\kappa = N/2$ rather than at the equilibrium position $\kappa = N$. In this case, the center of the wave packet remains stationary.}
\label{fig:ProtocolOld}
\end{figure}

First, one lets the wave function spread for a time $t_1 < \pi/\Omega$---if the expansion time is larger than half of an oscillation period, then the fastest way to decouple the spin and center-of-mass motion is to leave the system complete one full period. At this time, a microwave $\pi$-pulse is applied on the spins, causing them to flip their state, which for the Dicke states translates into the transformation $\kappa \rightarrow N - \kappa$. This reverts the force acting on each of the Dicke states, slowing down the expansion of the position variance and eventually leading to its compression, see Fig.~(\ref{fig:ProtocolOld}). A second $\pi$-pulse is required to guarantee that all the superposed wavepackets overlap not only in position but also in momentum. The spacing between the first and second $\pi$-pulses depends on the initial spreading time $t_1$ and is given by
\begin{equation}
t_2 (t_1) = \frac{2}{\Omega} \arcsin(\frac{\sin(\Omega t_1)}{\sqrt{5 - 4 \cos(\Omega t_1)}}).
\label{eq:t2}
\end{equation}
Finally, waiting for a time $t_1$ after the second $\pi$-pulse, all trajectories recombine and the position and spin degrees of freedom are decoupled again. Thus, for such a protocol the total time is $T_{\rm tot}(t_1) = 2 t_1 + t_2(t_1)$.

While at the end of the protocol, all the superposed trajectories overlap and the spin and motion decouple, the state of the system is not the same as that at the beginning of the protocol, because each trajectory has followed a different path in phase space and therefore acquired a different phase. The phase picked up by each trajectory $\kappa$, sometimes referred to as the geometric phase, is given by 
\begin{equation}
\phi_\kappa = (2\kappa-N)^2 [\left(\chi/2x_0\right)^2 \sin(\Omega t_1) + \frac{\eta}{\hbar} T_{\rm tot} ]
\label{eq:RelativePhase}
\end{equation} 
where $x_0=\sqrt{\hbar/(2 m \Omega)}$ is the width of the ground state. A detailed derivation of this phase is given in \cref{App:DecoherenceTwoPaths}.
Note that the phase is the same for trajectories that are symmetric about the origin, that is, for trajectories $\kappa=0$ and $\kappa=N$, for trajectories $\kappa=1$ and $\kappa=N-1$, and so on. Finally, provided that the whole process is coherent, at the end of the protocol,  the system is found in a state of the form $\hat{\rho}=\hat{\rho}_\mathrm{th}\otimes \ket{\psi_\mathrm{f}}\bra{\psi_\mathrm{f}}$ with
\begin{align}
    \ket{\psi_\mathrm{f}}=&\frac{1}{2^{N/2}}\sum_{\kappa=0}^N \sqrt{\binom{N}{\kappa}} e^{-i \phi_\kappa} \ket{\kappa}.
    \label{eq:FinalState1}
\end{align}
Due to the different phases picked up by each Dicke state, the spin ensemble at the end of the protocol is in general entangled. We emphasize that the proposed protocol remains valid regardless of whether the initial state is in the ground state or in a thermal state. This robustness is a general feature of Stern--Gerlach-type experiments and stems from the fact that the spin and motional degrees of freedom decouple at the end of the protocol~\cite{Bose2013, Kim2016}.

A breakdown of the linearity of quantum mechanics on the scales at which the protocol is performed would manifest as a deviation of the final state from that in \cref{eq:FinalState1}. This remains true even in cases where the thermal variance is large and dominates the total variance of the interferometer, since the spin degrees of freedom are sensitive to decoherence acting on each pure state that contributes to the mixed state.

Conversely, experimental verification of the final state in \cref{eq:FinalState1} would constitute proof of the validity of quantum mechanics at such scales. To provide a more quantitative account of such a possibility, in the next section, we compute the evolution of the system in the presence of, arguably, the most widely analyzed model of wavefunction collapse at macroscopic scales, namely, the Continuous Spontaneous Localization (CSL) model.  

\subsection{Test of the CSL model}\label{sec:Test of the CSL model}

Motivated by the aspiration to provide a unified framework to describe micro/quantum systems and macro/classical systems and to solve the so-called measurement problem in quantum mechanics, collapse models postulate modifications to the Schr\" odinger equation~\cite{Bassi2013}. These modifications are, typically, in the form of non-linear and stochastic extensions, whose effects manifest for systems of large mass and states of wide spatial delocalization. In contrast, for small masses and well-localized systems, these modifications are negligible, and spontaneous collapse models converge to the standard Schr\" odinger equation. In this sense, they provide a framework to describe a progressive breakdown of the linearity of quantum mechanics with increasing system mass.  Generally, the effect of these modifications is to induce spontaneous collapses of the wavefunction onto spatially well-localized states. The Continuous Spontaneous Localization (CSL) model is perhaps the most general and widely studied of the collapse models. It postulates a localization of the wavefunction in position space which depends on two free parameters, $\lambda_\mathrm{CSL}$ and $r_\mathrm{CSL}$. $\lambda_\mathrm{CSL}$ has units of s$^{-1}$ and relates to the frequency of the collapses, while $r_\mathrm{CSL}$ has units of m and relates to the length scale of the localization. The dynamics of the CSL model at the level of the density matrix, that is, after averaging over a large number of stochastically occurring collapses, is given, in coordinate representation, by~\cite{Bassi2013, Pearle1989, Ghirardi1990}

\begin{align}
    \dot\rho(q',q'')=&-\frac{i}{\hbar}\bra{q'} [\hat{H},\hat{\rho}]\ket{q''} \nonumber \\
    &-\xi \left(1-e^{-\frac{(q'-q'')^2}{4 r^2_\mathrm{CSL}}}\right)\rho(q',q''),
    \label{eq:CSLModelPosSpace}
\end{align}
where $ \rho(q',q'') = \bra{q'} \hat \rho \ket{q''}$ with $\ket{q}$ the position eigenstate, and $\xi$ is a coefficient that depends on the free parameters $\lambda_{\rm CSL}$ and $r_\mathrm{CSL}$, the mass and the shape of the system. For a spherical particle of radius $r$ and mass $m$, $\xi$ is given by~\cite{Nimmrichter2014}
 \begin{equation}
     \xi=\frac{m^2}{m_0^2}\lambda_\mathrm{CSL}f\left(\frac{r}{r_\mathrm{CSL}}\right) 
\end{equation}
with
\begin{equation}
    f\left(x\right)=\frac{6}{x^4}\left[ 1- \frac{2}{x^2} + \left( 1+\frac{2}{x^2} \right) e^{-x^2}\right],
 \end{equation}
and $m_0$ is a reference mass, typically, taken to be the mass of a nucleon.
 
 Equation \ref{eq:CSLModelPosSpace} indicates that the effect of the collapse model is to suppress the coherence between different position eigenstates. The coherence between spatial points $(q',q'')$ that are further away from each other decays faster than that between closer points. In the limit $|q'-q''|\ll r_\mathrm{CSL}$, the decay rate of the coherences tends to $\xi(q'-q'')^2/(4 r^2_\mathrm{CSL})$, while between points that are far away from each other, the decay rate converges to $\xi$.

\subsubsection{Interferometry in the presence of the CSL model}

Our goal is to use the interferometric protocol introduced in the previous section to provide bounds for the values that the free parameters of the CSL model, $\lambda_\mathrm{CSL}$ and $r_\mathrm{CSL}$, can take. To that end, we compute the state of the spin ensemble at the end of the protocol in the presence of the CSL model and determine how it differs from that of the noise-free case in \cref{eq:FinalState1}

It is noteworthy that the presence of CSL noise does not distort the trajectories that each Dicke state follows, but only the coherence between them. Thus, at the end of the protocol, all the trajectories overlap, as in the noise-free case, and the system is in a separable state of the form $\hat \rho = \hat \rho_{\rm CM}\otimes \hat \rho_{\rm S}$, where $\hat \rho_{\rm CM}$ denotes the center-of-mass state and $\hat \rho_{\rm S}$ that of the spin ensemble. Nevertheless, the loss of coherence between the different trajectories manifests as a reduction of the purity of the spin state, which can be computed to be
\begin{align}
    \hat{\rho}_\mathrm{S}=\frac{1}{2^{N}}\sum_{\kappa,\kappa^\prime=0}^N &\sqrt{\binom{N}{\kappa}\binom{N}{\kappa^\prime}} e^{-i (\phi_\kappa - \phi_{\kappa'})} \nonumber \\
    &\times e^{- \Lambda_{\kappa, \kappa'}} \ket{\kappa}\bra{\kappa^\prime},
    \label{eq:rhoSpin1}
\end{align}
where the relative phase $\phi_\kappa$ was introduced in \cref{eq:RelativePhase} and 
\begin{equation}
    \Lambda_{\kappa, \kappa'}= \xi \int_0^{T_\mathrm{tot}} (1-e^{-\frac{[\mathcal{X}_\kappa(t^\prime)-\mathcal{X}_{\kappa^\prime}(t^\prime)]^2}{4 r^2_\mathbf{CSL}}})\,\mathrm{d} t'\, ,
\end{equation}
as shown in \cref{App:DecoherenceTwoPaths}. Here, $\mathcal{X}_\kappa(t)=\chi \Re[\zeta_\kappa(t)]$ is the classical trajectory of the $\kappa^\mathrm{th}$ wave packet and is given in \cref{eq:Path1}. Since $\zeta_\kappa(t)$ is linear in $\kappa$, the decoherence rate between the $\kappa^\mathrm{th}$ and $\kappa^{\prime \mathrm{th}}$ trajectories is proportional to $(\kappa-\kappa^\prime)^2$. 

For matter-wave interferometers with a single spin, the standard procedure is to apply a $\pi/2$-pulse on the spin at the end of the protocol, and then measure the magnetic moment of the spin $(1/2) \hbar \gamma \sigma_z$. In the absence of noise, the pulse returns the spin to the initial state $\ket{\uparrow}$, and the expectation value of $\sigma_z$ is thus found to be one. A loss of coherence, for example, due to the presence of CSL noise, is then manifested as a reduction of the expectation value. Thus the expectation value of the magnetic moment of the spin accounts for the amount of noise observed by the system. The natural extension of such a protocol for the multi-spin case that we consider here is the application of a $\pi/2$-pulse to each of the spins in the ensemble and the subsequent measurement of the total magnetic moment of the nanoparticle
\begin{equation}
    \hat{M}=\frac{\hbar}{2}\gamma_p\sum_{n=1}^N\hat{\sigma}_z^{(n)},
\end{equation}
with $\hat{\sigma}_z^{(n)}$ the $z$ Pauli operator acting on the $n^\mathrm{th}$ spin. However, as we explain below, this will turn out not to be an optimal strategy.

As shown in \cref{App:Expectation value of M}, the expectation value of $\hat M$ after applying a $\pi/2$-pulse on all the spins at the end of the protocol, can be computed to be 
\begin{align}
    \frac{\langle\hat{M}\rangle(T_\mathrm{tot})}{\langle\hat{M}\rangle(0)}= e^{-\gamma(T_\mathrm{tot})}\cos^{N-1}(4 \phi),
    \label{eq:M1}
\end{align}
with $\gamma(T_\mathrm{tot})= \Lambda_{\kappa,\kappa\pm1} (T_{\rm tot})$ and $\phi=\phi_\kappa/(2\kappa-N)^2=(\chi/2 x_0)^2 \sin(\Omega t_1)+\eta T_\mathrm{tot}/\hbar$. As discussed, $\Lambda_{\kappa,\kappa'}$ is proportional to $(\kappa-\kappa')^2$, and therefore $\gamma(T_\mathrm{tot})$ is independent of $\kappa$. Assuming that the distance between nearest neighbor trajectories is much smaller than $r_{\rm CSL}$, which is not a particularly restrictive assumption, $\gamma(T_\mathrm{tot})$ has a closed form, although lengthy, solution which is given in \cref{App:Expectation value of M}.

Expression (\ref{eq:M1}) presents two notable drawbacks. Firstly, it fails to exploit the presence of multiple spins, resulting in a decay due to CSL noise that remains independent of the number of spins. Consequently, the sensitivity of the protocol mirrors that of the single-spin scenario, neglecting the potential enhancement from multiple spins and only leveraging decoherence between nearest-neighbor trajectories. This limitation stems from the nature of the operator $\hat M$, and can easily be understood under the following logic. The measurement of operator $\hat M$ after a collective $\pi/2$-pulse is equivalent to taking the expectation value of operator $(\hbar/2) \gamma_p \sum_{n=1}^N \hat \sigma^{(n)}_x$ over state~(\ref{eq:rhoSpin1}). Given that the $\hat \sigma_x^{(n)}$ operators, responsible for individual spin flips, induce transitions of the form $\hat \sigma_x^{(n)} \ket{\kappa} \rightarrow \{ \ket{\kappa -1}, \ket{\kappa + 1} \}$, only coherences between nearest-neighbor Dicke states contribute to the expectation value of $\hat M$.

Secondly, the presence of the term $\cos^{N-1}(4 \phi)$ in \cref{eq:M1} poses another challenge. This term arises from the relative phase accumulation among different trajectories and restricts the duration of the protocol to times that guarantee the condition $\phi = n \pi$, where $n$ is an integer. Any deviations from this condition result in the rapid vanishing of the expectation value of $\hat M$, regardless of the presence of CSL noise. Given that even the smallest nanoparticles harbor millions of nuclear spins, as indicated in Table~(\ref{tab:Overview}), minor deviations $\Delta \phi$ from $n \pi$ could prove detrimental. A rough estimate suggests that, since $\cos^N(n \pi + \Delta \phi) \approx 1 -N (\Delta \phi)^2/2 + \cdots$, determining $\phi$ with a precision below $1/\sqrt{N}$ radians is required. Nonetheless, with sufficient precision, one can always adjust the parameters and vary $t_1$ to meet this criterion. One example would be to set $t_1=\pi/\Omega$. Then, $t_2$ would be zero and the total time is $T_\mathrm{tot}=2 \pi/\Omega$. $\phi=l\pi$ then holds true for
\begin{equation}
    B'=\frac{\hbar \gamma_p^2 }{l V}\sqrt{\frac{\mu_0^3 \rho}{|\chi_V|^3 }},
    \label{eq:BpFringe}
\end{equation}
where $l$ is an integer. In general, the experiment can be conducted faster by finding the values of $t_1$ that meet the required conditions for a given set of system parameters.

We note that the variance of $\hat{M}$ does not provide a significant advantage. Since the variance is determined through single-shot measurements of $\hat{M}$, we normalize the quantity by $\langle \hat{M} \rangle (0)^2$ to enable a fair comparison between $\langle \hat{M} \rangle$ and $\langle \Delta\hat{M} \rangle^2 =\langle \hat{M}^2 \rangle - \langle \hat{M} \rangle^2$.
Analogous to \cref{App:Expectation value of M}, one can find $\langle \Delta\hat{M} \rangle^2$ to be
\begin{align}
   \frac{\langle \Delta \hat{M} \rangle ^2 (T_\mathrm{tot})}{\langle \hat{M} \rangle^2 (0)}=&\frac{1}{2N} \left[ 1-e^{-4 \gamma} \cos^N(8\phi) \right]  \nonumber \\
   &+\frac{1}{2} \big[ 1- 2 e^{-2\gamma} \cos^{2N-2}(4\phi)\nonumber \\
   & + e^{-4\gamma} \cos^{N-2}(8\phi) \big].
\end{align}
Notably, for large $N$, the term in the first line is suppressed. The rest is independent of $N$, therefore this observable does not show an improvement compared to \cref{eq:M1}. At this point, the argument of $\gamma=\gamma(T_\mathrm{tot})$ was omitted for readability.

\begin{figure}
    \centering
    \includegraphics{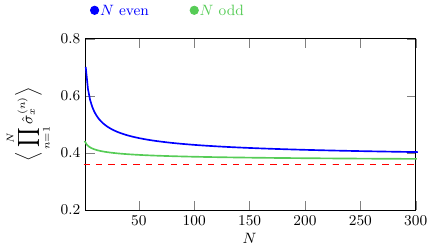}
    \caption{$\expval{\prod_{n=1}^N \hat \sigma_x^{(n)}}$ for an arbitrary set of system parameters and protocol duration. $\expval{\prod_{n=1}^N \hat \sigma_x^{(n)}}$ is greater for even values of $N$ and smaller for odd ones. This discrepancy arises from the fact that for even $N$, there exists a trajectory $\tilde{\kappa}=N/2$ for which $\tilde{\kappa}=N-\tilde{\kappa}$ and thus $\Lambda_{N/2,N/2}=0$, a condition not satisfied for odd $N$. Moreover, this term carries the greatest weight in the sum of \cref{eq:AlternativeObservable}. For large $N$, the difference between odd and even $N$ diminishes as expected and $\expval{\prod_{n=1}^N \hat \sigma_x^{(n)}}$ approaches the bound given by $\exp(-\xi T_\mathrm{tot})$, which is depicted in red.}
    \label{fig:AlternativeObservable}
\end{figure}

A more suitable observable for mitigating the impact of relative phases and harnessing the collective behavior of multiple spins is given by $ \propto \prod_{n=1}^N \hat \sigma_x^{(n)}$. This observable induces the transition $\prod_{n=1}^N \hat \sigma_x^{(n)} \ket{\kappa} = \ket{N-\kappa}$ between Dicke states. Notably, the two connected states accumulate the same relative phase throughout the protocol, that is $\phi_\kappa = \phi_{N-\kappa}$, effectively eliminating from the expectation value the dependence on the geometric phase $\phi$. Furthermore, the decay rate now relies on the total number of spins, as contributions to the expectation value stem not only from nearest-neighbor coherences. This can be computed as:
\begin{equation}
\expval{\prod_{n=1}^N \hat \sigma_x^{(n)} } = \frac{1}{2^N} \sum_{\kappa =0}^N \binom{N}{\kappa} e^{-\Lambda_{\kappa,N-\kappa}}\, .
\label{eq:AlternativeObservable}
\end{equation}
Figure~(\ref{fig:AlternativeObservable}) shows how \cref{eq:AlternativeObservable} behaves with increasing $N$ for an arbitrary set of system parameters and protocol duration. Notice that as $\Lambda_{\kappa,\kappa'}$ is bounded from below by $\xi T_\mathrm{tot}$, the expectation value $\langle\prod_{n=1}^N \hat \sigma_x^{(n)} \rangle$ at time $T_{\rm tot}$ converges to the value $\exp(-\xi\,T_\mathrm{tot})$ as $N$ increases. 

However, measuring such an observable presents greater challenges compared to measuring the magnetic moment of the nanoparticle. One way to do so would be to apply a global $\pi$-pulse on the spin ensemble that is controlled on the state of an ancillary qubit, that is, $U_{\rm c} = \ketbra{0}{0}_a \otimes \prod_{n=1}^N \hat \sigma_x^{(n)} + \ketbra{1}{1}_a\otimes \mathcal{I}_{2^N}$, where subscript $_a$ indicates the operators acting on the ancillary qubit, and $\mathcal{I}_{2^N}$ is the identity operator on the Hilbert space of the N spin ensemble. The expectation value of $\prod_{n=1}^N \hat \sigma_x^{(n)}$ can then be retrieved by measuring the off-diagonal element of the density matrix of the ancillary qubit  $\expval{\ketbra{0}{1}}_a$. Implementing such a controlled gate, however, may prove challenging in practice as it requires generating and maintaining coherence between the ancillary qubit and the N-spin ensemble for the duration of the readout. One could argue that were such an operation available to the experimentalist, they could instead use it to prepare an initial state that is a superposition of the two more distant Dicke states, $\ket{0}_a \otimes \ket{\kappa = 0} + \ket{1}_a \otimes \ket{\kappa = N}$, and in this way maximize the achievable position variance of the interferometer. Notice, however, that such a strategy would require maintaining the coherence between the ancilla and the spin ensemble for the duration of the experiment, while the use of such a gate only during readout may be significantly less challenging.
\\

\subsubsection{Modified protocol}
\label{subsec:modifiedProtocol}
Given the challenges faced by the standard matter-wave interferometry protocol when attempting to utilize the collective behavior of multiple spins, as highlighted above, here, we propose a modified version thereof. This adaptation aims to effectively leverage the presence of numerous spins while still relying on the measurement of the total magnetic moment of the nanoparticle, thus, circumventing the need for measuring correlations between the spins. 
\begin{figure}
    \centering
    \includegraphics[width=0.9\linewidth]{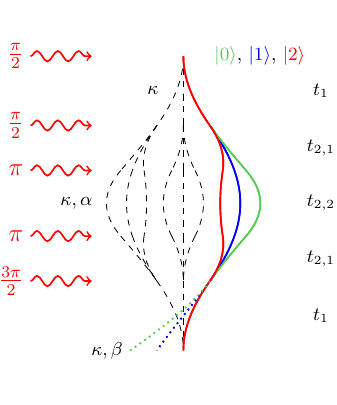}
    \caption{Example for the protocol to test the CSL model with two spins. A $\pi/2$ is applied in the beginning, bringing all spins into a superposition of up and down. Depending on the spin state, the particle starts oscillating around $d_\kappa$, the wave function expands in space. Here, the trajectory the particle takes if all spins are pointing up, in this case $\kappa=2$, is highlighted. After half a period at $t_1=\pi/\Omega$, a second $\pi/2$ pulse is applied and each trajectory again splits up into $N+1$ trajectories. Each of the trajectories is denoted by $\alpha$. For the $\kappa=2$ branch, the different spin states of the $\alpha^\mathrm{th}$ trajectory are color-coded. The times are given by $t_1=\pi/\Omega-2t_{2,1}-t_{2,2}$, and $t_{2,2}$ as in \cref{eq:t2}. In the unitary case, the particle continues traveling on the $\kappa^\mathrm{th}$ trajectory and all paths meet after another time $t_1$ in the origin with zero momentum. In the presence of noise, the coherence between the $\alpha^\mathrm{th}$ trajectories breaks down. This decoherence is mapped on the spin state when the $\alpha^\mathrm{th}$ trajectories meet. Therefore, there are also small contributions to Dicke states different from $\kappa$ after the $3\pi/2$ pulse is applied. The $N+1$ trajectories on the $\kappa\mathrm{th}$ branch after the $3\pi/2$ pulse are denoted with $\beta$. }
    \label{fig:ModifiedProtocol_resubmission}
\end{figure}
The modified protocol is depicted in \cref{fig:ModifiedProtocol_resubmission} for the minimal case of $2$ spins ($3$ Dicke states). In essence, the introduced modification is the application of additional $\pi/2$-pulses, which has the effect of multiplying the number of trajectories.  The detailed protocol is as follows. Initially, a  $\pi/2$-pulse is applied and the wave packet splits into $N+1$ wave packets each following a different trajectory, which are numbered consecutively by $\kappa\in[0,N]$. So far, the protocol is identical to the standard one. The novelty is introduced after a time $t_1=\pi/\Omega$, when a second $\pi/2$ pulse is applied. Each of the $N+1$ trajectories then further divides into an additional $N+1$ trajectories, numbered by $\alpha\in[0,N]$, and resulting in a total of $(N+1)^2$ trajectories that we label $(\kappa,\alpha)$. By repeating the protocol as in \cref{fig:ProtocolOld}, we ensure that the trajectories  $\alpha$ generated by the second $\pi/2$-pulse overlap.  If another $3\pi/2$-pulse is applied at this point, these trajectories recombine into the original $N+1$ trajectories, that is, $(\kappa, \alpha) \rightarrow (\kappa)$, see \cref{fig:ModifiedProtocol_resubmission}. If $t_1 = \pi/\Omega - 2t_{2,1} - t_{2,2}$ is satisfied, all trajectories overlap after another time $t_1$, and the spin ensemble is decoupled from the motion of the nanoparticle, provided that the process is coherent. The state of the spin ensemble at the end of the protocol is given by
\begin{align}
    \hat{\rho}_\mathrm{S}=\frac{1}{2^N}&\sum_{\kappa,\kappa^\prime=0}^N\sum_{\alpha,\alpha^\prime=0}^N\sum_{\beta,\beta^\prime=0}^N \sqrt{\binom{N}{\kappa}}\sqrt{\binom{N}{\kappa'}}\nonumber \\
    &\times \left(\frac{\pi}{2}\right)_{\kappa,\alpha}^{\otimes N}\left(\frac{3\pi}{2}\right)_{\alpha,\beta}^{\otimes N} \left(\frac{\pi}{2}\right)_{\alpha^\prime,\kappa^\prime}^{\otimes N} \left(\frac{3\pi}{2}\right)_{\beta^\prime,\alpha^\prime}^{\otimes N} \nonumber \\
    &\times e^{-i(\phi_{\kappa,\alpha,\beta}(T_\mathrm{tot})-\phi_{\kappa',\alpha',\beta'}(T_\mathrm{tot}))}  \nonumber \\
    &\times\tr{\hat{\rho}_{\kappa,\kappa\prime,\alpha,\alpha^\prime,\beta,\beta^\prime}(T_\mathrm{tot})} \ket{\beta}\bra{\beta'},
\end{align}
where $\left(\pi/2\right)_{\kappa,\alpha}^{\otimes N}$ and $\left(3\pi/2\right)_{\kappa,\alpha}^{\otimes N}$ are the matrix elements in Dicke basis of the $\pi/2$- and the $3\pi/2$-pulse, respectively, whose explicit expression is provided in \cref{App:pi/2 pulse in Dicke basis}. Provided the process was unitary, only the elements with $\{\beta,\beta'\}=\{\kappa,\kappa'\}$ survive and, furthermore, we find
\begin{align}
\tr{\hat{\rho}_{\kappa,\kappa\prime,\alpha,\alpha^\prime,\beta,\beta^\prime}(T_\mathrm{tot})} =
    \tr\Big\{\hat{D}(x)\hat{\rho}_0 \hat{D}^\dagger(x')\Big\},
      \label{eq:traceMotion}
\end{align}
where $x=\frac{\chi}{2 x_0}\zeta_{\kappa,\alpha, \beta}(T_\mathrm{tot})$ is used as the argument in the displacement operator $\hat{D}(x)=\exp(x \hat{a}^\dagger-x^*\hat{a})$ and $\zeta_{\kappa,\alpha, \beta}(t)$ are the classical trajectories. These, together with the geometric phase picked up by each of them, $\phi_{\kappa, \alpha, \beta}$, are given in \cref{App:Expectation value of M modified protocol}. $\hat{\rho}_0$ is the initial state of the mechanical degree of freedom. For the pulse timings described before, $\zeta_{\kappa, \alpha, \kappa}(T_\mathrm{tot}) =0$ and \cref{eq:traceMotion} reduces to $1$. On the other hand, the system parameters can be chosen to guarantee that $\phi_{\kappa, \alpha, \beta} = l \pi$, with $l$ an integer, which is the case for the examples given in \cref{tab:parameters}. Under these conditions, we recover $\hat{\rho}_\mathrm{S}=2^{-N}\sum_{\kappa,\kappa^\prime=0}^N \sqrt{\binom{N}{\kappa}\binom{N}{\kappa'}}\ket{\kappa}\bra{\kappa'}$. Thus, in this noiseless case, after applying a $\pi/2$-pulse and measuring $\hat{M}$, we find $\expval*{\hat M}(T_{\rm tot}) = \expval*{\hat M}(0)$, as expected.
\\
Now, we need to understand how the expectation value of $\hat M$ at the end of the protocol deviates from $\expval*{\hat M}(0)$ in the presence of CSL noise. When CSL noise acts on the system the coherences between the different $(\kappa,\alpha)$ paths decay at different rates. This decoherence manifests on $\langle \hat{M} \rangle$ in two ways: first, the loss of coherence between the $(\kappa,\alpha)$ and $(\kappa\pm 1,\alpha^\prime)$ trajectories results in the decoherence between the $\kappa^\mathrm{th}$ and $\kappa\pm 1^\mathrm{th}$ trajectories. This enhances the loss of coherence between neighboring trajectories at the end of the protocol, which is picked up by the expectation value of $\hat M$. An intuitive way to convince oneself why this scheme is so effective is that the $\pi/2$ and $3\pi/2$ pulses do not induce just a transition $\kappa \rightarrow \kappa\pm 1$, but also higher-order transitions. Second, the loss of coherence between trajectories $(\kappa, \alpha)$ and $(\kappa, \alpha')$ prevents the successful completion of the transformation $(\kappa, \alpha) \rightarrow (\kappa)$ after the third $3\pi/2$-pulse. The spin state does not return perfectly to the $\kappa^\mathrm{th}$ Dicke state but to a superposition of Dicke states, inducing a transformation of trajectories of the form $(\kappa, \alpha) \rightarrow (\kappa, \beta)$. Therefore, not all trajectories overlap at the end of the protocol and, thus, the spin and mechanical degrees of freedom do not become separable. This, however, is not detrimental as it enhances the decoherence due to the CSL noise on the spin ensemble,  making it more easily detectable. The full expression of the state of the spin ensemble at the end of the protocol in the presence of CSL noise is given in~\cref{App:Expectation value of M modified protocol}.

A second advantage of the modified protocol is that it allows us to tune the spatial maximal delocalization of the wave function by varying $t_1$, which we define as
\begin{equation}
    \Delta x(t)^2 = \sum_{\kappa, \alpha = 0}^N \chi_{\kappa,\alpha}^2(t) \, P(\kappa,\alpha).
\end{equation}
$P(\kappa,\alpha)$ is the probability for the $(\kappa,\alpha)^\mathrm{th}$ trajectory being realised and is given by
\begin{equation}
    P(\kappa, \alpha) = \binom{N}{\kappa} \left(\left(\frac{\pi}{2}\right)_{\kappa, \alpha}^{\otimes N}\right)^2.
\end{equation}
$\chi_{\kappa,\alpha}^2(t)$ is the corresponding trajectory, which is of the form $\chi_{\kappa,\alpha}^2(t)=c_\kappa(t) \kappa+c_\alpha(t) \alpha$. In \cref{App:Analysis of Wave Functions Structure}, we find $\Delta x(t)^2 = c_\kappa^2 \frac{N}{4} + c_\alpha^2 \frac{N + N^2}{8}$. While we do not have a rigorous proof of this relation, it agrees perfectly with all numerical tests we have performed. As long as the terms proportional to $c_\kappa$ dominate, the wave function retains a Gaussian shape. When this is no longer the case, the wave function begins to develop two distinct, widely separated peaks. In general, we look to remain within the Gaussian regime of the interferometer in order to guarantee that most of the trajectories occur in a region where the magnetic field gradient remains linear.

As an example, for $r = 44\,\mathrm{nm}$, $t_1 \Omega = 3.1$, $\Omega t_{2,1} = 0.0022$, and $\Omega t_{2,2} = 0.0044$, we find at the time of maximal delocalization: $\chi_{\kappa,\alpha}(t_1 + t_{2,1} + t_{2,2}/2) = 5.2 \cdot 10^{-6} \kappa + 1.3 \cdot 10^{-11} \alpha.$
A particle of this size contains approximately $1.5 \cdot 10^7$ hydrogen nuclear spins. The wave function remains Gaussian, and we find that its standard deviation spreads over a variation of the  magnetic field of $\Delta B = B' \Delta x = 10.4\,\mathrm{mT}$. Moreover, if the magnetic field gradient remains constant over a range of $100\,\mathrm{mT}$, then more than $99.99\%$ of all trajectories stay within the linear region at all times.
Another possible parameter set is $r = 105\,\mathrm{nm}$, $\Omega t_1 = 3$, $\Omega t_{2,1} = 0.03$, and $\Omega t_{2,2} = 0.061$. For these values, we obtain
$\chi_{\kappa,\alpha}(t_1 + t_{2,1} + t_{2,2}/2) = 3.9 \cdot 10^{-7} \kappa + 1.8 \cdot 10^{-10} \alpha$.
Such a particle contains around $2.1 \cdot 10^8$ hydrogen spins. The wave function remains Gaussian, and its width spread over a magnetic field variation of $\Delta B = 14\,\mathrm{mT}$. These parameter sets are given in \cref{tab:Overview}.

\subsection{Evaluation of \texorpdfstring{$\langle\hat{M}\rangle(T_\mathrm{tot})/\langle\hat{M}\rangle(0)$}{TEXT}}

\begin{figure}
\centering
\includegraphics[width=1\columnwidth]{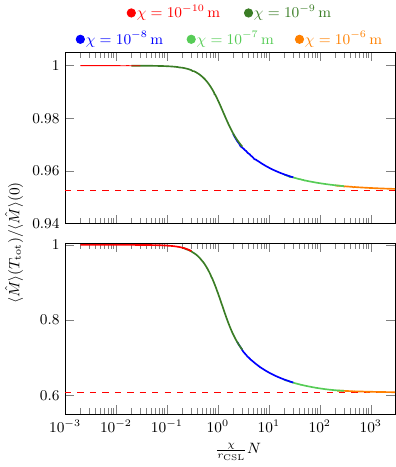}
\caption{Illustration of the behavior of $\langle\hat{M}\rangle(T_\mathrm{tot})/\langle\hat{M}\rangle(0)$ for the parameters in \cref{tab:Overview} a) (upper panel) and \cref{tab:Overview} c) (lower panel) and for fixed $\chi$. 
Due to limited computation times, $\langle\hat{M}\rangle(T_\mathrm{tot})/\langle\hat{M}\rangle(0)$ could not be evaluated numerically for $N\approx10^8$ as required, but only for $N<300$. Therefore, additionally to varying $N$, $\chi$ was varied as this is equivalent to varying the gyromagnetic ratio of the spins. In different colors, one sees $\langle\hat{M}\rangle(T_\mathrm{tot})/\langle\hat{M}\rangle(0)$ for different, fixed $\chi$ and $N\in[10,300]$. As one can see, the plots strongly indicate that $\langle\hat{M}\rangle(T_\mathrm{tot})/\langle\hat{M}\rangle(0)$ is a function of $(\chi/r_\mathrm{CSL}) N$. Furthermore,  for $(\chi/r_\mathrm{CSL}) N>10^2$, it approaches its lower limit given by $\exp(-\xi T_\mathrm{tot})$ as indicated by the red, dashed line. For all parameters in \cref{tab:Overview}, $(\chi/r_\mathrm{CSL}) N\gg 10^2$. Therefore we can assume that $\langle\hat{M}\rangle(T_\mathrm{tot})/\langle\hat{M}\rangle(0)$ is very close to its lower bound.} 
\label{fig:ResFinalProtocol}
\end{figure}

Now we have to evaluate the expectation value of $\hat{M}$ in the presence of the CSL model.
Unfortunately, $\langle\hat{M}\rangle(T_\mathrm{tot})/\langle\hat{M}\rangle(0)$ has no closed form expression and needs to be evaluated numerically. Moreover, its numerical evaluation becomes increasingly hard with a growing number of spins $N$. Our simulations are limited to values of $N \approx 200$. In \cref{fig:ResFinalProtocol} we plot the value $\langle\hat{M}\rangle(T_\mathrm{tot})/\langle\hat{M}\rangle(0)$ as a function of $(\chi / r_{\rm CSL}) N$. Even though the numerical results clearly suggest that $\langle\hat{M}\rangle(T_\mathrm{tot})/\langle\hat{M}\rangle(0)$ is a function of $(\chi / r_{\rm CSL}) N$, we have been unable to prove this analytically.
To explore the full range of the function we combine curves corresponding to different values of $\chi$ where for each curve $N$ ranges from $10$ to $250$. It is evident that the expectation value of $\hat M$ now depends on the number of spins as desired, and can, thus, outperform the sensitivity of a single spin interferometer. It is noteworthy to mention, that this observable is lower bounded by $\langle\hat{M}\rangle(T_\mathrm{tot})/\langle\hat{M}\rangle(0)\geq \exp(-\xi\,T_\mathrm{tot})$, with the bound corresponding to the case where the coherence between any pair of Dicke states decays at the maximum decay rate, that is, at $\xi$. We observe that as $(\chi / r_{\rm CSL}) N$ grows, the magnetic moment of the nanoparticle approaches this limit. This lower bound can be used to restrict the values that the free parameters of the CSL model can take. For instance, if a given experiment measures at the end of the protocol a magnetization $\expval*{\hat M}_{\rm exp}$, an upper bound on $\xi$ is given by
\begin{equation}
\xi < \frac{\ln{\expval*{\hat M}_{\rm exp}}}{T_{\rm tot}}\, ,
\end{equation}
as any larger value of $\xi$ would result in stronger decoherence and therefore a lower value of $\expval*{\hat M}$ at the end of the protocol. As $\xi$ depends on $r_\mathrm{CSL}$ and $\lambda_\mathrm{CSL}$, they can be bounded by measuring $\expval*{\hat M}_{\rm exp}$.

\begin{figure}
    \centering
    \includegraphics[width=1\columnwidth]{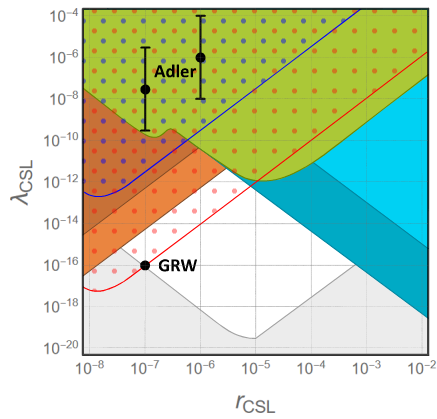}
    \caption{Exclusion plot of the parameter space of the CSL model, with $\lambda_\mathrm{CSL}$ in the vertical axis and $r_\mathrm{CSL}$ in the horizontal one. Various experimentally excluded areas are depicted: the green area corresponds to cantilever-based experiments~\cite{Vinante2020}, the blue region represents experiments conducted with the gravitational wave detectors LIGO and LISA Pathfinder~\cite{Carlesso2016, Carlesso2018, Helou2017}, the brown bounds are derived from spontaneous X-ray emission tests~\cite{Donadi2021, Arnquist2022}, while the gray area is theoretically excluded~\cite{Toros2017}. The values for $\lambda_\mathrm{CSL}$ and $r_\mathrm{CSL}$ proposed by Adler~\cite{Adler2007} and by Ghirardi, Rimini and Weber~\cite{Ghirardi1986} are highlighted in black.  The red line bounds the exclusions region obtained from the modified protocol (\cref{subsec:modifiedProtocol}) using pentacene-doped naphthalene and the parameters listed in~\cref{tab:Overview}a. %Comparable exclusion regions are obtained for the other two parameter sets presented in the table. 
    For comparison the dark blue line marks the border of exclusion region using the \textit{minimal} protocol described in \cref{fig:ProtocolOld}, under the same parameter assumptions. 
    %The minimal protocol, does not exploit the presence of multiple spins and lacks the sensitivity required to impose meaningful bounds. 
    The modified protocol allows for the exploration of previously untested regions of the parameter space.}
    \label{fig:CSL}
\end{figure}

\subsection{Bounds on the parameter space of the CSL model}
In \cref{fig:CSL}, the parameter space of the CSL model is shown, marking the areas already excluded by various other experiments. In the same figure, we show the areas that could, in principle, be excluded by our protocol if operated with the set of parameters in \cref{tab:Overview} (a). Based on this analysis, the parameters proposed by GRW could be potentially excluded. Comparable bounds could be achieved with the other two parameter sets in \cref{tab:Overview}. However, achieving complete exclusion of the parameter space would likely necessitate significantly longer experimental times or heavier particles. For comparison, we also show a dark blue line delineating the region corresponding to the \textit{``minimal'' protocol} introduced in \cref{subsec:minimalProtocol}, in which the multi-spin character of the interferometer is not exploited. It is evident that the modified interferometric scheme, incorporating an additional sequence of control pulses, substantially enhances the achievable bounds on the CSL model. In particular, it may enable the demonstration of more stringent constraints than those reported in previous experimental efforts.

The question arises as to whether the scheme can be further enhanced by adding additional $\pi/2$ and $3\pi/2$ pulses. This would, on the one hand, increase the duration of the interferometric protocol as well as the maximal width of the delocalization of the wave function, both of which would increase the sensitivity to possible collapses of the wave function. However, as error sources due to the increased number of pulses would also increase, it is difficult to assess whether further improvements can be obtained and we leave the detailed analysis to future work. In any case, we do not expect a significant improvement for the parameter regime investigated in this work, as $\langle\hat{M}\rangle(T_\mathrm{tot})/\langle\hat{M}\rangle(0)$ already approaches its lower bound with the scheme presented here.

\subsection{Measurement schemes} \label{sec:Measurement schemes}
We now move on to the question of the measurement of $\langle \hat{M} \rangle$. One possibility could be to leverage the presence of the photo-excited triplet state of the embedded pentacene molecule and its magnetic interaction with the nuclear spin ensemble. In particular, the polarization of the nuclear spin ensemble would have two notable effects on the pentacene molecule. On the one hand, it would shift the energy splitting between the magnetic triplet states $\ket{+1}$ and $\ket{-1}$. On the other hand, the polarization of the spin-lattice could be transferred to the triplet state of the pentacene molecule, provided that this is prepared in a suitable initial state. Both of these effects are, in principle, measurable and proportional to the polarization of the spin ensemble. However, they are bound to be more sensitive to the state of the spins near the pentacene molecule, while underestimating the contribution from spins sitting further from it. We thus turn to alternative measurement schemes that can be sensitive to the global nuclear spin polarization. In particular, we propose to exploit the spin-motion coupling of the nanoparticle to map expectation values of the spin ensemble onto measurements of the center-of-mass position.

The Hamiltonian of our system is of the form
\begin{equation}
\hat H = \hbar \Omega \hat a^\dag \hat a + \hbar \lambda \hat Z (\hat a + {\hat a}^\dag),
\end{equation}
where $\hat Z = \sum_{n=1}^N \sigma_z^{(n)}$ is the observable of interest and $\lambda = (\gamma_p B'/2)\sqrt{\hbar/ 2 m \Omega}$. $\langle \hat{Z} \rangle$ is directly related to $\langle \hat{M}\rangle$ by $\langle \hat{M} \rangle = \hbar/2 \gamma \langle \hat{Z} \rangle$. For the particle diameters below $200\,$nm that we consider, it is safe to assume that the magnetic field gradient is constant across the particle. Consequently, a flip of any of the nuclear spins yields the same momentum kick which, in turn, can then be measured by the displacement. The evolution under such a Hamiltonian has an exact solution, which in the interaction picture w.r.t. $H_0 = \hbar \Omega a^\dag a$, is given by the unitary-evolution operator
\begin{equation}
\hat U = e^{\lambda \hat Z [\alpha(t) {\hat a}^\dag  - \alpha^*(t) \hat a ]} e^{ i \frac{\lambda^2}{\Omega} {\hat Z}^2 [t - \frac{\sin(\Omega t)}{\Omega}]},
\end{equation}
where $\alpha (t) = (1 - e^{i \Omega t})/ \Omega$. 

To see how such an evolution maps the operator $\hat Z$ onto the position, we consider the time evolution of the dimensionless position operator $\hat X = \hat a + {\hat a}^\dag$ in the Heisenberg picture
\begin{equation}
\hat X (t) = {\hat U}^\dag \hat X(0) \hat U = \hat X(0) + \frac{4 \lambda \hat Z}{\Omega} \sin^2(\Omega t/2).
\end{equation}
In this form, it becomes evident that, for any state of the spin-motion composite system, the expectation value of the $\hat Z$ operator can be retrieved from the expectation value of the position operator as
\begin{equation}
    \expval*{\hat Z} = \frac{\Omega}{4 \lambda \sin^2(\Omega t /2)} [\expval*{\hat X}(t) - \expval*{\hat X}(0)].
\end{equation}
Here, the coefficient $\Omega/(4 \lambda)$ captures the required precision in the position measurement. While this will, in general, depend on the parameters of the system, we find that the position measurement precision required to make meaningful measurements of the spin polarization is not that great. For example, for the case of a particle of radius $r= 100$~nm and a magnetic field gradient of $B'=10^5$~T/m, we find that $\Omega/(4 \lambda) =4.3$. This suggests that a position displacement of the ground state width, which in this example is $x_0 = 4\cdot10^{-10}$~m, would provide enough sensitivity to distinguish a change in $\expval*{\hat M}$ of 4 units, corresponding to the change induced by four nuclear spin flips. This is quite a remarkable measurement precision, taking into account that optical position measurements of levitated nanoparticles with $\approx 10^{-10}$~m resolution have been demonstrated~\cite{Aspelmeyer2021} and as low as $10^{-12}$~m with charged particles~\cite{Dania2022}. For an initially fully polarized particle, $\langle \hat{Z} \rangle=N$. The change in the displacement of a naphthalene crystal which is initially fully polarized is then given by
\begin{equation}
    \Delta x_\mathrm{dis}=x_0 \frac{4\lambda}{\Omega}\left(1- \frac{\langle \hat{M} \rangle (T_\mathrm{tot})}{\langle \hat{M} \rangle (0)} \right).
\end{equation}
In \cref{tab:Overview}, $\Omega/(4 \lambda)$ as well as $\Delta x_\mathrm{dis}$ for three different examples are given.

The presence of the pentacene molecule presents an additional method to read out the position of the naphthalene particle. By probing the Zeeman splitting of the photo-excited triplet state of pentacene, which depends on the position of the particle due to the presence of the magnetic field gradient, information on the position of the particle can be retrieved. Such techniques are standard in magnetometry with NV centers~\cite{Vershovskii2020}. In this approach, the lifetime of the photo-excited triplet state limits the achievable position measurement resolution, as this determines the linewidth the optical transition between the ground state and the excited, triplet state. In naphthalene, the lifetime of the excited triplet states is in the order of $\tau=50\,\mathrm{\mu s}$~\cite{Eichhorn2013}, corresponding to a line width $\Delta \omega_\mathbf{pent} = \tau^{-1} = (2\pi)3.1\,$kHz. Interactions with the surrounding nuclear spins would in general broaden this line width, but as magic angle spinning suppresses dipole-dipole interactions, this broadening is also suppressed in our setup. To rapidly resolve the change in displacement due to the CSL model, this should ideally induce a larger Zeeman shift than $\Delta \omega_\mathbf{pent}$.

For the example discussed above, with a magnetic field gradient of $B'=10^5$~T/m, we find the following relation between position displacement $\Delta x$ and the shift of the transition frequency $\Delta \omega$ in SI units, $\Delta \omega \approx 1.8 \cdot 10^{13} \Delta x$. This suggests that a displacement of the order of the zero-point motion of the nanoparticle, which in this case, for a particle of $r=100$~nm, is $x_0 = 7 \cdot 10^{-11}$~m, induces a frequency shift on the order of kHz, and corresponds to a magnetic field change on the order of $\mu$T. Hence, displacements a few times the zero-point motion could, in principle, be resolved with this method. To see the effect of the CSL model, for the numbers as in \cref{tab:Overview}, displacements on the order of $10^{-5}\,$m and larger would have to be resolved. This results in Zeeman shifts in the order of MHz and higher, which is well above the limit imposed by the line width. 

In reality, the nuclear spins in the nanoparticle will not be fully polarized. When the nuclear spins of the nanoparticle are polarized by a percentage $p$, the initial magnetization and consequently the final magnetization too will be reduced by a factor of $p$. Also, the polarization may vary from run to run. However, $\langle \hat{M}\rangle(T_\mathrm{tot})/\langle\hat{M}\rangle(0)$ does not depend on the initial polarization. Therefore, if the polarization is measured before each run, one can compensate for variations in the initial polarization.

\subsection{Pulsed control in the presence of magic-angle spinning}
\label{sec:Pulse}

The protocol suggested in the previous subsection operates under the assumption that the particle is undergoing magic angle spinning. Due to this rotation, the magnetic field observed by the nuclear spins will in general be time-dependent, and thus, it prompts the question of whether this affects our ability to deliver with sufficient precision the microwave pulses required for the protocol. In this section, we look into answering this question.

\begin{figure}
    \centering
    \includegraphics{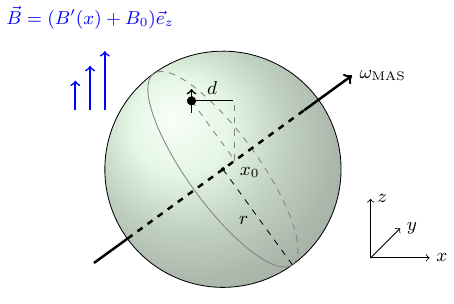}
    \caption{Here, a spherical nanoparticle with radius $r$ in a magnetic field gradient is depicted. The particle rotates with $\omega_\mathrm{MAS}$ around an axis that is tided with respect to the magnetic field. A spin in this particle will see an oscillating magnetic field $\vec{B}_\mathrm{eff}=\{B_0+B'[x_0+d \sin(\omega_\mathrm{MAS} t)]\}\vec{e}_z$. $x_0$ is the x coordinate of the point on the rotation axes closest to the spin and d is the distance between the spin and this point along the x-axis.}
    \label{fig:MAS}
\end{figure}

The particle is sitting in a magnetic field $B(x)=B_0+B'x$ that points in the z-direction and is rotating with frequency $\omega_\mathrm{MAS}$ around an axis that is tilted with respect to the magnetic field by an angle $\phi_{MAS}=\arccos(1/\sqrt{3})\approx54.74$°. In this arrangement, a microwave pulse is applied to the system with frequency $\omega_{\rm M}$ and Rabi frequency $\Omega_{\rm M}$. The Hamiltonian describing such a configuration for a given spin $i$ is provided by 

\begin{align}
    \hat{H}=&\frac{\hbar}{2} \gamma_p \{ B_0+B' [ x_0+d \sin(\omega_\mathrm{MAS} t)] \}\hat{\sigma}_z^{(i)}\nonumber \\
    &+\frac{\hbar}{2}\Omega_{\rm M} \cos(\omega_{\rm M} t) \hat{\sigma}^{(i)}_x.
\end{align}
Here, $x_0$ is the $x$ coordinate of the point on the rotation axis closest to the spin and $d$ is the distance between the spin and this point along the $x$-axis, as depicted in \cref{fig:MAS}. We define the parameters $\Omega_\mathrm{MAS}=\gamma_p B'd$ and $\omega_0=\gamma_p (B_0+B'x_0)$. Notice that the latter corresponds to the Larmor frequency of a spin sitting on the rotation axis at $x_0$. To elucidate the effect of the magic angle spinning on our ability to perform the desired spin rotation, we move to an interaction picture with respect to $H_0 = (\hbar/2) \Omega_\mathrm{MAS} \sin(\omega_\mathrm{MAS} t) \hat{\sigma}_z^{(i)}$, resulting in the Hamiltonian
\begin{align}
    &\hat{H}_\mathrm{int}=\frac{\hbar}{2} \omega_0\hat{\sigma}_z +\frac{\hbar}{2}\Omega_{\rm M} \cos (\omega_{\rm M}t) \times \nonumber \\
    &\Big(e^{-i\frac{\Omega_\mathrm{MAS}}{\omega_\mathrm{MAS}}\cos(\omega_\mathrm{MAS} t)}\sigma_-^{(i)} +e^{i\frac{\Omega_\mathrm{MAS}}{\omega_\mathrm{MAS}}\cos(\omega_\mathrm{MAS} t)}\sigma_+^{(i)}\Big).
\end{align}
We identify that the required regime to suppress the effect of magic angle spinning is given by $\omega_\mathrm{MAS} \gg \Omega_\mathrm{MAS}$, where we recover the Hamiltonian corresponding to a static spin with energy splitting $\omega_0$ that is driven at frequency $\omega_{\rm M}$, that is $\hat{H}_\mathrm{int}=\hbar/2(\omega_0 \hat{\sigma}_z+ \Omega_\mathrm{M} \cos(\omega_{\rm M}) \hat{\sigma}_x)$. Thus, in this regime, all spins in the body of the naphthalene nanoparticle behave as if they were sitting on the rotation axis, with each spin taking the position on the axis that is the closest to its actual position in the crystal. Since $d$ cannot be larger than the radius of the particle, assuming a spherical particle of radius $r$, an upper bound on $\Omega_\mathrm{MAS}$ is given by $\Omega_\mathrm{MAS, max}=\gamma_p B' r $. For the parameters given in \cref{tab:Overview} c), we find $\Omega_\mathrm{MAS, max} \lessapprox (2 \pi) 1.2\,$MHz, which is several orders of magnitude below the rotation frequencies on the order of hundreds of MHz considered in our protocol to attain efficient magic angle spinning. We, thus, do not expect the rotation of the particle to impose limitations on our ability to deliver the required spin rotation pulses.

However, since the body has a spatial extent in the direction of the magnetic field gradient and the rotation axis is not orthogonal to the magnetic field gradient, not all spins on the axis will have the same energy splitting. Notice that the rotation axis could be arranged to lie in the $y-z$ plane,
therefore, giving all the spins on the axis the same $x$ coordinate. However, this would not solve the problem, as any magnetic field exhibiting a gradient along the $x$ direction would also have to exhibit a gradient along the $y$ direction to satisfy the Maxwell equations. For a particle with a radius 70\,nm that sits in a magnetic field gradient of $B'=5\cdot10^4\,\mathrm{T/m}$, a magnetic field variation from one end of the axis to the other of $\Delta B= 7\cdot10^{-4}\,$ T will occur, assuming that the rotation axis is fully contained in the $x-z$ plane. Thus, the energy splittings of the spins sitting on the axis are contained in a frequency band of width $\Delta\omega_\mathrm{particle}=(2 \pi) 30\,$kHz. As radio frequency pulses on hydrogen can achieve Rabi frequencies above $100\,$kHz$\gg \Delta\omega_\mathrm{particle}$, composite pulse techniques (see e.g. Ref.~\cite{Levitt1986, Levitt2007}) offer a potential path to compensate for detunings and Rabi frequency errors that are up to approximately half of the Rabi frequency. 

During the protocol, the delocalization of the center-of-mass wavepacket will further broaden the resonance peak of the spin ensemble.
In the modified protocol, the pulses are applied while the position state's width is given by \cref{eq:PositionVariance}. When neglecting the position variance of the initial state, the width is given by $\langle \Delta x \rangle \approx 2 \chi \sqrt{N}$, where $\chi$ represents the spacing between the equilibrium positions around which the different trajectories oscillate, and is defined in \cref{eq:HOparameters}. The frequency band associated with a given position width is $\Delta \omega_\mathrm{wp} = \gamma_p \langle \Delta x \rangle B'$. As $\chi$ is inversely proportional to $B'$, $\Delta \omega_\mathrm{wp}$ is independent of the magnetic field gradient, and only depends on system constants and the size of the particle. In particular, one finds $\Delta \omega_\mathrm{wp}=(2\pi) 2r^{-3/2}\cdot10^{-4}$. For a particle with $r=45\,$nm, the spectrum width is approximately $\Delta\omega_\mathrm{wp}\approx20\,$MHz.
We would like to point out that composite pulses alone will not suffice to compensate for such detunings. Other potential approaches might involve the use of chirped pulses, which can implement broadband rotations~\cite{Sarkar2021}. Modified schemes of the protocol that momentarily turn off the magnetic-field gradient during the duration of the pulses could also provide a viable means to efficiently apply the required spin rotations to the entire spin ensemble. While this aspect is crucial to the protocol, the detailed design of such pulses is considered beyond the scope of this work and will be presented elsewhere.

\section{Possible noise sources}
\label{sec:Noise}
In this section, we analyze some of the most relevant sources of noise acting on the interferometer without being 
exhaustive. A more complete analysis will be required for concrete experimental setups.

\subsection{Coherence time of the nuclear spins}

As already discussed, for the $T_1$ time of nuclear spins in naphthalene, with approximately 920\,h at $25\,$K ~\cite{Henstra1990, Eichhorn2013,Eichhorn2013a, Can2015, Quan2019}, and the short timescales of the proposed interferometry protocols ($< 1$~ms), even for 
the large number of nuclear spins in the considered nanoparticles, $N = 10^6 - 10^8$, the probability that at least one of the spins in the ensemble flips 
during the protocol is small. However, as in the proposed experiment with its rapid rotation, the $T_1$ time may turn out to be shortened
somewhat and it is therefore pertinent to analyze the effect of stochastic spin flips on the performance of the protocol.

\begin{figure}
    \centering
    \includegraphics{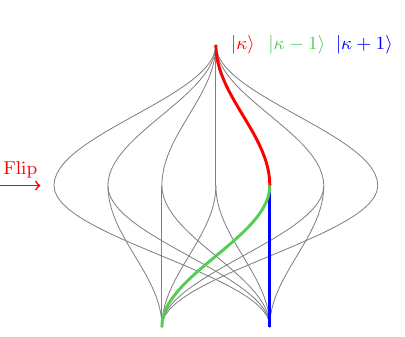}
    \caption{An example of the simplest case of the interferometric protocol without $\pi$ pulses involves the trajectories recombining at the center after a total time of $T_\mathrm{tot}=2\pi/\Omega$. If a $T_1$ flip occurs during the protocol, for all states except $\kappa = 0$ and $\kappa = N$, the $\kappa^\mathrm{th}$ Dicke state transitions into an incoherent superposition of $\ket{\kappa+1}$ and $\ket{\kappa-1}$. Consequently, the trajectories split into an incoherent superposition of two trajectories oscillating around the $\kappa+1^\mathrm{th}$ and $\kappa-1^\mathrm{th}$ equilibrium points, respectively. Thus, the trajectories will recombine at two different points. In phase space, the maximum distance between these two points is given by $|\Delta \zeta|=4\chi/x_0$, as shown in \cref{App:Spin Flips}. }
    \label{fig:T1Flip}
\end{figure}

When a spin flips, it changes the equilibrium position around which each state of the spin ensemble is oscillating due to the presence of the magnetic field gradient. For a joint eigenstate of all the individual $\sigma_z^{(i)}$ operators, e.g. $\ket{\uparrow \uparrow \downarrow \uparrow \cdots}$, the equilibrium position will jump to the right or the left depending on whether the spin involved in the rotation undergoes a spin flip from $\ket{\downarrow}$ to $\ket{\uparrow}$ or the opposite. Since the trajectories of the interferometer are associated with Dicke states, and Dickes states are superpositions of spin states with the same number of spin up, we expect that all trajectories, except for  $\kappa=0$ and $\kappa=N$, will split into an incoherent superposition of two trajectories in the event of a spin flip. As a consequence, trajectories do not recombine in a single point, as would be the case in the noise-free scenario. Instead, two recombination points emerge, one to the left and one to the right of the original recombination point, such that half of the trajectories recombine on one of these points and half on the other. In \cref{fig:T1Flip}, this process is depicted for a six-spin case. In \cref{App:Spin Flips}, an analytical derivation of this phenomenon is provided. For each additional spin flip, the number of recombination points is doubled, such that if two spin flips occur then trajectories recombine in 4 different points, and so on. The result of this process is that at the end of the protocol spins and motion will not end up in a product state. This contributes to the loss of purity of the spin ensemble upon tracing the mechanical degrees of freedom, and is, therefore, not desirable.

While an exhaustive numerical simulation of this form of noise is unattainable for any realistic number of spins, it is useful to calculate the spatial spread of the recombination points to understand how much decoherence can be expected from such stochastic spin flips. If the recombination points are spread over a length scale that is below the zero-point motion of the nanoparticle, then one can expect the spin and motion to be almost separable, and the induced decoherence to be small. In general, the distance between the recombination points will depend on the parameters of the system and the instant at which the spin-flip occurs. For the protocol described in this work, the maximal distance between the two recombination points in dimensionless position space is given by 
\begin{equation}
|\Delta\zeta|=\frac{4\chi}{x_0},
\end{equation}
corresponding to the case where the flip happens while the wave packet is maximally expanded.
$\chi$ is the spacing between the spin-dependent equilibrium points around which the trajectories oscillate and is given in \cref{eq:HOparameters}. For the parameters as in \cref{tab:Overview}, $\chi/x_0$ will be in the order of 0.1-1. Moreover, $T_1$ flips could be even more destructive if the flip happens in between the $\pi/2$ and $3 \pi/2$ pulses in the modified protocol, as the fanned-out trajectories will not recombine in one point anymore. Although the full extent of decoherence induced by an individual spin flip is not accurately quantifiable at this point, we aim to operate in a regime where such flips are highly improbable.

Assuming that the spin-flips occur independently from each other, at a rate given by $N/T_1$, and follow a Poisson distribution, the probability that during an experimental run of duration $T_\mathrm{tot}$, at least one spin flips is given by
\begin{equation}
    P_\mathrm{flip}=1-\exp{-\frac{T_\mathrm{tot}}{T_1}\cdot N}.
\end{equation}
With the parameters in \cref{tab:Overview} b), the run time of the protocol to test the CSL model is given by $T_\mathrm{tot}=78\,$ms. The total number of nuclear spins is $N=1.5\cdot10^7$. We then find that the probability for at least one spin to flip is $30\,\%$. By increasing the magnetic field gradient, the protocol can be conducted faster and this probability reduced.
For the parameters in \cref{tab:Overview} a), $T_\mathrm{tot}$ is given by 7.8\,ms, and the total number of spins is $1.5\cdot 10^7$. For these values, the probability for one or more spin flips is reduced to $4\,\%$. These values might be further reduced if the experiment is conducted with the bulk temperature of the naphthalene nanoparticle below 25\,K, as required to avoid other forms of noise such as that coming from black body radiation.

Regarding the $T_2$ time of the nuclear spins, we refer the reader to \cref{sec:Magic Angle Spinning}, where we describe the use of magic angle spinning. By rotating the particle sufficiently fast around an axis tilted by the magic angle with respect to the magnetic field, $T_2$ times of hundreds of milliseconds or even seconds can be achieved. For example, if the particle is rotating at a frequency of 23\,MHz, a $T_2$ time of 1\,s can be expected.
To check to what extent $T_2$ processes can be tolerated, we describe them using the Lindblad equation
\begin{equation}
    \hat{\rho}=-\frac{i}{\hbar}[\hat{H},\hat{\rho}]+\frac{\Gamma}{2}\sum_{n=1}^N\left(\hat{\sigma}^{(n)}_z \hat{\rho} \hat{\sigma}^{(n)}_z-\hat{\rho}\right)
    \label{eq:MasterEquationDephasing}
\end{equation}
where $\Gamma=T_2^{-1}$.
Since the Hamiltonian in \cref{eq:Hamiltonian} commutes with the Lindbladian, the two processes can be described independently. Pure dephasing can also be described using Kraus operators
\begin{align}
    \hat{M}_0&=e^{-\frac{\Gamma}{2} t} \mathbf{1}\, , \nonumber \\
    \hat{M}_1&=\sqrt{1-e^{\Gamma t}}\,\frac{\mathbf{1+\hat{\sigma}_z}}{2}\, , \nonumber \\
    \hat{M}_2&=\sqrt{1-e^{\Gamma t}}\,\frac{\mathbf{1-\hat{\sigma}_z}}{2}.
\end{align}
Pure dephasing of $N$ spins can therefore be written as
\begin{equation}
    \mathcal{S}[\hat{\rho},t]=\sum_{i_1,...,i_N=0}^2 M_{i_1}\otimes ... \otimes M_{i_N} \hat{\rho}  M_{i_1}^\dagger\otimes... \otimes M_{i_N}^\dagger\, .
\end{equation}
Bringing everything together, we find for the full-time evolution under \cref{eq:MasterEquationDephasing}
\begin{equation}
    \hat{\rho}(t+\Delta t)=\mathcal{M}_\mathrm{dp}[\hat{\rho},\Delta t]=\hat{U}(\Delta t)\mathcal{S}[\hat{\rho}, \Delta t]\hat{U}^\dagger(\Delta t)
\end{equation}
where $\hat{U}$ is given by $\exp[-(i/\hbar)\, \hat{H} \, \Delta t]$.

Nevertheless, the pulses do not commute with $S[\hat{\rho},t]$. We will look at a minimal version of the protocol as in  \cref{fig:ModifiedProtocol_resubmission} for simplicity, where $t_1=\pi/\Omega, 2t_{2,1}=2 \pi/\Omega$ and $t_{2,2}=0$.
Therefore, the full-time evolution of the protocol is given by
\begin{align}
    \hat{\rho}\left(\frac{\pi}{\Omega}\right)&= \mathcal{M}_\mathrm{dp}\left[\hat{\rho}_0,\frac{\pi}{\Omega}\right]\, , \nonumber \\
    \hat{\rho}\left(\frac{3\pi}{\Omega}\right)&=\mathcal{M}_\mathrm{dp}\left[ U_{\pi/2}^{\otimes N} \hat{\rho}\left(\frac{\pi}{\Omega}\right) U_{\pi/2}^{\dagger\otimes N},\frac{2\pi}{\Omega}\right]\, , \nonumber \\
    \hat{\rho}\left(\frac{4\pi}{\Omega}=T_\mathrm{tot}\right)&=\mathcal{M}_\mathrm{dp}\left[ U_{3\pi/2}^{\otimes N} \hat{\rho}\left(\frac{3\pi}{\Omega}\right) U_{3\pi/2}^{\dagger\otimes N},\frac{\pi}{\Omega}\right]\, ,
    \label{eq:rhoDephasing}
\end{align}
where $\hat{\rho}_0$ is as in \cref{eq:InitialState}.
The average magnetization is then given by $\langle \hat{M}(T_\mathrm{tot}) \rangle = \tr\{\hat{M} U_{\pi/2}^{\otimes N} \hat{\rho}(T_\mathrm{tot}) U_{\pi/2}^{\dagger\otimes N}\}$.
While we did not find an analytical expression for $\langle \hat{M}(T_\mathrm{tot}) \rangle$, we evaluated it explicitly for $N\in[2,20]$. 
In all cases, we found
\begin{align}
    \frac{\langle \hat{M}(T_\mathrm{tot}) \rangle}{\langle \hat{M}(0) \rangle} = \frac{1}{2} \Big( & e^{-4 \Gamma T_\mathrm{tot}} - e^{-4 \Gamma T_\mathrm{tot} - 64 \left( \frac{1}{2} + \bar{n} \right) \left( \frac{\chi}{x_0}\right)^2} \nonumber \\
   + &e^{-2 \Gamma T_\mathrm{tot}} + e^{-2 \Gamma T_\mathrm{tot} - 64 \left( \frac{1}{2} + \bar{n} \right) \left( \frac{\chi}{x_0}\right)^2}  \Big)\, ,
   \label{eq:MDephasing}
\end{align}
where $\bar{n}$ is the mean occupation number of the initial thermal position state.
For our choice of parameters, $\chi/x_0$  is on the order of one. Also, in general, $\bar{n}$ can be very large, so we assume  $( 1/2 + \bar{n}  ) (\chi/x_0 )^2 \gg 1$. Under this assumption, $\langle \hat{M}(T_\mathrm{tot}) \rangle/\langle \hat{M}(0) \rangle$ becomes
\begin{equation}
    \frac{\langle \hat{M}(T_\mathrm{tot}) \rangle}{\langle \hat{M}(0) \rangle} = \frac{1}{2} \left( e^{-4 \Gamma T_\mathrm{tot}} + e^{-2 \Gamma T_\mathrm{tot}} \right)
\end{equation}
which also constitutes a lower bound on $\langle \hat{M}(T_\mathrm{tot}) \rangle/\langle \hat{M}(0) \rangle$. For $T_\mathrm{tot}=1\,$ms and $\Gamma^{-1} = T_2=1\,$s, we find $\langle \hat{M}(T_\mathrm{tot}) \rangle/\langle \hat{M}(0) \rangle\approx0.997.$

\subsection{Collisions with air molecules}
Since no experiment can be conducted in a perfect vacuum, there will always be some gas molecules that might collide with the nanoparticle. 
For a spherical particle in an ideal gas containing gas particles of mass $m_g$, temperature $T_g$, and pressure $P$, the collision frequency is given by
\begin{equation}
    \gamma_g=\frac{8 \pi P r^2}{m_g \overline{v}_g},
\end{equation}
where $\overline{v}_g=\sqrt{8 k_\mathrm{B} T_g/(\pi m_g)}$ is the mean velocity of the gas molecules. We assume cryogenic temperatures of $T_g=4\,$K and a gas composition that is mainly dominated by H$_2$ molecules with $m_g=2\,$a.m.u. We are interested in the probability of finding no collision, as, in general, a single collision would be enough to destroy the visibility of the interferometer. For $P=5\cdot10^{-11}\,$P and the parameters as in \cref{tab:Overview} b), the probability of having no collisions with air molecules is $\exp(-\gamma_g\,T_\mathrm{tot}/3)\approx0.91$. For the parameter set as in \cref{tab:Overview} a) and c), a pressure of $P=5\cdot10^{-10}\,$ Pa suffices to ensure, that the probability of no collisions happening are higher than $90\,$\%. Notice that gas pressures as low as $10^{-16}$\, Pa have been demonstrated~\cite{Sellner2017}. Therefore, the conditions to guarantee no collisions with air molecules, while demanding, are, in principle, achievable. 

\subsection{Black body radiation}

Another possible noise source is induced by thermal photons that are either scattered or emitted by the nanoparticle.
Just as the noise induced by the CSL model, the noise due to black body radiation is well described by a master equation
of the form \cite{Schlosshauer2010}
\begin{align}
    \dot{\rho}(q',q'')=&-\frac{i}{\hbar}\bra{q'}[\hat{H},\hat{\rho}]\ket{q''}\nonumber \\
    &-\xi_{bb}\left(1-e^{-\frac{(q'-q'')^2}{4 r_\mathrm{bb}^2}}\rho(q',q'')\right).
\end{align}
Here, $r_\mathrm{bb}$ is called the localization distance, and $\xi_\mathrm{bb}$ represents the localization strength. Both parameters depend on the specific decoherence process. The localization distance of thermal photons is related to the thermal wavelength by $r_\mathrm{bb}=\lambda_\mathrm{th}/2$ \cite{RomeroIsart2011}. Therefore, for thermal photons, we find $r_\mathrm{bb}=\pi^{2/3}\hbar c (2 k_B T_{b/e})$, where $T_{b/e}$ is the temperature of the bulk and environment, respectively. For scattered photons, the localization strength is given by \cite{Schlosshauer2010, RomeroIsart2011}
\begin{equation}
    \frac{\xi_\mathrm{bb,sc}}{4 r_\mathrm{bb}^2}=\frac{8!~8 \zeta(9) c~ r^6}{9 \pi}\left(\frac{k_B T_e}{\hbar c}\right)^9 \Re\left[\frac{\epsilon-1}{\epsilon+2}\right]^2.
\end{equation}
Here, $\zeta(x)$ is the Riemann function.
For the absorption and emission of thermal photons, the localization strength is given by
\begin{equation}
    \frac{\xi_\mathrm{bb,em/ab}}{4 r_\mathrm{bb}^2}=\frac{16 \pi^5 c~ r^3}{189}\left(\frac{k_B T_{b/e}}{\hbar c}\right)^6 \Im\left[\frac{\epsilon-1}{\epsilon+2}\right].
\end{equation}
Here, $\epsilon$ is the dielectric constant of naphthalene in the spectrum range of the thermal photons. Due to the lack of available data, we assumed a conservative value of $\epsilon=1.1+ 0.15 i$.
For $r_\mathrm{bb} \gg (q'-q'')$, the noise term approaches $\xi_\mathrm{bb,sc}(q'-q'')^2/(4 r_\mathrm{bb}^2) $, while for $r_\mathrm{bb} \ll (q'-q'')$, it saturates to $\xi_\mathrm{bb,sc}$. In any case, both limits are upper bounds for all $q',q''$. At $4\,$K, we find $r_\mathrm{bb}\approx0.5\,$mm which, in general, is neither significantly larger nor smaller than the delocalization of the particle in our interferometric protocol. Nonetheless, we will consider $\xi_\mathrm{bb,sc}$ as an upper bound for this analysis.
To ensure that the noise induced by possible spontaneous collapses is greater than the noise induced by interactions with thermal photons, one must ensure $\xi_\mathrm{bb}=\xi_\mathrm{bb,sc}+\xi_\mathrm{bb,em}+\xi_\mathrm{bb, ab}<\xi$, where $\xi$ is the localization strength of the CSL model as in \cref{eq:CSLModelPosSpace}. For a particle of $r=100\,$nm and at cryogenic temperatures, the noise due to emission and absorption dominates over the noise induced by the scattering of thermal photons, as one can easily convince oneself. Thus, it is justified to only ensure that $\xi_\mathrm{bb, em}+\xi_\mathrm{bb, ab}=2\xi_\mathrm{bb, em/ab}<\xi$. 
In \cref{tab:Overview}, the maximum internal or bulk temperature that can be tolerated to ensure $\xi_\mathrm{bb}<\xi$ for different parameters is shown. Note, that $\xi_\mathrm{bb}$ already represents an upper limit for the noise induced by thermal photons; hence, it may be possible to tolerate slightly higher temperatures.

\section{Conclusion}
\label{sec:Conclusions}
We propose the levitation of a class of materials that are characterized by having embedded, non-permanent electron spins that serve to effectively hyperpolarize the nuclear spins in the crystal. We show that levitation of these materials opens the door to performing magic-angle spinning with unprecedented rotation frequencies, which can suppress spin-spin interactions to previously unattainable degrees. This, combined with the exceptionally long lifetimes of nuclear spin polarization demonstrated in such materials, paves the way for the use of multi-spin-dependent forces with a wide range of applications in levitated optomechanics. We believe that the use of such novel materials, potentially tailored for specific applications, and the associated control techniques can open new avenues in the field of levitated optomechanics. In particular, our investigation underscores the promise of pentacene-doped naphthalene in overcoming several limitations inherent in Stern-Gerlach-type matter-wave interferometers relying on materials hosting color centers. On the one hand, the transient nature of the photo-excited triplet state used for polarization ensures the disappearance of the associated decoherence channels post-polarization, in contrast to the constant presence of NV centers in diamonds. On the other hand, the remarkably prolonged polarization lifetime in naphthalene suggests minimal magnetic impurities, removing a major source of decoherence present in diamond-based interferometers. Moreover, the homogeneous distribution of nuclear spins, and the absence of a preferential axis of magnetization, mitigates the emergence of undesirable torques, which are challenging to compensate for in other types of interferometers and typically lead to a loss of interferometric feasibility. Building upon these advantages, we have developed a multi-spin matter-wave interferometry protocol and assessed its capacity to establish bounds for the free parameters of collapse models. Importantly, our noise analysis indicates that our interferometer can improve upon existing bounds using state-of-the-art or near-term technology. Beyond testing collapse models, such a sensitive matter-wave interferometer could also prove useful for detecting extremely weak forces or fields that couple to the spin degrees of freedom. This may open the possibility of probing new physics, such as in searches for QCD axions.

Besides the applications in matter-wave interferometry, we believe that levitated naphthalene may be used to push the boundaries of NMR applications, by implementing magic angle spinning at unparalleled rotational frequencies, and thus reach unprecedented nuclear spin coherence times. Moreover, our proposal for detecting spin ensemble polarization with exceptional accuracy through position displacement measurements further augments the catalog of potential NMR applications made possible by levitation. 

It is pertinent to note that pentacene-doped naphthalene serves as just one member of a broader family of optically polarizable materials, to which much of our analysis is transferable. Examples would be p-dibromobenzene doped with p-dichlorobenzene~\cite{Deimling1980} or p-terphenyl doped with pentacene~\cite{Iinuma2000} which do, for example, exhibit much lower particle losses.

Finally, it is noteworthy to mention that our analysis relies on properties of naphthalene, such as its refractive index or heat capacity, that were measured in bulk samples and in restricted ranges of temperature and pressure conditions. Thus, their behavior outside these regimes, such as in nanoscale particles, remains incompletely understood. While this naturally limits the extent of our conclusions, it also opens an unprecedented opportunity to explore these material properties in a levitated arrangement, going beyond the already charted regimes.

Taken together, these factors make a compelling case for the exploration of naphthalene in the context of levitation and promise exciting advancements for the quantum applications of levitated optomechanical platforms.

\section{Acknowledgments}
We thank Benjamin Stickler for insightful discussions about levitated systems. Furthermore, we would like to thank Tim R. Eichhorn, Markus Schwörer and Jens Pflaum for fruitful discussions about naphthalene. This work was supported by the ERC Synergy grant HyperQ (Grant No. 856432), the QuantERA Project LEMAQUME (Grant No. 500314265) and the BMBF under the funding program ‘quantum technologies—from basic research to market’ in the project Spinning (project no. 13N16215).

\bibliographystyle{quantum_modified}
\bibliography{Paper}

@Article{Quan2019,
  author    = {Y. Quan and B. van den Brandt and J. Kohlbrecher and W.Th. Wenckebach and P. Hautle},
  journal   = {Nuclear Instruments and Methods in Physics Research Section A},
  title     = {A transportable neutron spin filter},
  year      = {2019},
  month     = {mar},
  pages     = {22--26},
  volume    = {921},
  doi       = {10.1016/j.nima.2018.12.047},
  publisher = {Elsevier {BV}},
}

@InCollection{Hennel2004,
  author    = {Jacek W. Hennel and Jacek Klinowski},
  booktitle = {Topics in Current Chemistry},
  publisher = {Springer Berlin Heidelberg},
  title     = {Magic-Angle Spinning: a Historical Perspective},
  year      = {2004},
  month     = {oct},
  doi       = {10.1007/b98646},
}

@Article{Schuck2018,
  author    = {Marcel Schuck and Daniel Steinert and Thomas Nussbaumer and Johann W. Kolar},
  journal   = {Science Advances},
  title     = {Ultrafast rotation of magnetically levitated macroscopic steel spheres},
  year      = {2018},
  month     = {jan},
  number    = {1},
  volume    = {4},
  pages     = {e170151},
  doi       = {10.1126/sciadv.1701519},
  publisher = {American Association for the Advancement of Science ({AAAS})},
}

@Article{Bassi1983,
  author    = {P. S. Bassi and N. K. Sharma and M. K. Sharma},
  journal   = {Crystal Research and Technology},
  title     = {Mode of solidification and strength properties({II}). Naphthalene-benzil binary system},
  year      = {1983},
  number    = {9},
  pages     = {1191--1197},
  volume    = {18},
  doi       = {10.1002/crat.2170180922},
  publisher = {Wiley},
}

@Article{Zorin2006,
  author    = {Vadim E. Zorin and Steven P. Brown and Paul Hodgkinson},
  journal   = {The Journal of Chemical Physics},
  title     = {Origins of linewidth in H1 magic-angle spinning {NMR}},
  year      = {2006},
  month     = {oct},
  number    = {14},
  volume    = {125},
  pages     = {144508},
  doi       = {10.1063/1.2357602},
  publisher = {{AIP} Publishing},
}

@Article{Tennakone1978,
  author    = {K. Tennakone and M. G. C. Peiris},
  journal   = {American Journal of Physics},
  title     = {Sublimation of moth balls},
  year      = {1978},
  month     = {apr},
  number    = {4},
  pages     = {418--419},
  volume    = {46},
  doi       = {10.1119/1.11338},
  publisher = {American Association of Physics Teachers ({AAPT})},
}

@Book{Haynes2014,
  editor    = {William M. Haynes},
  publisher = {{CRC} Press},
  title     = {{CRC} Handbook of Chemistry and Physics},
  year      = {2014},
  month     = {jun},
  doi       = {10.1201/b17118},
}

@Article{Arita2013,
  author    = {Yoshihiko Arita and Michael Mazilu and Kishan Dholakia},
  journal   = {Nature Communications},
  title     = {Laser-induced rotation and cooling of a trapped microgyroscope in vacuum},
  year      = {2013},
  month     = {aug},
  number    = {1},
  volume    = {4},
  pages     = {2374},
  doi       = {10.1038/ncomms3374},
  publisher = {Springer Science and Business Media {LLC}},
}

@Article{Bhagavantam1929,
  author    = {S. Bhagavantam},
  journal   = {Proceedings of the Royal Society of London. Series A},
  title     = {The magnetic anisotropy of naphthalene crystals},
  year      = {1929},
  month     = {jul},
  number    = {795},
  pages     = {545--554},
  volume    = {124},
  doi       = {10.1098/rspa.1929.0137},
  publisher = {The Royal Society},
}

@Article{Nimmrichter2014,
  author    = {Nimmrichter, Stefan and Hornberger, Klaus and Hammerer, Klemens},
  journal   = {Physical Review Letters},
  title     = {Optomechanical Sensing of Spontaneous Wave-Function Collapse},
  year      = {2014},
  issn      = {1079-7114},
  month     = jul,
  number    = {2},
  pages     = {020405},
  volume    = {113},
  doi       = {10.1103/physrevlett.113.020405},
  publisher = {American Physical Society (APS)},
}

@Article{Ghirardi1986,
  author    = {Ghirardi, G. C. and Rimini, A. and Weber, T.},
  journal   = {Physical Review D},
  title     = {Unified dynamics for microscopic and macroscopic systems},
  year      = {1986},
  issn      = {0556-2821},
  month     = jul,
  number    = {2},
  pages     = {470--491},
  volume    = {34},
  doi       = {10.1103/physrevd.34.470},
  publisher = {American Physical Society (APS)},
}

@Article{RomeroIsart2011,
  author    = {Romero-Isart, Oriol},
  journal   = {Physical Review A},
  title     = {Quantum superposition of massive objects and collapse models},
  year      = {2011},
  issn      = {1094-1622},
  month     = nov,
  number    = {5},
  pages     = {052121},
  volume    = {84},
  doi       = {10.1103/physreva.84.052121},
  publisher = {American Physical Society (APS)},
}

@Book{Schlosshauer2010,
  author    = {Schlosshauer, Maximilian A.},
  publisher = {Springer Berlin Heidelberg},
  title     = {Decoherence: and the Quantum-To-Classical Transition},
  year      = {2010},
  isbn      = {3642071422},
  month     = nov,
  ean       = {9783642071423},
  pagetotal = {436},
  doi = {https://doi.org/10.1007/978-3-540-35775-9}
}

@Article{Delic2020,
  author    = {Delić, Uroš and Reisenbauer, Manuel and Dare, Kahan and Grass, David and Vuletić, Vladan and Kiesel, Nikolai and Aspelmeyer, Markus},
  journal   = {Science},
  title     = {Cooling of a levitated nanoparticle to the motional quantum ground state},
  year      = {2020},
  issn      = {1095-9203},
  month     = feb,
  number    = {6480},
  pages     = {892--895},
  volume    = {367},
  doi       = {10.1126/science.aba3993},
  publisher = {American Association for the Advancement of Science (AAAS)},
}

@Article{GonzalezBallestero2021,
  author    = {Gonzalez-Ballestero, Carlos and Aspelmeyer, Markus and Novotny, Lucas and Quidant, Roman and Romero-Isart, Oriol},
  journal   = {Science},
  title     = {Levitodynamics: Levitation and control of microscopic objects in vacuum},
  year      = {2021},
  issn      = {1095-9203},
  pages    = {168},
  volume    = {374},
  doi       = {10.1126/science.abg3027},
  publisher = {American Association for the Advancement of Science (AAAS)},
}

@Article{Weiss2021,
  author    = {Weiss, T. and Roda-Llordes, M. and Torrontegui, E. and Aspelmeyer, M. and Romero-Isart, O.},
  journal   = {Physical Review Letters},
  title     = {Large Quantum Delocalization of a Levitated Nanoparticle Using Optimal Control: Applications for Force Sensing and Entangling via Weak Forces},
  year      = {2021},
  issn      = {1079-7114},
  month     = jul,
  number    = {2},
  pages     = {023601},
  volume    = {127},
  doi       = {10.1103/physrevlett.127.023601},
  publisher = {American Physical Society (APS)},
}

@Article{Cosco2021,
  author    = {Cosco, F. and Pedernales, J. S. and Plenio, M. B.},
  journal   = {Physical Review A},
  title     = {Enhanced force sensitivity and entanglement in periodically driven optomechanics},
  year      = {2021},
  issn      = {2469-9934},
  month     = jun,
  number    = {6},
  pages     = {l061501},
  volume    = {103},
  doi       = {10.1103/physreva.103.l061501},
  publisher = {American Physical Society (APS)},
}

@Article{Yin2013,
  author    = {Yin, Zhang-qi and Li, Tongcang and Zhang, Xiang and Duan, L. M.},
  journal   = {Physical Review A},
  title     = {Large quantum superpositions of a levitated nanodiamond through spin-optomechanical coupling},
  year      = {2013},
  issn      = {1094-1622},
  month     = sep,
  number    = {3},
  pages     = {033614},
  volume    = {88},
  doi       = {10.1103/physreva.88.033614},
  publisher = {American Physical Society (APS)},
}

@Article{Eichhorn2013,
  author    = {Eichhorn, T. R. and Haag, M. and van den Brandt, B. and Hautle, P. and Wenckebach, W.Th.},
  journal   = {Chemical Physics Letters},
  title     = {High proton spin polarization with DNP using the triplet state of pentacene-d14},
  year      = {2013},
  issn      = {0009-2614},
  month     = jan,
  pages     = {296--299},
  volume    = {555},
  doi       = {10.1016/j.cplett.2012.11.007},
  publisher = {Elsevier BV},
}

@Article{Can2015,
  author    = {Can, T.V. and Ni, Q.Z. and Griffin, R.G.},
  journal   = {Journal of Magnetic Resonance},
  title     = {Mechanisms of dynamic nuclear polarization in insulating solids},
  year      = {2015},
  issn      = {1090-7807},
  month     = apr,
  pages     = {23--35},
  volume    = {253},
  doi       = {10.1016/j.jmr.2015.02.005},
  publisher = {Elsevier BV},
}

@Article{Eichhorn2013a,
  author    = {Eichhorn, T. R. and Brandt, B. van den and Hautle, P. and Henstra, A. and Wenckebach, W. Th.},
  journal   = {Molecular Physics},
  title     = {Dynamic nuclear polarisation via the integrated solid effect II: experiments on naphthalene-h8 doped with pentacene-d14},
  year      = {2013},
  issn      = {1362-3028},
  month     = dec,
  number    = {13},
  pages     = {1773--1782},
  volume    = {112},
  doi       = {10.1080/00268976.2013.863405},
  publisher = {Informa UK Limited},
}

@Article{Henstra1990,
  author    = {Henstra, A. and Lin, T.-S. and Schmidt, J. and Wenckebach, W.Th.},
  journal   = {Chemical Physics Letters},
  title     = {High dynamic nuclear polarization at room temperature},
  year      = {1990},
  issn      = {0009-2614},
  month     = jan,
  number    = {1},
  pages     = {6--10},
  volume    = {165},
  doi       = {10.1016/0009-2614(90)87002-9},
  publisher = {Elsevier BV},
}

@Article{Rudolph2021,
  author    = {Rudolph, Henning and Schäfer, Jonas and Stickler, Benjamin A. and Hornberger, Klaus},
  journal   = {Physical Review A},
  title     = {Theory of nanoparticle cooling by elliptic coherent scattering},
  year      = {2021},
  issn      = {2469-9934},
  month     = apr,
  number    = {4},
  pages     = {043514},
  volume    = {103},
  doi       = {10.1103/physreva.103.043514},
  publisher = {American Physical Society (APS)},
}

@Article{Komatsu1993,
  author    = {Komatsu, Koichi and Murata, Yasujiro and Sugita, Nobuyuki and Takeuchi, Ken’ichi and Wan, Terence S.M.},
  journal   = {Tetrahedron Letters},
  title     = {Use of naphthalene as a solvent for selective formation of the 1:1 diels-alder adduct of C60 with anthracene},
  year      = {1993},
  issn      = {0040-4039},
  month     = jan,
  number    = {52},
  pages     = {8473--8476},
  volume    = {34},
  doi       = {10.1016/s0040-4039(00)61362-x},
  publisher = {Elsevier BV},
}

@book{Wang2010,
  author    = {Wang, Zerong},
  title     = {Comprehensive Organic Name Reactions and Reagents},
  year      = {2010},
  doi       = {10.1002/9780470638859},
  isbn      = {9780470638859},
  publisher = {John Wiley \& Sons, Inc}
}

@Article{Dash2015,
  author    = {Dash, Smruti Rekha and Sarkar, Ritwik and Bhattacharyya, Santanu},
  journal   = {Ceramics International},
  title     = {Gel casting of hydroxyapatite with naphthalene as pore former},
  year      = {2015},
  issn      = {0272-8842},
  month     = apr,
  number    = {3},
  pages     = {3775--3790},
  volume    = {41},
  doi       = {10.1016/j.ceramint.2014.11.053},
  publisher = {Elsevier BV},
}

@Article{Wood2021,
  author        = {B. D. Wood and S. Bose and G. W. Morley},
  title         = {Spin dynamical decoupling for generating macroscopic superpositions of a free-falling nanodiamond},
  year          = {2021},
  journal     = {Physical Review A},
  volume = {105},
  pages = {012824},
  doi           = {10.1103/PhysRevA.105.012824}
}

@Article{Berry1996,
  author    = {Berry, Michael Victor},
  journal   = {Proceedings of the Royal Society of London. Series A},
  title     = {The Levitron: an adiabatic trap for spins},
  year      = {1996},
  issn      = {1471-2946},
  month     = dec,
  number    = {1948},
  pages     = {1207--1220},
  volume    = {452},
  doi       = {10.1098/rspa.1996.0062},
  publisher = {The Royal Society},
}

@Article{Gov1999,
  author    = {Gov, S. and Shtrikman, S. and Thomas, H.},
  journal   = {Physica D: Nonlinear Phenomena},
  title     = {On the dynamical stability of the hovering magnetic top},
  year      = {1999},
  issn      = {0167-2789},
  month     = feb,
  number    = {3–4},
  pages     = {214--224},
  volume    = {126},
  doi       = {10.1016/s0167-2789(98)00282-6},
  publisher = {Elsevier BV},
}

@article{Barnes2024,
title = {Towards optical MAS magnetic resonance using optical traps},
journal = {Journal of Magnetic Resonance Open},
volume = {18},
pages = {100145},
year = {2024},
issn = {2666-4410},
doi = {https://doi.org/10.1016/j.jmro.2023.100145},
url = {https://www.sciencedirect.com/science/article/pii/S2666441023000535},
author = {Lea Marti and Nergiz {Şahin Solmaz} and Michal Kern and Anh Chu and Reza Farsi and Philipp Hengel and Jialiang Gao and Nicholas Alaniva and Michael A. Urban and Ronny Gunzenhauser and Alexander Däpp and Daniel Klose and Jens Anders and Giovanni Boero and Lukas Novotny and Martin Frimmer and Alexander B. Barnes}, 
}

@article{Novotny2018,
  author    = {Reimann, René and Doderer, Michael and Hebestreit, Erik and Diehl, Rozenn and Frimmer, Martin and Windey, Dominik and Tebbenjohanns, Felix and Novotny, Lukas},
  journal   = {Physical Review Letters},
  title     = {GHz Rotation of an Optically Trapped Nanoparticle in Vacuum},
  year      = {2018},
  issn      = {1079-7114},
  month     = jul,
  number    = {3},
  pages     = {033602},
  volume    = {121},
  doi       = {10.1103/physrevlett.121.033602},
  publisher = {American Physical Society (APS)},
}

@article{Zhang2021,
author = {Yuanbin Jin and Jiangwei Yan and Shah Jee Rahman and Jie Li and Xudong Yu and Jing Zhang},
journal = {Photon. Res.},
number = {7},
pages = {1344--1350},
publisher = {Optica Publishing Group},
title = {6 GHz hyperfast rotation of an optically levitated nanoparticle in vacuum},
volume = {9},
month = {Jul},
year = {2021},
url = {https://opg.optica.org/prj/abstract.cfm?URI=prj-9-7-1344},
doi = {10.1364/PRJ.422975},
}

@article{Moore2018,
  title = {Optical rotation of levitated spheres in high vacuum},
  author = {Monteiro, Fernando and Ghosh, Sumita and van Assendelft, Elizabeth C. and Moore, David C.},
  journal = {Phys. Rev. A},
  volume = {97},
  issue = {5},
  pages = {051802},
  numpages = {5},
  year = {2018},
  month = {May},
  publisher = {American Physical Society},
  doi = {10.1103/PhysRevA.97.051802},
  url = {https://link.aps.org/doi/10.1103/PhysRevA.97.051802}
}

@article{Ramamoorthy2023,
author = {Nishiyama, Yusuke and Hou, Guangjin and Agarwal, Vipin and Su, Yongchao and Ramamoorthy, Ayyalusamy},
title = {Ultrafast Magic Angle Spinning Solid-State NMR Spectroscopy: Advances in Methodology and Applications},
journal = {Chemical Reviews},
volume = {123},
number = {3},
pages = {918-988},
year = {2023},
doi = {10.1021/acs.chemrev.2c00197}
}

@article{Ruzicka2005,
author = {Růžička, Květoslav and Fulem, Michal and Růžička, Vlastimil},
title = {Recommended Vapor Pressure of Solid Naphthalene},
journal = {Journal of Chemical \& Engineering Data},
volume = {50},
number = {6},
pages = {1956-1970},
year = {2005},
doi = {10.1021/je050216m},
URL = {https://doi.org/10.1021/je050216m}
}

@Article{Oddershede2004,
  author    = {Oddershede, Jette and Larsen, Sine},
  journal   = {The Journal of Physical Chemistry A},
  title     = {Charge Density Study of Naphthalene Based on X-ray Diffraction Data at Four Different Temperatures and Theoretical Calculations},
  year      = {2004},
  issn      = {1520-5215},
  month     = jan,
  number    = {6},
  pages     = {1057--1063},
  volume    = {108},
  doi       = {10.1021/jp036186g},
  publisher = {American Chemical Society (ACS)},
}

@Article{Bateman2014,
  author    = {Bateman, James and Nimmrichter, Stefan and Hornberger, Klaus and Ulbricht, Hendrik},
  journal   = {Nature Communications},
  title     = {Near-field interferometry of a free-falling nanoparticle from a point-like source},
  year      = {2014},
  issn      = {2041-1723},
  month     = sep,
  number    = {1},
  volume    = {5},
  pages     = {4788},
  doi       = {10.1038/ncomms5788},
  publisher = {Springer Science and Business Media LLC},
}

@Article{Chirico2002,
  author    = {Chirico, R.D and Knipmeyer, S.E and Steele, W.V},
  journal   = {The Journal of Chemical Thermodynamics},
  title     = {Heat capacities, enthalpy increments, and derived thermodynamic functions for naphthalene between the temperatures 5K and 440K},
  year      = {2002},
  issn      = {0021-9614},
  month     = nov,
  number    = {11},
  pages     = {1873--1884},
  volume    = {34},
  doi       = {10.1016/s0021-9614(02)00262-8},
  publisher = {Elsevier BV},
}

@Book{Schwoerer2006a,
  author    = {Schwoerer, Markus and Wolf, Hans Christoph},
  publisher = {Wiley-VCH, Berlin},
  title     = {Organic Molecular Solids},
  year      = {2006},
  isbn      = {978-3-527-40540-4},
  note      = {page 145},
 doi        = {10.1002/9783527618651}
}

@Article{Port1978,
  author    = {Port, H. and Rund, D.},
  journal   = {Journal of Molecular Structure},
  title     = {High resolution excitation spectroscopy on triplet excitons in organic molecular crystals comparison of naphthalene and anthracene},
  year      = {1978},
  issn      = {0022-2860},
  month     = jan,
  pages     = {455--464},
  volume    = {45},
  doi       = {10.1016/0022-2860(78)87089-6},
  publisher = {Elsevier BV},
}

@Book{Bohren1998,
  author    = {Bohren, Craig F. and Huffman, Donald R.},
  publisher = {Wiley},
  title     = {Absorption and Scattering of Light by Small Particles},
  year      = {1998},
  isbn      = {9783527618156},
  month     = apr,
  doi       = {10.1002/9783527618156},
}

@Article{Haag2012,
  author    = {Haag, M. and van den Brandt, B. and Eichhorn, T. R. and Hautle, P. and Wenckebach, W.Th.},
  journal   = {Nuclear Instruments and Methods in Physics Research Section A},
  title     = {Spin filtering neutrons with a proton target dynamically polarized using photo-excited triplet states},
  year      = {2012},
  issn      = {0168-9002},
  month     = jun,
  pages     = {91--97},
  volume    = {678},
  doi       = {10.1016/j.nima.2012.03.014},
  publisher = {Elsevier BV},
}

@Article{Eichhorn2022,
  author    = {Eichhorn, Tim R. and Parker, Anna J. and Josten, Felix and Müller, Christoph and Scheuer, Jochen and Steiner, Jakob M. and Gierse, Martin and Handwerker, Jonas and Keim, Michael and Lucas, Sebastian and Qureshi, Mohammad Usman and Marshall, Alastair and Salhov, Alon and Quan, Yifan and Binder, Jan and Jahnke, Kay D. and Neumann, Philipp and Knecht, Stephan and Blanchard, John W. and Plenio, Martin B. and Jelezko, Fedor and Emsley, Lyndon and Vassiliou, Christophoros C. and Hautle, Patrick and Schwartz, Ilai},
  journal   = {Journal of the American Chemical Society},
  title     = {Hyperpolarized Solution-State NMR Spectroscopy with Optically Polarized Crystals},
  year      = {2022},
  issn      = {1520-5126},
  month     = feb,
  number    = {6},
  pages     = {2511--2519},
  volume    = {144},
  doi       = {10.1021/jacs.1c09119},
  publisher = {American Chemical Society (ACS)},
}

@Article{Schwartz2018,
  author    = {Schwartz, Ilai and Scheuer, Jochen and Tratzmiller, Benedikt and Müller, Samuel and Chen, Qiong and Dhand, Ish and Wang, Zhen-Yu and Müller, Christoph and Naydenov, Boris and Jelezko, Fedor and Plenio, Martin B.},
  journal   = {Science Advances},
  title     = {Robust optical polarization of nuclear spin baths using Hamiltonian engineering of nitrogen-vacancy center quantum dynamics},
  year      = {2018},
  issn      = {2375-2548},
  number    = {8},
  volume    = {4},
  pages     = {eaat8978},
  doi       = {10.1126/sciadv.aat8978},
  publisher = {American Association for the Advancement of Science (AAAS)},
}

@Article{Sarkar2021,
  author    = {Sarkar, Sambeda and Purusottam, Rudra N. and Kumar, Ashutosh and Khaneja, Navin},
  journal   = {Journal of Magnetic Resonance},
  title     = {Chirp pulse sequences for broadband $\pi$ rotation},
  year      = {2021},
  issn      = {1090-7807},
  month     = jul,
  pages     = {107002},
  volume    = {328},
  doi       = {10.1016/j.jmr.2021.107002},
  publisher = {Elsevier BV},
}

@Article{Vershovskii2020,
  author    = {Vershovskii, A. K. and Dmitriev, A. K.},
  journal   = {Technical Physics},
  title     = {A Weak Magnetic Field Sensor Based on Nitrogen-Vacancy Color Centers in a Diamond Crystal},
  year      = {2020},
  issn      = {1090-6525},
  month     = aug,
  number    = {8},
  pages     = {1301--1306},
  volume    = {65},
  doi       = {10.1134/s1063784220080216},
  publisher = {Pleiades Publishing Ltd},
}

@Article{Martinetz2020,
  author    = {Martinetz, Lukas and Hornberger, Klaus and Millen, James and Kim, M. S. and Stickler, Benjamin A.},
  journal   = {npj Quantum Information},
  title     = {Quantum electromechanics with levitated nanoparticles},
  year      = {2020},
  number    = {1},
  volume    = {6},
  pages = {101},
  doi       = {10.1038/s41534-020-00333-7}
}

@Article{Vinante2020,
  author    = {Vinante, A. and Carlesso, M. and Bassi, A. and Chiasera, A. and Varas, S. and Falferi, P. and Margesin, B. and Mezzena, R. and Ulbricht, H.},
  journal   = {Physical Review Letters},
  title     = {Narrowing the Parameter Space of Collapse Models with Ultracold Layered Force Sensors},
  year      = {2020},
  issn      = {1079-7114},
  month     = sep,
  number    = {10},
  pages     = {100404},
  volume    = {125},
  doi       = {10.1103/physrevlett.125.100404},
  publisher = {American Physical Society (APS)},
}

@Article{Carlesso2018,
  author    = {Carlesso, Matteo and Paternostro, Mauro and Ulbricht, Hendrik and Vinante, Andrea and Bassi, Angelo},
  journal   = {New Journal of Physics},
  title     = {Non-interferometric test of the continuous spontaneous localization model based on rotational optomechanics},
  year      = {2018},
  issn      = {1367-2630},
  month     = aug,
  number    = {8},
  pages     = {083022},
  volume    = {20},
  doi       = {10.1088/1367-2630/aad863},
  publisher = {IOP Publishing},
}

@Article{Helou2017,
  author    = {Helou, Bassam and Slagmolen, B. J. J. and McClelland, David E. and Chen, Yanbei},
  journal   = {Physical Review D},
  title     = {LISA pathfinder appreciably constrains collapse models},
  year      = {2017},
  issn      = {2470-0029},
  month     = apr,
  number    = {8},
  pages     = {084054},
  volume    = {95},
  doi       = {10.1103/physrevd.95.084054},
  publisher = {American Physical Society (APS)},
}

@Article{Donadi2021,
  author    = {Donadi, Sandro and Piscicchia, Kristian and Del Grande, Raffaele and Curceanu, Catalina and Laubenstein, Matthias and Bassi, Angelo},
  journal   = {The European Physical Journal C},
  title     = {Novel CSL bounds from the noise-induced radiation emission from atoms},
  year      = {2021},
  issn      = {1434-6052},
  month     = aug,
  number    = {8},
  volume    = {81},
  pages = {773},
  doi       = {10.1140/epjc/s10052-021-09556-0},
  publisher = {Springer Science and Business Media LLC},
}

@Article{Arnquist2022,
  author    = {Arnquist, I. J. and Avignone, F. T. and Barabash, A. S. and Barton, C. J. and Bhimani, K. H. and Blalock, E. and Bos, B. and Busch, M. and Buuck, M. and Caldwell, T. S. and Chan, Y-D. and Christofferson, C. D. and Chu, P.-H. and Clark, M. L. and Cuesta, C. and Detwiler, J. A. and Efremenko, Yu. and Ejiri, H. and Elliott, S. R. and Giovanetti, G. K. and Green, M. P. and Gruszko, J. and Guinn, I. S. and Guiseppe, V. E. and Haufe, C. R. and Henning, R. and Hervas Aguilar, D. and Hoppe, E. W. and Hostiuc, A. and Kim, I. and Kouzes, R. T. and Lannen V., T. E. and Li, A. and Lopez, A. M. and López-Castaño, J. M. and Martin, E. L. and Martin, R. D. and Massarczyk, R. and Meijer, S. J. and Oli, T. K. and Othman, G. and Paudel, L. S. and Pettus, W. and Poon, A. W. P. and Radford, D. C. and Reine, A. L. and Rielage, K. and Ruof, N. W. and Tedeschi, D. and Varner, R. L. and Vasilyev, S. and Wilkerson, J. F. and Wiseman, C. and Xu, W. and Yu, C.-H. and Zhu, B. X.},
  journal   = {Physical Review Letters},
  title     = {Search for Spontaneous Radiation from Wave Function Collapse in the Majorana Demonstrator},
  year      = {2022},
  issn      = {1079-7114},
  month     = aug,
  number    = {8},
  pages     = {080401},
  volume    = {129},
  doi       = {10.1103/physrevlett.129.080401},
  publisher = {American Physical Society (APS)},
}

@Article{Toros2017,
  author    = {Toroš, Marko and Gasbarri, Giulio and Bassi, Angelo},
  journal   = {Physics Letters A},
  title     = {Colored and dissipative continuous spontaneous localization model and bounds from matter-wave interferometry},
  year      = {2017},
  issn      = {0375-9601},
  month     = dec,
  number    = {47},
  pages     = {3921--3927},
  volume    = {381},
  doi       = {10.1016/j.physleta.2017.10.002},
  publisher = {Elsevier BV},
}

@Article{Carlesso2016,
  author    = {Carlesso, Matteo and Bassi, Angelo and Falferi, Paolo and Vinante, Andrea},
  journal   = {Physical Review D},
  title     = {Experimental bounds on collapse models from gravitational wave detectors},
  year      = {2016},
  issn      = {2470-0029},
  month     = dec,
  number    = {12},
  pages     = {124036},
  volume    = {94},
  doi       = {10.1103/physrevd.94.124036},
  publisher = {American Physical Society (APS)},
}

@Article{Adler2007,
  author    = {Adler, Stephen L},
  journal   = {Journal of Physics A: Mathematical and Theoretical},
  title     = {Lower and upper bounds on CSL parameters from latent image formation and IGM~heating},
  year      = {2007},
  issn      = {1751-8121},
  month     = oct,
  number    = {44},
  pages     = {13501--13501},
  volume    = {40},
  doi       = {10.1088/1751-8121/40/44/c01},
  publisher = {IOP Publishing},
}

@Article{Bassi2013,
  author    = {Bassi, Angelo and Lochan, Kinjalk and Satin, Seema and Singh, Tejinder P. and Ulbricht, Hendrik},
  journal   = {Reviews of Modern Physics},
  title     = {Models of wave-function collapse, underlying theories, and experimental tests},
  year      = {2013},
  issn      = {1539-0756},
  month     = apr,
  number    = {2},
  pages     = {471--527},
  volume    = {85},
  doi       = {10.1103/revmodphys.85.471},
  publisher = {American Physical Society (APS)},
}

@Article{Pearle1989,
  author    = {Pearle, Philip},
  journal   = {Physical Review A},
  title     = {Combining stochastic dynamical state-vector reduction with spontaneous localization},
  year      = {1989},
  issn      = {0556-2791},
  month     = mar,
  number    = {5},
  pages     = {2277--2289},
  volume    = {39},
  doi       = {10.1103/physreva.39.2277},
  publisher = {American Physical Society (APS)},
}

@Article{Ghirardi1990,
  author    = {Ghirardi, Gian Carlo and Pearle, Philip and Rimini, Alberto},
  journal   = {Physical Review A},
  title     = {Markov processes in Hilbert space and continuous spontaneous localization of systems of identical particles},
  year      = {1990},
  issn      = {1094-1622},
  month     = jul,
  number    = {1},
  pages     = {78--89},
  volume    = {42},
  doi       = {10.1103/physreva.42.78},
  publisher = {American Physical Society (APS)},
}

@Article{Dania2022,
  author    = {Dania, Lorenzo and Heidegger, Katharina and Bykov, Dmitry S. and Cerchiari, Giovanni and Araneda, Gabriel and Northup, Tracy E.},
  journal   = {Physical Review Letters},
  title     = {Position Measurement of a Levitated Nanoparticle via Interference with Its Mirror Image},
  year      = {2022},
  issn      = {1079-7114},
  month     = jun,
  number    = {1},
  pages     = {013601},
  volume    = {129},
  doi       = {10.1103/physrevlett.129.013601},
  publisher = {American Physical Society (APS)},
}

@article{Ashkin1970,
  title = {Acceleration and Trapping of Particles by Radiation Pressure},
  author = {Ashkin, A.},
  journal = {Phys. Rev. Lett.},
  volume = {24},
  issue = {4},
  pages = {156--159},
  numpages = {0},
  year = {1970},
  month = {Jan},
  publisher = {American Physical Society},
  doi = {10.1103/PhysRevLett.24.156},
  url = {https://link.aps.org/doi/10.1103/PhysRevLett.24.156}
}

@article{Cirac2010,
title={Toward quantum superposition of living organisms},
author={Romero-Isart, Oriol and Juan, Mathieu L and Quidant, Romain and Cirac, J Ignacio},
journal= {New Journal of Physics},
volume={12},
pages={033015},
year={2010},
doi = {10.1088/1367-2630/12/3/033015}
}

@article{Zoller2010,
author = {D. E. Chang  and C. A. Regal  and S. B. Papp  and D. J. Wilson  and J. Ye  and O. Painter  and H. J. Kimble  and P. Zoller },
title = {Cavity opto-mechanics using an optically levitated nanosphere},
journal = {Proceedings of the National Academy of Sciences},
volume = {107},
number = {3},
pages = {1005-1010},
year = {2010},
doi = {10.1073/pnas.0912969107},
URL = {https://www.pnas.org/doi/abs/10.1073/pnas.0912969107}}

@article{Schneider2010,
  title = {Cavity cooling of an optically trapped nanoparticle},
  author = {Barker, P. F. and Shneider, M. N.},
  journal = {Phys. Rev. A},
  volume = {81},
  issue = {2},
  pages = {023826},
  numpages = {6},
  year = {2010},
  month = {Feb},
  publisher = {American Physical Society},
  doi = {10.1103/PhysRevA.81.023826},
  url = {https://link.aps.org/doi/10.1103/PhysRevA.81.023826}
}

@article{Aspelmeyer2021,
	author = {Magrini, Lorenzo and Rosenzweig, Philipp and Bach, Constanze and Deutschmann-Olek, Andreas and Hofer, Sebastian G. and Hong, Sungkun and Kiesel, Nikolai and Kugi, Andreas and Aspelmeyer, Markus},
	date = {2021/07/01},
	doi = {10.1038/s41586-021-03602-3},
	id = {Magrini2021},
	isbn = {1476-4687},
	journal = {Nature},
	number = {7867},
	pages = {373--377},
	title = {Real-time optimal quantum control of mechanical motion at room temperature},
	url = {https://doi.org/10.1038/s41586-021-03602-3},
	volume = {595},
	year = {2021}}

@article{Novotny2021,
	author = {Tebbenjohanns, Felix and Mattana, M. Luisa and Rossi, Massimiliano and Frimmer, Martin and Novotny, Lukas},
	date = {2021/07/01},
	doi = {10.1038/s41586-021-03617-w},
	id = {Tebbenjohanns2021},
	isbn = {1476-4687},
	journal = {Nature},
	number = {7867},
	pages = {378--382},
	title = {Quantum control of a nanoparticle optically levitated in cryogenic free space},
	url = {https://doi.org/10.1038/s41586-021-03617-w},
	volume = {595},
	year = {2021}}

@article{Marquardt2014,
  title = {Cavity optomechanics},
  author = {Aspelmeyer, Markus and Kippenberg, Tobias J. and Marquardt, Florian},
  journal = {Rev. Mod. Phys.},
  volume = {86},
  issue = {4},
  pages = {1391--1452},
  numpages = {62},
  year = {2014},
  month = {Dec},
  publisher = {American Physical Society},
  doi = {10.1103/RevModPhys.86.1391},
  url = {https://link.aps.org/doi/10.1103/RevModPhys.86.1391}
}

@article{Li2018,
  title = {Optically Levitated Nanodumbbell Torsion Balance and GHz Nanomechanical Rotor},
  author = {Ahn, Jonghoon and Xu, Zhujing and Bang, Jaehoon and Deng, Yu-Hao and Hoang, Thai M. and Han, Qinkai and Ma, Ren-Min and Li, Tongcang},
  journal = {Phys. Rev. Lett.},
  volume = {121},
  issue = {3},
  pages = {033603},
  numpages = {5},
  year = {2018},
  month = {Jul},
  publisher = {American Physical Society},
  doi = {10.1103/PhysRevLett.121.033603},
  url = {https://link.aps.org/doi/10.1103/PhysRevLett.121.033603}
}

@article{Li2020,
	author = {Ahn, Jonghoon and Xu, Zhujing and Bang, Jaehoon and Ju, Peng and Gao, Xingyu and Li, Tongcang},
	date = {2020/02/01},
	doi = {10.1038/s41565-019-0605-9},
	id = {Ahn2020},
	isbn = {1748-3395},
	journal = {Nature Nanotechnology},
	number = {2},
	pages = {89--93},
	title = {Ultrasensitive torque detection with an optically levitated nanorotor},
	url = {https://doi.org/10.1038/s41565-019-0605-9},
	volume = {15},
	year = {2020}}

@article{Millen2017,
author = {Stefan Kuhn and Alon Kosloff and Benjamin A. Stickler and Fernando Patolsky and Klaus Hornberger and Markus Arndt and James Millen},
journal = {Optica},
number = {3},
pages = {356--360},
publisher = {Optica Publishing Group},
title = {Full rotational control of levitated silicon nanorods},
volume = {4},
month = {Mar},
year = {2017},
url = {https://opg.optica.org/optica/abstract.cfm?URI=optica-4-3-356},
doi = {10.1364/OPTICA.4.000356},
}

@article{Kim2021,
	author = {Stickler, Benjamin A. and Hornberger, Klaus and Kim, M. S.},
	date = {2021/08/01},
	doi = {10.1038/s42254-021-00335-0},
	id = {Stickler2021},
	isbn = {2522-5820},
	journal = {Nature Reviews Physics},
	number = {8},
	pages = {589--597},
	title = {Quantum rotations of nanoparticles},
	url = {https://doi.org/10.1038/s42254-021-00335-0},
	volume = {3},
	year = {2021}}

@article{Vamivakas2015,
	author = {Neukirch, Levi P. and von Haartman, Eva and Rosenholm, Jessica M. and Nick Vamivakas, A.},
	date = {2015/10/01},
	doi = {10.1038/nphoton.2015.162},
	id = {Neukirch2015},
	isbn = {1749-4893},
	journal = {Nature Photonics},
	number = {10},
	pages = {653--657},
	title = {Multi-dimensional single-spin nano-optomechanics with a levitated nanodiamond},
	url = {https://doi.org/10.1038/nphoton.2015.162},
	volume = {9},
	year = {2015}}

@article{Li2016,
	author = {Hoang, Thai M. and Ahn, Jonghoon and Bang, Jaehoon and Li, Tongcang},
	date = {2016/07/19},
	doi = {10.1038/ncomms12250},
	id = {Hoang2016},
	isbn = {2041-1723},
	journal = {Nature Communications},
	number = {1},
	pages = {12250},
	title = {Electron spin control of optically levitated nanodiamonds in vacuum},
	url = {https://doi.org/10.1038/ncomms12250},
	volume = {7},
	year = {2016}}

@article{Quidant2018,
	author = {Conangla, Gerard P. and Schell, Andreas W. and Rica, Ra{\'u}l A. and Quidant, Romain},
	date = {2018/06/13},
	doi = {10.1021/acs.nanolett.8b01414},
	isbn = {1530-6984},
	journal = {Nano Letters},
	journal1 = {Nano Letters},
	journal2 = {Nano Lett.},
	month = {06},
	number = {6},
	pages = {3956--3961},
	publisher = {American Chemical Society},
	title = {Motion Control and Optical Interrogation of a Levitating Single Nitrogen Vacancy in Vacuum},
	type = {doi: 10.1021/acs.nanolett.8b01414},
	url = {https://doi.org/10.1021/acs.nanolett.8b01414},
	volume = {18},
	year = {2018},
	year1 = {2018}}

@article{Lukin2020,
  title = {Single-Spin Magnetomechanics with Levitated Micromagnets},
  author = {Gieseler, J. and Kabcenell, A. and Rosenfeld, E. and Schaefer, J. D. and Safira, A. and Schuetz, M. J. A. and Gonzalez-Ballestero, C. and Rusconi, C. C. and Romero-Isart, O. and Lukin, M. D.},
  journal = {Phys. Rev. Lett.},
  volume = {124},
  issue = {16},
  pages = {163604},
  numpages = {7},
  year = {2020},
  month = {Apr},
  publisher = {American Physical Society},
  doi = {10.1103/PhysRevLett.124.163604},
  url = {https://link.aps.org/doi/10.1103/PhysRevLett.124.163604}
}

@article{Ulbricht2020,
  title = {Ultralow Mechanical Damping with Meissner-Levitated Ferromagnetic Microparticles},
  author = {Vinante, A. and Falferi, P. and Gasbarri, G. and Setter, A. and Timberlake, C. and Ulbricht, H.},
  journal = {Phys. Rev. Appl.},
  volume = {13},
  issue = {6},
  pages = {064027},
  numpages = {13},
  year = {2020},
  month = {Jun},
  publisher = {American Physical Society},
  doi = {10.1103/PhysRevApplied.13.064027},
  url = {https://link.aps.org/doi/10.1103/PhysRevApplied.13.064027}
}

@article{Budker2019,
  title = {Dynamics of a Ferromagnetic Particle Levitated over a Superconductor},
  author = {Wang, Tao and Lourette, Sean and O'Kelley, Sean R. and Kayci, Metin and Band, Y.B. and Kimball, Derek F. Jackson and Sushkov, Alexander O. and Budker, Dmitry},
  journal = {Phys. Rev. Appl.},
  volume = {11},
  issue = {4},
  pages = {044041},
  numpages = {9},
  year = {2019},
  month = {Apr},
  publisher = {American Physical Society},
  doi = {10.1103/PhysRevApplied.11.044041},
  url = {https://link.aps.org/doi/10.1103/PhysRevApplied.11.044041}
}

@article{Harris2023,
      title={Superfluid Helium Drops Levitated in High Vacuum}, 
      author={C. D. Brown and Y. Wang and M. Namazi and G. I. Harris and M. T. Uysal and J. G. E. Harris},
      journal = {Physical Review Letters},
      volume = {130}, 
      pages = {216001},
      year = {2023},
      doi = {https://doi.org/10.1103/PhysRevLett.130.216001}
}

@article{Hetet2022,
  title = {Angle Locking of a Levitating Diamond Using Spin Diamagnetism},
  author = {Perdriat, M. and Huillery, P. and Pellet-Mary, C. and H\'etet, G.},
  journal = {Phys. Rev. Lett.},
  volume = {128},
  issue = {11},
  pages = {117203},
  numpages = {6},
  year = {2022},
  month = {Mar},
  publisher = {American Physical Society},
  doi = {10.1103/PhysRevLett.128.117203},
  url = {https://link.aps.org/doi/10.1103/PhysRevLett.128.117203}
}

@article{Hetet2021,
  title = {Magnetic torque enhanced by tunable dipolar interactions},
  author = {Pellet-Mary, C. and Huillery, P. and Perdriat, M. and H\'etet, G.},
  journal = {Phys. Rev. B},
  volume = {104},
  issue = {10},
  pages = {L100411},
  numpages = {5},
  year = {2021},
  month = {Sep},
  publisher = {American Physical Society},
  doi = {10.1103/PhysRevB.104.L100411},
  url = {https://link.aps.org/doi/10.1103/PhysRevB.104.L100411}
}

@article{Hetet2020,
	author = {Delord, T. and Huillery, P. and Nicolas, L. and H{\'e}tet, G.},
	date = {2020/04/01},
	doi = {10.1038/s41586-020-2133-z},
	id = {Delord2020},
	isbn = {1476-4687},
	journal = {Nature},
	number = {7801},
	pages = {56--59},
	title = {Spin-cooling of the motion of a trapped diamond},
	url = {https://doi.org/10.1038/s41586-020-2133-z},
	volume = {580},
	year = {2020}}

@article{Bose2013,
  title = {Matter-Wave Interferometry of a Levitated Thermal Nano-Oscillator Induced and Probed by a Spin},
  author = {Scala, M. and Kim, M. S. and Morley, G. W. and Barker, P. F. and Bose, S.},
  journal = {Phys. Rev. Lett.},
  volume = {111},
  issue = {18},
  pages = {180403},
  numpages = {5},
  year = {2013},
  month = {Oct},
  publisher = {American Physical Society},
  doi = {10.1103/PhysRevLett.111.180403},
  url = {https://link.aps.org/doi/10.1103/PhysRevLett.111.180403}
}

@article{Kim2016,
  title = {Free Nano-Object Ramsey Interferometry for Large Quantum Superpositions},
  author = {Wan, C. and Scala, M. and Morley, G. W. and Rahman, ATM. A. and Ulbricht, H. and Bateman, J. and Barker, P. F. and Bose, S. and Kim, M. S.},
  journal = {Phys. Rev. Lett.},
  volume = {117},
  issue = {14},
  pages = {143003},
  numpages = {6},
  year = {2016},
  month = {Sep},
  publisher = {American Physical Society},
  doi = {10.1103/PhysRevLett.117.143003},
  url = {https://link.aps.org/doi/10.1103/PhysRevLett.117.143003}
}

@article{Plenio2020,
  title = {Motional Dynamical Decoupling for Interferometry with Macroscopic Particles},
  author = {Pedernales, Julen S. and Morley, Gavin W. and Plenio, Martin B.},
  journal = {Phys. Rev. Lett.},
  volume = {125},
  issue = {2},
  pages = {023602},
  numpages = {6},
  year = {2020},
  month = {Jul},
  publisher = {American Physical Society},
  doi = {10.1103/PhysRevLett.125.023602},
  url = {https://link.aps.org/doi/10.1103/PhysRevLett.125.023602}
}

@article{Plenio2021,
  title = {Ground-State Cooling of Levitated Magnets in Low-Frequency Traps},
  author = {Streltsov, Kirill and Pedernales, Julen S. and Plenio, Martin B.},
  journal = {Phys. Rev. Lett.},
  volume = {126},
  issue = {19},
  pages = {193602},
  numpages = {6},
  year = {2021},
  month = {May},
  publisher = {American Physical Society},
  doi = {10.1103/PhysRevLett.126.193602},
  url = {https://link.aps.org/doi/10.1103/PhysRevLett.126.193602}
}

@article{Plenio2020a,
  title = {Decoherence-Free Rotational Degrees of Freedom for Quantum Applications},
  author = {Pedernales, J. S. and Cosco, F. and Plenio, M. B.},
  journal = {Phys. Rev. Lett.},
  volume = {125},
  issue = {9},
  pages = {090501},
  numpages = {7},
  year = {2020},
  month = {Aug},
  publisher = {American Physical Society},
  doi = {10.1103/PhysRevLett.125.090501},
  url = {https://link.aps.org/doi/10.1103/PhysRevLett.125.090501}
}

@article{Albrecht2014,
  title={Testing quantum gravity by nanodiamond interferometry with nitrogen-vacancy centers},
  author={Albrecht, Andreas and Retzker, Alex and Plenio, Martin B},
  journal={Physical Review A},
  volume={90},
  number={3},
  pages={033834},
  year={2014},
  doi = {10.1103/PhysRevA.90.033834}
}

@article{march2023long,
  title={Long spin coherence and relaxation times in nanodiamonds milled from polycrystalline 12C diamond},
  author={March, James E and Wood, Benjamin D and Stephen, Colin J and Fervenza, Laura Dur{\'a}n and Breeze, Ben G and Mandal, Soumen and Edmonds, Andrew M and Twitchen, Daniel J and Markham, Matthew L and Williams, Oliver A and others},
  journal={Physical Review Applied},
  volume={20},
  number={4},
  pages={044045},
  year={2023},
  doi = {10.1103/PhysRevApplied.20.044045}
}

@article{pedernales2023origin,
  title={On the origin of force sensitivity in tests of quantum gravity with delocalised mechanical systems},
  author={Pedernales, Julen S and Plenio, Martin B},
  journal={Contemporary Physics},
  volume={64},
  number={2},
  pages={147--163},
  year={2023},
  doi = {https://doi.org/10.1080/00107514.2023.2286074}
}

@article{Black2004,
title = {Toward quantum-limited position measurements using optically levitated microspheres},
journal = {Physics Letters A},
volume = {321},
number = {2},
pages = {99-102},
year = {2004},
issn = {0375-9601},
doi = {https://doi.org/10.1016/j.physleta.2003.12.022},
url = {https://www.sciencedirect.com/science/article/pii/S0375960103018413},
author = {Kenneth G. Libbrecht and Eric D. Black}
}

@Article{Levitt1986,
  author    = {Levitt, Malcolm H.},
  journal   = {Progress in Nuclear Magnetic Resonance Spectroscopy},
  title     = {Composite pulses},
  year      = {1986},
  issn      = {0079-6565},
  month     = jan,
  number    = {2},
  pages     = {61--122},
  volume    = {18},
  doi       = {10.1016/0079-6565(86)80005-x},
  publisher = {Elsevier BV},
}

@Article{Levitt2007,
  author    = {Levitt, Malcolm H.},
  month     = mar,
  title     = {Composite Pulses},
  year      = {2007},
  doi       = {10.1002/9780470034590.emrstm0086},
  isbn      = {9780470034590},
  journal   = {Encyclopedia of Magnetic Resonance},
  publisher = {John Wiley \& Sons, Ltd},
}

@Article{Deimling1980,
  author    = {Deimling, M and Brunner, H and Dinse, K.P and Hausser, K.H and Colpa, J.P},
  journal   = {Journal of Magnetic Resonance (1969)},
  title     = {Microwave-induced optical nuclear polarization (MI-ONP)},
  year      = {1980},
  issn      = {0022-2364},
  month     = may,
  number    = {2},
  pages     = {185--202},
  volume    = {39},
  doi       = {10.1016/0022-2364(80)90128-6},
  publisher = {Elsevier BV},
}

@Article{Iinuma2000,
  author    = {Iinuma, M. and Takahashi, Y. and Shaké, I. and Oda, M. and Masaike, A. and Yabuzaki, T. and Shimizu, H. M.},
  journal   = {Physical Review Letters},
  title     = {High Proton Polarization by Microwave-Induced Optical Nuclear Polarization at 77 K},
  year      = {2000},
  issn      = {1079-7114},
  month     = jan,
  number    = {1},
  pages     = {171--174},
  volume    = {84},
  doi       = {10.1103/physrevlett.84.171},
  publisher = {American Physical Society (APS)},
}

@Article{VanStrien1980,
  author    = {Van Strien, A.J. and Schmidt, J.},
  journal   = {Chemical Physics Letters},
  title     = {An EPR study of the triplet state of pentacene by electron spin-echo techniques and laser flash excitation},
  year      = {1980},
  issn      = {0009-2614},
  month     = mar,
  number    = {3},
  pages     = {513--517},
  volume    = {70},
  doi       = {10.1016/0009-2614(80)80115-1},
  publisher = {Elsevier BV},
}

@Article{Sakamoto2023,
  author    = {Sakamoto, Keita and Hamachi, Tomoyuki and Miyokawa, Katsuki and Tateishi, Kenichiro and Uesaka, Tomohiro and Kurashige, Yuki and Yanai, Nobuhiro},
  journal   = {Proceedings of the National Academy of Sciences},
  title     = {Polarizing agents beyond pentacene for efficient triplet dynamic nuclear polarization in glass matrices},
  year      = {2023},
  number    = {44},
  volume    = {120},
  pages     = {e2307926120},
  doi       = {10.1073/pnas.2307926120},
  publisher = {Proceedings of the National Academy of Sciences},
}

@Article{Sellner2017,
  author    = {Sellner, S and Besirli, M and Bohman, M and Borchert, M J and Harrington, J and Higuchi, T and Mooser, A and Nagahama, H and Schneider, G and Smorra, C and Tanaka, T and Blaum, K and Matsuda, Y and Ospelkaus, C and Quint, W and Walz, J and Yamazaki, Y and Ulmer, S},
  journal   = {New Journal of Physics},
  title     = {Improved limit on the directly measured antiproton lifetime},
  year      = {2017},
  issn      = {1367-2630},
  month     = aug,
  number    = {8},
  pages     = {083023},
  volume    = {19},
  doi       = {10.1088/1367-2630/aa7e73},
  publisher = {IOP Publishing},
}

@Article{Wasielewski2020,
  author    = {Wasielewski, Michael R. and Forbes, Malcolm D. E. and Frank, Natia L. and Kowalski, Karol and Scholes, Gregory D. and Yuen-Zhou, Joel and Baldo, Marc A. and Freedman, Danna E. and Goldsmith, Randall H. and Goodson, Theodore and Kirk, Martin L. and McCusker, James K. and Ogilvie, Jennifer P. and Shultz, David A. and Stoll, Stefan and Whaley, K. Birgitta},
  journal   = {Nature Reviews Chemistry},
  title     = {Exploiting chemistry and molecular systems for quantum information science},
  year      = {2020},
  issn      = {2397-3358},
  month     = jul,
  number    = {9},
  pages     = {490--504},
  volume    = {4},
  doi       = {10.1038/s41570-020-0200-5},
  publisher = {Springer Science and Business Media LLC},
}

@article{Pedernales2022,
  title = {Robust macroscopic matter-wave interferometry with solids},
  author = {Pedernales, Julen S. and Plenio, Martin B.},
  journal = {Phys. Rev. A},
  volume = {105},
  issue = {6},
  pages = {063313},
  numpages = {8},
  year = {2022},
  month = {Jun},
  publisher = {American Physical Society},
  doi = {10.1103/PhysRevA.105.063313},
  url = {https://link.aps.org/doi/10.1103/PhysRevA.105.063313}
}

@Article{Faltermeier2006,
  author    = {Faltermeier, Daniel and Gompf, Bruno and Dressel, Martin and Tripathi, Ashutosh K. and Pflaum, Jens},
  journal   = {Physical Review B},
  title     = {Optical properties of pentacene thin films and single crystals},
  year      = {2006},
  issn      = {1550-235X},
  month     = sep,
  number    = {12},
  pages     = {125416},
  volume    = {74},
  doi       = {10.1103/physrevb.74.125416},
  publisher = {American Physical Society (APS)},
}

@Article{Naito2024,
  author    = {Naito, Tomoya and Suzuki, Tomoaki and Ikezoe, Yasuhiro},
  journal   = {Applied Physics Letters},
  title     = {Diamagnetic levitation of water realized with a simple device consisting of ordinary permanent magnets},
  year      = {2024},
  issn      = {1077-3118},
  month     = dec,
  number    = {26},
  volume    = {125},
  doi       = {10.1063/5.0241203},
  publisher = {AIP Publishing},
}

@Article{Rusconi2017,
  author    = {Rusconi, C. C. and Pöchhacker, V. and Cirac, J. I. and Romero-Isart, O.},
  journal   = {Physical Review B},
  title     = {Linear stability analysis of a levitated nanomagnet in a static magnetic field: Quantum spin stabilized magnetic levitation},
  year      = {2017},
  issn      = {2469-9969},
  month     = oct,
  number    = {13},
  pages     = {134419},
  volume    = {96},
  doi       = {10.1103/physrevb.96.134419},
  publisher = {American Physical Society (APS)},
}

@Article{Rugar2004,
  author    = {Rugar, D. and Budakian, R. and Mamin, H. J. and Chui, B. W.},
  journal   = {Nature},
  title     = {Single spin detection by magnetic resonance force microscopy},
  year      = {2004},
  issn      = {1476-4687},
  month     = jul,
  number    = {6997},
  pages     = {329--332},
  volume    = {430},
  doi       = {10.1038/nature02658},
  publisher = {Springer Science and Business Media LLC},
}

\onecolumn
\appendix

\section{Spin dynamics in a quadrupole magnetic field}\label{App:Spin dynamics in physical magnetic field}

While we assumed a magnetic field of the form $\vec{B}=(B' \hat{x}+B_0)\vec{k}$ in the main text, where $\vec{k}$ defines the quantization axis of the spins. This field does not satisfy Maxwell's equations which dictate that the magnetic field has to be free of divergences, namely $\nabla\cdot\vec{B}=0$. This requires at least one other component perpendicular to $\vec{k}$. One simple variant that fulfills Maxwell's equations is given by the 2D quadrupole field $\vec{B}=(-B' \hat{z}, 0, B'\hat{x}+B_0)$. This field provides a confinement in one additional dimension and could, in principle, induce spin transitions.
The corresponding Hamiltonian of this configuration is given by
\begin{align}
    \hat{H}=\frac{\hat{p}^2_z+ \hat{p}_x^2}{2m}-\frac{\chi_V V}{2 \mu_0}\left[(B' \hat{z})^2+(B' \hat{x})^2\right]+\frac{\hbar}{2} B'\gamma_p \sum_{n=1}^N \left(-\hat{z} \hat{\sigma}_x^{(n)} + \hat{x} \hat{\sigma}_z^{(n)} \right) + \frac{\hbar}{2}B_0 \gamma_p \sum_{n=1}^N\hat{\sigma}_z^{(n)}.
\end{align}
By switching to ladder operators defined by $\hat{x}_i=\sqrt{\hbar/(2m\omega)}(\hat{a}_i^\dagger+\hat{a}_i), \hat{p}_i=i \sqrt{\hbar m \omega/2}(\hat{a}_i^\dagger-\hat{a}_i)$, the Hamiltonian can be rewritten to find
\begin{equation}
    \hat{H}=\hbar \Omega(\hat{a}_z^\dagger\hat{a}_z+\hat{a}_x^\dagger\hat{a}_x)+ \frac{\hbar}{2} \Gamma \left[ -(\hat{a}_z^\dagger+\hat{a}_z)\sum_{n=1}^N\hat{\sigma}_x^{(n)} + (\hat{a}_x^\dagger+\hat{a}_x)\sum_{n=1}^N\hat{\sigma}_z^{(n)} \right]+\frac{\hbar}{2}\omega_0 \sum_{n=1}^N\hat{\sigma}_z^{(n)}
\end{equation}
where we defined 
\begin{equation}
    \Omega=\sqrt{\frac{|\chi_V|}{\rho \mu_0}}B', \quad \Gamma=\gamma_p B' \sqrt{\frac{\hbar}{2 m \omega}} \quad \mathrm{and} \quad \omega_0=\gamma_p B_0.
\end{equation}
Here, $\Omega$ is the mode frequency, and $\Gamma$ the coupling strength between the modes and the spins. We write the Pauli matrices in spin-ladder operators as $\hat{\sigma}_x=\hat{\sigma}_+ + \hat{\sigma}_-$ and by switching into a rotating picture with respect to $\hbar \Omega(\hat{a}_z^\dagger\hat{a}_z+\hat{a}_x^\dagger\hat{a}_x)+ \frac{\hbar}{2} \omega_0 \sum_{n=1}^N\hat{\sigma}_z^{(n)}$ we find
\begin{equation}
    \hat{H}_\text{int}^I= \frac{\hbar}{2} \Gamma \sum_{n=1}^N \left[-(\hat{a}_z^\dagger e^{i \Omega t}+\hat{a}_z e^{- i \Omega t})(\hat{\sigma}_+^{(n)} e^{i \omega_0 t} + \sigma_-^{(n)}e^{-i \omega_0 t}) + (\hat{a}_x^\dagger e^{i\Omega t}+\hat{a}_xe^{-i \Omega t})\hat{\sigma}_z^{(n)} \right]
\end{equation}
For $B'=10^4\mathrm{T/m}$, the trap frequency is given by $ \Omega = 130(2\pi)\,$Hz. For a particle of $100\,$nm radius, the coupling strength is given by $\Gamma=50(2\pi)\,$Hz. As the eigenfrequency of the  nuclear spins is given by $\omega_0 =\gamma_p B_0$, already at a magnetic field of $B_0=1$ mT, $\omega_0 = 43(2\pi)\,$kHz, ensuring that the spins are far detuned from $\Omega$. Thus, in most experimentally relevant configurations, we have that $\omega_0 \gg \{ \Gamma, \Omega \}$, and we can safely discard terms rotating at frequencies $\omega_0 \pm \Omega \gg \Gamma$
\begin{equation}
    \hat{H}=\hbar \Omega(\hat{a}_z^\dagger\hat{a}_z+\hat{a}_x^\dagger\hat{a}_x)+ \frac{\hbar}{2} \Gamma (\hat{a}_x^\dagger+\hat{a}_x)\sum_{n=1}^N\hat{\sigma}_z^{(n)} +\frac{\hbar}{2}\omega_0 \sum_{n=1}^N\hat{\sigma}_z^{(n)}.
\end{equation}
That is, the coupling of the spins to the $x_1$ mode will generally be highly non-resonant and its effect negligible. This justifies the use of a one dimensional gradient in the analysis provided in the main text.

\section{Decoherence between \texorpdfstring{$\kappa^\mathrm{th}$}{kappa-th} and \texorpdfstring{$\kappa^{\prime \mathrm{th}}$}{kappa'-th} wave packet} \label{App:DecoherenceTwoPaths}

In this section, we want to find an expression for $\tr{\hat{\rho}_{\kappa,\kappa'}(T_\mathrm{tot})}$ in the presence of the CSL model. Using \cref{eq:CSLModelPosSpace}, together with \cref{eq:rhoKappaKappapWithoutNoise}, the master equation governing the time evolution of $\rho_{\kappa,\kappa'}(q',q'')=\bra{q'}\hat{\rho}_{\kappa,\kappa'}\ket{q''}$ is given by
\begin{align}
    \bra{q'}\dot{\hat{\rho}}_{\kappa,\kappa^\prime}\ket{q''}=&-\frac{i}{\hbar}\bra{q'}\hat{H}_{\kappa} \hat{\rho}_{\kappa,\kappa^\prime} -  \hat{\rho}_{\kappa,\kappa^\prime} \hat{H}_{\kappa^\prime} + \eta_{\kappa, \kappa'} \hat{\rho}_{\kappa,\kappa^\prime}\ket{q''} -\xi \left( 1- e^{-\frac{(q'-q'')^2}{4 r_\mathrm{CSL}^2}}\right)\rho_{\kappa,\kappa'}(q',q'').
\end{align}
It can be easily shown that
\begin{align}
    &e^{-\frac{(q'-q'')^2}{4 r_\mathrm{CSL}^2}}\rho(q',q'')=\frac{1}{\sqrt{\pi} r_\mathrm{CSL}}\bra{q'}\int_{-\infty}^{\infty} \,\mathrm{d}x\, e^{-\frac{(\hat{q}-x)^2}{2 r^2_\mathrm{CSL}}} \hat{\rho} e^{-\frac{(\hat{q}-x)^2}{2 r^2_\mathrm{CSL}}}\ket{q''}.
\end{align}
We define
\begin{equation}
    T[\hat{\rho}]\equiv\frac{1}{\sqrt{\pi} r_\mathrm{CSL}}\,\int_{-\infty}^{\infty} \,\mathrm{d}x\,  e^{-\frac{(\hat{q}-x)^2}{2 r^2_\mathrm{CSL}}} \hat{\rho} e^{-\frac{(\hat{q}-x)^2}{2 r^2_\mathrm{CSL}}}
\end{equation}
and write the master equation as
\begin{align}
   \bra{q'}\dot{\hat{\rho}}_{\kappa,\kappa^\prime}\ket{q''}= &-\frac{i}{\hbar}\bra{q'} \hat{H}_{\kappa} \hat{\rho}_{\kappa,\kappa^\prime} -  \hat{\rho}_{\kappa,\kappa^\prime} \hat{H}_{\kappa^\prime}+ \eta_{\kappa, \kappa'} \hat{\rho}_{\kappa,\kappa^\prime} -\xi \left( \hat{\rho}-T[\hat{\rho}] \right) \ket{q''}. \nonumber \\
\end{align}
First, we go into the rotating picture to eliminate the unitary time evolution by defining
\begin{equation}
    \hat{\rho}^S_{\kappa,\kappa^\prime}=\hat{U}^\dagger_\kappa(t) \hat{\rho}_{\kappa,\kappa^\prime}\hat{U}_{\kappa^\prime}(t) e^{\frac{i}{\hbar} \eta_{\kappa,\kappa'}}.
\end{equation}
With
\begin{align}
    \hat{U}_\kappa(t)&=e^{-\frac{i}{\hbar}\hat{H}_k t}
\end{align}
the equation governing the evolution of $\hat{\rho}^S_{\kappa,\kappa^\prime}$ is given by
\begin{align}
    \dot{\hat{\rho}}^S_{\kappa,\kappa^\prime}&=-\xi \left(\hat{\rho}^S_{\kappa,\kappa^\prime}-T^S[\hat{\rho}^S_{\kappa,\kappa^\prime}]\right) \nonumber
\end{align}
where
\begin{align}
    T^S[\hat{\rho}^S_{\kappa,\kappa^\prime}]&=\nonumber \hat{U}^\dagger_\kappa(t) T[\hat{\rho}_{\kappa,\kappa^\prime}] \hat{U}_{\kappa^\prime}(t).
\end{align}
$T^S[\hat{\rho}^S_{\kappa,\kappa^\prime}]$ can be further simplified by using the unitary nature of $\hat{U}_\kappa(t)$. With $\hat{q}_\kappa\equiv \hat{U}^\dagger_\kappa(t) \hat{q} \hat{U}_{\kappa}(t)$ we find
\begin{align}
    T^S[\hat{\rho}^S_{\kappa,\kappa^\prime}]&=\frac{1}{\sqrt{\pi}r_\mathrm{CSL}}\,\int_{-\infty}^{\infty} \,\mathrm{d}x\, e^{-\frac{(\hat{q}_\kappa-x)^2}{2 r^2_\mathrm{CSL}}} \hat{\rho}^S_{\kappa,\kappa^\prime} e^{-\frac{(\hat{q}_{\kappa^\prime}-x)^2}{2r^2_\mathrm{CSL}}}.
\end{align}

To evaluate $\hat{q}_\kappa$, one first has to find an expression for the time evolution of the $\kappa^\mathrm{th}$ wave packet. We realize that we can write $\hat{H}_\kappa$ in terms of the ladder operators $\hat{a},\hat{a}^\dagger$ defined by $\hat{q}=x_0(\hat{a}^\dagger+\hat{a})$ and $\hat{p}=i p_0(\hat{a}^\dagger-\hat{a})$. $x_0$ and $p_0$ are given by $x_0=\sqrt{\hbar/(2 m \Omega)}$ and $p_0=\sqrt{\hbar m \Omega/2}$. By using the properties of the displacement operator $\hat{D}(\zeta)=\exp(\zeta \hat{a}^\dagger-\zeta^*\hat{a})$, we can write
\begin{equation}
    e^{-\frac{i}{\hbar} \hat{H}_\kappa t}=e^{i \Omega t \hat{a}^\dagger \hat{a}} \hat{D}\left(\frac{\chi}{2 x_0} (2\kappa-N)\right) e^{-i \Omega t \hat{a}^\dagger \hat{a}}.
\end{equation}
Each $\pi$ pulse flips all spins, therefore $\kappa$ goes to $N-\kappa$. We can use the properties of the displacement operator to find the full-time evolution along the $\kappa^\mathrm{th}$ trajectory, which can be written as
\begin{align}
    \hat{U}_\kappa(t)=e^{-i\left(\frac{\chi}{2 x_0}\right)^2\phi_\kappa(t)}\hat{D}\left(\frac{\chi}{2 x_0}\zeta_\kappa(t)\right) e^{-i\Omega t\hat{a}^\dagger\hat{a}}.
\end{align}
$\zeta_\kappa(t)$ represents the classical phase space trajectory of the $\kappa^\mathrm{th}$ wave packet. The position of the center of this wave packet is given by $\mathcal{X}_\kappa(t)=\chi\Re[\zeta_\kappa(t)]$. $\phi_\kappa(t)=\phi_\kappa^\prime+(2\kappa-N)^2 \eta T_\mathrm{tot}/\hbar$ as well as  $\zeta_\kappa(t)$ are given by

\begin{align}
\zeta_\kappa=(2\kappa-N)\begin{cases} 
      1-e^{-i \Omega t } & 0 \leq t < t_1 \\
      e^{-i \Omega  (t_1+t)}-2 e^{-i t_1 \Omega }+1 & t_1 \leq t < t_1+t_2 \\
      2 e^{-i \Omega  (t_1+t_2)}-e^{-i \Omega  (t_1+t_2+t)} -2 e^{-i t_1 \Omega }+1 & t_1+t_2 \leq t < T_\mathrm{tot}
   \end{cases}
   \label{eq:Path1}
\end{align}
and
\begin{align}
\phi^\prime_\kappa(t)=
&-(2\kappa-N)^2\begin{cases} 
      \sin (t \Omega ) & 0 \leq t < t_1 \\
      2 \left[\sin (t \Omega )+\sin (t_2 \Omega )\right] -\sin (\Omega  (t_1+t)) & t_1 \leq t < t_1+t_2 \\
      2 \sin( t\Omega)+2 \sin(t_1 \Omega) +4 \sin(t_2 \Omega) - \\ 
      \;\;2 \big[\sin((t+t_2)\Omega)+\sin((t_1+t_2)\Omega)\big]\\
      \;\;+\sin((t+t_1+t_2)\Omega)& t_1+t_2 \leq t < T_\mathrm{tot}
   \end{cases}.
   \label{eq:Phase1}
\end{align}

By using the properties of the displacement operator and the Baker–Campbell–Hausdorff formula, $\hat{q}_\kappa$ can be written as
\begin{align}
    \hat{U}^\dagger_\kappa(t) \hat{q} \hat{U}_\kappa(t) &= \cos(\Omega t) \hat{q}+\sin(\Omega t) \hat{p} \frac{x_0}{p_0} + \mathcal{X}_\kappa(t).
    \label{eq:AppxShift}
\end{align}

As we are interested in $\tr(\hat{\rho}_{\kappa,\kappa^\prime})$, we use the linearity and the cyclic properties of the trace and $[\hat{q}_\kappa,\hat{q}_{\kappa^\prime}]=0$ to find
\begin{align}
    \tr{T^S[\hat{\rho}^S]}&=\frac{1}{\sqrt{\pi} r_\mathrm{CSL}}\,\tr{\int_{-\infty}^{\infty}\mathrm{d}x\,  e^{\frac{(\hat{q}_\kappa-x)^2-(\hat{q}_{\kappa^\prime}-x)^2}{2 r^2_\mathrm{CSL}}} \hat{\rho}^S } 
    =\tr{e^{-\frac{(\hat{q}_\kappa-\hat{q}_{\kappa^\prime})^2}{4 r^2_\mathrm{CSL}}} \hat{\rho}^S }.
\label{eq:AppRhoTildeIntermediate}
\end{align}
When plugging \cref{eq:AppxShift} into \cref{eq:AppRhoTildeIntermediate}, all terms involving $\hat{q}$ and $\hat{p}$ cancel out and one finds
\begin{align}
    \pdv{\tr{\hat{\rho}^S_{\kappa,\kappa^\prime}}}{t}=-\xi \left[1-e^{-\frac{(\mathcal{X}_\kappa(t)-\mathcal{X}_{\kappa^\prime}(t))^2}{4 r^2_\mathrm{CSL}}}\right] \tr{\hat{\rho}^S_{\kappa,\kappa^\prime}}.
\end{align}
This differential equation can be easily solved by
\begin{align}
    \tr{\hat{\rho}^S_{\kappa,\kappa^\prime}(T_\mathrm{tot})}=&e^{-\Lambda_{\kappa,\kappa'}}\tr{\hat{\rho}^S_{\kappa,\kappa^\prime}(0)}
\end{align}
with
\begin{equation}
    \Lambda_{\kappa, \kappa'} =   \xi \int_0^{T_\mathrm{tot}} (1-e^{-\frac{[\mathcal{X}_\kappa(t^\prime)-\mathcal{X}_{\kappa^\prime}(t^\prime)]^2}{4 r^2_\mathbf{CSL}}})\,\mathrm{d} t'.
\end{equation}
The protocol is designed such that the $\kappa^\mathrm{th}$ and the $\kappa^{\prime\,\mathrm{th}}$ wave packets overlap, meaning that  $\zeta_\kappa(T_\mathrm{tot})=\zeta_{\kappa^\prime}(T_\mathrm{tot})$. Hence, $\hat{U}_\kappa(T_\mathrm{tot})$ and $\hat{U}_{\kappa^\prime}(T_\mathrm{tot})$ are the same up to a phase. Therefore, $\tr{\hat{\rho}^S_{\kappa,\kappa^\prime}(T_\mathrm{tot})}=\tr{\hat{\rho}_{\kappa,\kappa^\prime}(T_\mathrm{tot})}\exp(-i \Delta \phi_{\kappa,\kappa^\prime}(T_\mathrm{tot}))$ holds true, where we defined $\Delta\phi_{\kappa,\kappa'} \equiv \phi_{\kappa}(T_\mathrm{tot})-\phi_{\kappa'}(T_\mathrm{tot})$. At $t=0$, the wave packets did not split up yet. Therefore, we know $\tr{\hat{\rho}_{\kappa,\kappa^\prime}(0)}=\tr{\hat{\rho}(0)}=1$. $\tr{\hat{\rho}_{\kappa,\kappa^\prime}(T_\mathrm{tot})}$ is then given by
\begin{align}
    \tr{\hat{\rho}_{\kappa,\kappa^\prime}}=&e^{-i \Delta\phi_{\kappa,\kappa'}}e^{-\Lambda_{\kappa,\kappa'}}.
    \label{eq:DGLFullDecoherence}
\end{align}
If $(\mathcal{X}_\kappa(t^\prime)-\mathcal{X}_{\kappa^\prime}(t^\prime) ) \ll r_\mathrm{CSL}$, $\Lambda_{\kappa,\kappa'}$ simplifies to 
\begin{align}
    \Lambda_{\kappa,\kappa'}\approx (T_\mathrm{tot}) & \frac{\xi}{4 r^2_\mathrm{CSL}} \int_0^{T_\mathrm{tot}} (\mathcal{X}_\kappa(t^\prime)-\mathcal{X}_{\kappa^\prime}(t^\prime))^2\,\mathrm{d} t'.
\end{align}

\section{Expectation value of \texorpdfstring{$\langle\hat{M}\rangle$}{Lg} } \label{App:Expectation value of M}
We want to find the expectation value of $\langle\hat{M}\rangle(T_\mathrm{tot})=\tr{\hat{U}_{\pi/2}^{\otimes N}\hat{\rho}_\mathrm{s}(T_\mathrm{tot})\hat{U}^{\dagger \otimes N}_{\pi/2}\hat{M}}$, where $\hat{U}_{\pi/2}$ is a $\pi/2$ pulse acting on the spins and $\hat{\rho}_\mathrm{S}$ is given by \cref{eq:rhoSpin1}. Due to the cyclic properties of the trace and since $\hat{U}_{\pi/2}\hat{\sigma}_z\hat{U}^\dagger_{\pi/2}=\hat{\sigma}_x$, $\langle \hat{M}\rangle(T_\mathrm{tot})$ is given by

\begin{align}
    \langle\hat{M}\rangle(T_\mathrm{tot})&=\frac{\hbar\gamma_p}{2^{N+1}}\sum_{\kappa,\kappa^\prime=0}^N \sqrt{\binom{N}{\kappa}\binom{N}{\kappa^\prime}}  e^{-i \Delta \phi_{\kappa,\kappa^\prime}}e^{-\Lambda_{\kappa,\kappa'}}~\tr{\hat{X}\ket{\kappa}\bra{\kappa^\prime}}.
\end{align}
Here, $\hat{X}=\sum_{n=1}^N \hat{\sigma}_x^{(n)}$ is the sum over all $\hat{\sigma}_x$ operators. 

To find an expression for $\hat{X}\ket{\kappa}$, one can to use combinatorics and the relation $\hat{\sigma}_x\ket{\uparrow/\downarrow}=\ket{\downarrow/\uparrow}$. It can be shown that
\begin{align}
    \sqrt{ \binom{N}{\kappa} }\hat{X}\ket{\kappa}=& \left[ N-(\kappa-1) \right]\sqrt{\binom{N}{\kappa-1}}\ket{\kappa-1} +(\kappa+1)\sqrt{\binom{N}{\kappa+1}}\ket{\kappa+1}.
\end{align}

Since Dicke states are orthonormal, we find that for $\langle \hat{M} \rangle(T_\mathrm{tot})$ only terms like $\kappa^\prime=\kappa\pm 1$ contribute to the double sum. Furthermore, $\mathcal{X}_\kappa(t)$ is linear in $\kappa$, therefore only the nearest neighbors' decoherences contribute and $\mathcal{X}_\kappa(t^\prime)-\mathcal{X}_{\kappa\pm1}(t^\prime)$ is independent of $\kappa$ and can be taken out of the sum. The left sum has a closed form solution that is given by
\begin{align}
    \langle\hat{M}\rangle(T_\mathrm{tot})=\langle\hat{M}\rangle(0) e^{-\gamma}\cos^{N-1}(4 \phi)
\end{align}
with 
\begin{align}
    &\phi=(\chi/2 x_0)^2 \sin(\Omega t_1)+\eta T_\mathrm{tot}/\hbar, \quad \gamma=\Lambda_{\kappa,\kappa\pm1}.
\end{align}
If $\mathcal{X}_{\kappa,\kappa\pm1}(t^\prime) \ll r_\mathrm{CSL}$ at all times $\gamma$ simplifies to $\gamma=\xi/(2r_\mathrm{CSL}^2) \int_0^{T_\mathrm{tot}} \mathcal{X}_{\kappa,\kappa\pm1}(t^\prime)^2\,\mathrm{d} t'$, which has a closed form solution given by
\begin{align}
    \gamma&=\frac{\xi}{4 r_\mathrm{CSL}^2}\frac{\chi^2}{4\Omega} \Big( 14\Omega(2 t_1 + t_2) - 8\Omega(t_1 + t_2) \cos(\Omega t_1) + 8\Omega t_1(-2\cos(\Omega t_2) + \cos(\Omega(t_1 + t_2))) \nonumber \\
    & \hspace{2.2cm} + 8\sin(\Omega t_1) - 11\sin(2\Omega t_1) + 4\sin(3\Omega t_1) - \sin(4\Omega t_1) + 24\sin(2\Omega(t_1 + t_2)) \nonumber \\
    & \hspace{2.2cm} - 4\sin(3\Omega(t_1 + t_2)) + 3\sin(4\Omega(t_1 + t_2)) - 8\sin(\Omega(2t_1 + t_2)) + 5\sin(2\Omega(2t_1 + t_2)) \nonumber \\
    & \hspace{2.2cm} - 16\sin(\Omega(t_1 + 2t_2)) - 4\sin(2\Omega(t_1 + 2t_2)) - 12\sin(3\Omega t_1 + 2\Omega t_2) \nonumber \\
    & \hspace{2.2cm} + 8\sin(2\Omega t_1 + 3\Omega t_2) - 8\sin(4\Omega t_1 + 3\Omega t_2) + 4\sin(5\Omega t_1 + 3\Omega t_2) \nonumber \\
    & \hspace{2.2cm} + 4\sin(3\Omega t_1 + 4\Omega t_2) - 4\sin(5\Omega t_1 + 4\Omega t_2) + \sin(6\Omega t_1 + 4\Omega t_2) \Big).
\end{align}

\section{\texorpdfstring{$\pi/2$}{Lg} pulse in Dicke basis}\label{App:pi/2 pulse in Dicke basis}

To write down the explicit expression for $\langle\hat{M}\rangle(T_\mathrm{tot})/\langle\hat{M}\rangle(0)$ for the modified protocol, the matrix elements of a $\pi/2$ and $3\pi/2$ pulse in the Dicke basis are needed.

To do that, we first look at the closely related Hadamard gate, i.e., we want to know what $H^{\otimes N}\ket{\kappa}$ looks like on the Dicke basis. To do so, let us look at the eigenvectors of the Hadamard gate $H$ on the spin basis. In this section, ket vectors in the spin basis will be denoted by $\ket{\cdot}_S$
\begin{equation}
H^{\otimes N}\ket{k}_S=\frac{1}{2^{N/2}} \sum_{a\in\{0,1\}}(-1)^{a\cdot k}\ket{a}_S
\end{equation}
where "$1$" corresponds to "$\uparrow$" and "$0$" corresponds to "$\downarrow$".
By definition, the Dicke state $\ket{\kappa}$ can be written in the spin basis as
\begin{equation}
\sqrt{\binom{N}{\kappa}}\ket{\kappa}=\sum_{k\in \pi(\underbrace{1,...,1}_\kappa,\underbrace{0,...,0}_{N-\kappa})} \ket{k}_S=\sum_{k\in \pi(k_0)} \ket{k}_S
\end{equation}
where $\pi$ is the permutation operator and $k_0$ is the vector $k_0=\{1\}^{\otimes \kappa}\{0\}^{\otimes (N-\kappa)}$. Hence, $H^{\otimes N}\ket{\kappa}$ can be written as
\begin{equation}
H^{\otimes N}\ket{\kappa}=\frac{1}{2^\frac{N}{2} \sqrt{\binom{N}{\kappa}}}\sum_{a\in\{0,1\}}\sum_{k\in\pi(k_0)} (-1)^{k\cdot a}\ket{a}_S.
\end{equation}

\begin{figure}
\centering
\includegraphics{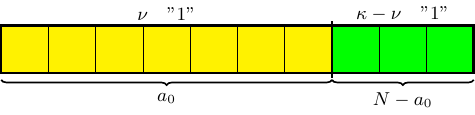}
\caption{There are $\binom{a_0}{\nu}$ possibilities to place $\nu$ ones in the yellow entries of $k'$ and $\binom{N-a_0}{\kappa-\nu}$ possibilities  to place $\kappa-\nu$ ones in the green entries.}
\label{fig:Findpi/2}
\end{figure}

Lets look at the vectors of the form $\ket{a_0}_S=\ket{\{1 \}^{\otimes a_0}\{0\}^{\otimes (N-a_0)}}$ and find the weighting factor of $\ket{a_0}_S$ i.e. $\sum_{k\in\pi(k_0)} (-1)^{k\cdot a_0}$. More explicitly, we want to find an expression for
\begin{equation}
\sum_{k\in\pi(k_0)} (-1)^{k\cdot a_0}=\sum_{k\in\pi(k_0)} (-1)^{\sum_{i=1}^{a_0} k_i}\, .
\end{equation}
To evaluate this expression, let us look at \cref{fig:Findpi/2}. 
We sum over all permutations of $k_0$ that contain $\kappa$ ones in total. For a given permutation that contains $\nu$ ones in the first $a_0$ entries, there are $\kappa-\nu$ ones in the last $N-a_0$ entries. 
There are $\binom{a_0}{\nu}\binom{N-a_0}{\kappa-\nu}$ possibilities to distribute the $\nu$ and $N-\nu$ ones that way. Now, $\nu$ goes from zero to $\kappa$.
Hence, we find
\begin{equation}
\sum_{k\in\pi(k_0)} (-1)^{\sum_{i=1}^{a_0} k_i}=\sum_{\nu=0}^\kappa (-1)^\nu \binom{a_0}{\nu}\binom{N-a_0}{\kappa-\nu}.
\end{equation}

Since we sum over all permutations of $k_0$, the weighting factor of all $\ket{a}_S$ with the same number of spins up will be the same. We can summarize all $\ket{a}_S$ with with same numbers of spin up in the Dicke state $\sqrt{\binom{N}{\alpha}}\ket{\alpha}$ and find
\begin{equation}
H^{\otimes N}\ket{\kappa}=\frac{1}{2^\frac{N}{2}} \sum_{\alpha=0}^N \sqrt{\frac{\binom{N}{\alpha}}{\binom{N}{\kappa}}}\sum_{\nu=0}^\kappa (-1)^\nu \binom{\alpha}{\nu}\binom{N-\alpha}{\kappa-\nu} \ket{\alpha}.
\end{equation}
The $N$-particle Hadamard gate in Dicke basis is therefore given by
\begin{equation}
H^{\otimes N}_{\kappa,\alpha}=\frac{1}{2^\frac{N}{2}}\sqrt{\frac{\binom{N}{\alpha}}{\binom{N}{\kappa}}}\sum_{\nu=0}^\kappa (-1)^\nu \binom{\alpha}{\nu}\binom{N-\alpha}{\kappa-\nu}.
\end{equation}

What is left is to retrieve the $\pi/2$ pulse from the Hadamard gate. We can do that by applying the $X$ operator on the Hadamard gate as one can check easily for the one spin case
\begin{equation}
X H =\frac{1}{\sqrt{2}}\begin{pmatrix}
0 & 1 \\ 1 & 0
\end{pmatrix}
\begin{pmatrix}
1 & 1 \\ 1 & -1
\end{pmatrix}
=\frac{1}{\sqrt{2}}
\begin{pmatrix}
1 & -1 \\ 1 & 1
\end{pmatrix} = \left(\frac{\pi}{2}\right).
\end{equation}
A $X$ gate flips $\ket{\uparrow}_S$ to $\ket{\downarrow}_S$ and the other way around. Therefore, $X^{\otimes N}$ acts on a Dicke state like $X^{\otimes N}\ket{\alpha}=\ket{N-\alpha}$. We find with a few arithmetic transformations
\begin{align}
U_{\pi/2}^{\otimes N}\ket{\kappa}&=X^{\otimes N} H^{\otimes N}\ket{\kappa} =\frac{1}{2^\frac{N}{2}} \sum_{\alpha=0}^N \sqrt{\frac{\binom{N}{\alpha}}{\binom{N}{\kappa}}}\sum_{\nu=0}^\kappa (-1)^\nu \binom{N-\alpha}{\nu}\binom{\alpha}{\kappa-\nu} \ket{\alpha}.
\end{align}
Therefore, the matrix elements of $U_{\pi/2}^{\otimes N}\ket{\kappa}$ in Dicke basis are
\begin{align}
\left(\frac{\pi}{2}\right)_{\kappa,\alpha}^{\otimes N}&=\bra{\alpha}U_{\pi/2}^{\otimes N}\ket{\kappa} =\frac{1}{2^\frac{N}{2}}\sqrt{\frac{\binom{N}{\alpha}}{\binom{N}{\kappa}}}\sum_{\nu=0}^\kappa (-1)^\nu \binom{N-\alpha}{\nu}\binom{\alpha}{\kappa-\nu}.
\label{eq:pi/2MatrixElements}
\end{align}

Furthermore, a $3\pi/2$ pulse is given by $U_{3\pi/2}=U_\pi U_{\pi/2}$. One can easily convince oneself, that $U_{\pi}\ket{\alpha}=(-1)^{\alpha+1}\ket{N-\alpha}$. Therefore, we find

\begin{equation}
    \left(\frac{3\pi}{2}\right)_{\kappa,\alpha}^{\otimes N}=\frac{(-1)^{\alpha+1}}{2^\frac{N}{2}}\sqrt{\frac{\binom{N}{\alpha}}{\binom{N}{\kappa}}}\sum_{\nu=0}^\kappa (-1)^\nu \binom{\alpha}{\nu}\binom{N-\alpha}{\kappa-\nu}.
\end{equation}

\section{Expectation value of \texorpdfstring{$\langle \hat{M}\rangle$}{Lg}, modified protocol}\label{App:Expectation value of M modified protocol}

For the protocol as depicted in \cref{fig:ModifiedProtocol_resubmission}, the classical trajectories of the $(\kappa,\alpha,\beta)^\mathrm{th}$ trajectory is given by
\begin{align}
&\zeta_{\kappa,\alpha,\beta}(t)=\\
&\begin{cases} 
      \Tilde{\kappa}\left(1-e^{-i \Omega t }\right) & 0 \leq t < T_1 \\
      \tilde{\alpha}\left(1 - e^{-i (t - T_1) \Omega}\right) - \tilde{\kappa}e^{-i t \Omega} \left(-1 + e^{i T_1 \Omega}\right)  & T_1 \leq t < T_{2,1} \\
     \tilde{\alpha} e^{-i t \Omega} \left(e^{i t \Omega} + e^{i T_1 \Omega} - 2 e^{i T_{2,1} \Omega}\right) 
- \tilde{\kappa}e^{-i t \Omega} \left(-1 + e^{i T_1 \Omega}\right) 
 & T_{2,1} \leq t < T_{2,2} \\
      \tilde{\alpha}e^{-i t \Omega}\!\left(e^{i t \Omega} - e^{i T_1 \Omega} + 2\,e^{i T_{2,1}) \Omega}  - 2\,e^{i T_{2,2}\Omega}
\right) - \tilde{\kappa}e^{-i t \Omega}\!\left(-1 + e^{i T_1 \Omega}\right)
 & T_{2,2} \leq t < T_{2,3}\\
      \tilde{\beta}\left(-1 + e^{-i (t - T_{2,3}) \Omega} \right)
+ \tilde{\kappa}e^{-i t \Omega} \left(1 - e^{i T_1 \Omega} \right) 
 & T_{2,3} \leq t < T_\mathrm{tot}
   \end{cases}.
   \label{eq:PatheModifiedProtocolResubmission}
\end{align}
with $\tilde{\kappa}=2\kappa-N$, $\tilde{\alpha}=2\alpha-N, \tilde{\beta}=2\beta-N$, $T_1=t_1, T_{2,1}=t_1+t_{2,1}$, $T_{2,2}=t_1+t_{2,1}+t_{2,2}$, $T_{2,3}=t_1+2t_{2,1}+t_{2,2}$ and $T_\mathrm{tot}=2t_1+2t_{2,1}+t_{2,2}$. The phase each trajectory accumulates is given by
\begin{align}
&\phi_{\kappa,\alpha,\beta}(t)=\\
&\begin{cases} 
      \Tilde{\kappa}^2\sin(t \Omega) + \tilde{\kappa}^2\frac{\eta}{\hbar} t & 0 \leq t < T_1 \\
      \tilde{\alpha}^2 \sin\left((t - T_1) \Omega\right) + \tilde{\kappa}^2 \sin\left(T_1 \Omega\right)+ \tilde{\alpha} \tilde{\kappa} \left[ -\sin(t \Omega) + \sin\left((t - T_1) \Omega\right) + \sin\left(T_1 \Omega\right) \right]\\
      \quad + \frac{\eta}{\hbar} \left[\tilde{\kappa}^2 T_1 + \tilde{\alpha}^2 (t-T_1)\right]
  & T_1 \leq t < T_{2,1} \\
     \tilde{\kappa}^2 \sin(T_1 \Omega)
+ \tilde{\alpha}^2 \left\{
    -\sin((t - T_1) \Omega)
    + 2\left[
        \sin((t - T_{2,1}) \Omega)
        + \sin(t_{2,1} \Omega)
    \right]
\right\}\\
\quad+ \tilde{\alpha} \tilde{\kappa} \left[
    \sin(t \Omega)
    - \sin((t - T_1) \Omega)
    + \sin(T_1 \Omega)
    + 2\sin(t_{21} \Omega)
    - 2\sin(T_{21} \Omega)
\right]\\
\quad+ \frac{\eta}{\hbar} \left[\tilde{\kappa}^2 T_1 + \tilde{\alpha}^2 (t-T_1)\right]
 & T_{2,1} \leq t < T_{2,2} \\
      \tilde{\kappa}^2 \sin(t_1 \Omega)
+ \tilde{\alpha}^2 \{
    \sin((t - T_1) \Omega)
    + 2 [
        -\sin((t - T_{2,1}) \Omega)
        + \sin(t_{2,1} \Omega)
        \\ \quad+ \sin((t - T_{2,2}) \Omega)
        + 2\sin(t_{2,2} \Omega)
        - \sin((t_{2,1} + t_{2,2}) \Omega)
    ]
\}\\
\quad+ \tilde{\alpha} \tilde{\kappa} \{
    -\sin(t \Omega)
    + \sin((t - t_1) \Omega)
    + \sin(t_1 \Omega)
    + 2 [
        \sin(t_{21} \Omega)\\
        \quad- \sin((T_{2,1}) \Omega)
        - \sin((t_{21} + t_{22}) \Omega)
        + \sin((T_{2,2}) \Omega)
    ]
\}\\
\quad+ \frac{\eta}{\hbar} \left[\tilde{\kappa}^2 T_1 + \tilde{\alpha}^2 (t-T_1)\right]
 & T_{2,2} \leq t < T_{2,3}\\
4 \tilde{\alpha}^2 \sin(t_{2,1} \Omega)
+ \tilde{\beta}^2 \frac{
     \left(
        -5 \sin(t \Omega)
        + 2 \sin(t \Omega - t_{2,1} \Omega)
        + 2 \sin(t \Omega + t_{2,1} \Omega)
    \right)
}{
    -5 + 4 \cos(t_{2,1} \Omega)
}\\
\quad+ \tilde{\kappa}^2\frac{
     \left(
        -5 \sin(t_1 \Omega)
        + 2 \sin( (t_1 - t_{2,1}) \Omega)
        + 2 \sin( (t_1  + t_{2,1}) \Omega)
    \right)
}{
    -5 + 4 \cos(t_{2,1} \Omega)
}\\
\quad+  \frac{\tilde{\beta} \tilde{\kappa}}{-5 + 4 \cos(t_{2,1} \Omega)}
[
        \sin(t \Omega)
        + \sin(t_1 \Omega)
        + 4 \sin(t_{2,1} \Omega)
        - 4 \sin(2 t_{2,1} \Omega)\\
        \quad- \sin( (t  + t_1) \Omega)
        - 4 \sin( (t + t_{2,1}) \Omega)
        - 4 \sin( (t_1+ t_{2,1}) \Omega)
        + 4 \sin((t + T_{2,1} )\Omega)\\
        \quad+ 4 \sin( (t + 2 t_{2,1}) \Omega)
        + 4 \sin((t_1 + 2 t_{2,1}) \Omega)
        - 4 \sin( (t + t_1+ 2 t_{2,1}) \Omega)
]\\
\quad\frac{\eta}{\hbar} \left[\tilde{\kappa}^2 T_1 + \tilde{\alpha}^2 (T_{2,3}-T_1) + \tilde{\beta}^2 (t-T_{2,3})\right]
 & T_{2,3} \leq t < T_\mathrm{tot}
   \end{cases}.
   \label{eq:PhaseModifiedProtocolResubmission}
\end{align}
As one can easily check, the phase each trajectory accumulates for all $(\kappa,\alpha,\beta=\kappa)$ is a multiple of $\pi$ if the parameters are chosen according to \cref{tab:parameters}.
For an initial state as in \cref{eq:InitialState} and 
\begin{equation}
    \hat{\rho}_{\kappa,\kappa\prime,\alpha,\alpha^\prime,\beta,\beta^\prime}=\hat{U}_{\kappa,\alpha,\beta}\hat{\rho}_\mathrm{th}\hat{U}^\dagger_{\kappa',\alpha',\beta'},
\end{equation}
the final spin state is given by
\begin{align}
    \hat{\rho}_\mathrm{S}=\frac{1}{2^N}\sum_{\kappa,\kappa^\prime=0}^N\sum_{\alpha,\alpha^\prime=0}^N\sum_{\beta,\beta^\prime=0}^N & \sqrt{\binom{N}{\kappa}}\sqrt{\binom{N}{\kappa'}} \left(\frac{\pi}{2}\right)_{\kappa,\alpha}^{\otimes N}\left(\frac{3\pi}{2}\right)_{\alpha,\beta}^{\otimes N} \left(\frac{\pi}{2}\right)_{\alpha^\prime,\kappa^\prime}^{\otimes N} \left(\frac{3\pi}{2}\right)_{\beta^\prime,\alpha^\prime}^{\otimes N} \nonumber \\
    &\times\tr{\hat{\rho}_{\kappa,\kappa\prime,\alpha,\alpha^\prime,\beta,\beta^\prime}(T_\mathrm{tot})} \ket{\beta}\bra{\beta}.
\end{align}

For the trajectories that recombine, the trace over $\hat{\rho}_{\kappa,\kappa\prime,\alpha,\alpha^\prime,\beta,\beta^\prime}(T_\mathrm{tot})$ is given by

\begin{equation}
    e^{-\xi \int_0^T [1-e^{-[\mathcal{X}_{\kappa,\alpha,\beta}(t^\prime)-\mathcal{X}_{\kappa^\prime,\alpha^\prime,\beta^\prime}(t^\prime)]^2/(4 r^2_\mathbf{CSL})}]\,\mathrm{d} t'},
\end{equation}
where $\mathcal{X}_{\kappa,\alpha,\beta}(t)$ is given by $\mathcal{X}_{\kappa,\alpha,\beta}(t)=\chi\,\Re[\zeta{\kappa,\alpha,\beta}(t)]$. 

If the trajectories do not recombine, the trace over $\hat{\rho}_{\kappa,\kappa\prime,\alpha,\alpha^\prime,\beta,\beta^\prime}(T_\mathrm{tot})$ goes to zero mainly because the wave packets do not perfectly overlap. The overlap between two thermal states, where one is displaced by $\zeta_i$ and the other by $\zeta_j$, is given by $\tr{\hat{D}(\zeta_i)\hat{\rho}_\mathrm{th}\hat{D}^\dagger(\zeta_j)}$.
To find an expression for this overlap, we expand $\hat{\rho}_\mathrm{th}$ in coherent states
\begin{equation}
\hat{\rho}_\mathrm{th}=\int\,\frac{\mathrm{d}\alpha^2}{\pi} \, \frac{e^{-\frac{|\alpha|^2}{\bar{n}}}}{\bar{n}}\ket{\alpha}\bra{\alpha}
\end{equation}
where $\bar{n}$ is the occupation number.
To evaluate the trace, we use the definition of a coherent state $\ket{\alpha}=\hat{D}(\alpha)\ket{0}$ and the properties of the displacement operator. One then finds
\begin{align}
\tr{\hat{D}(\zeta_i)\ket{\alpha}\bra{\alpha}\hat{D}^\dagger(\zeta_j)}&=e^{\phi}e^{\Delta\zeta \alpha^*-\Delta\zeta^*\alpha}e^{\frac{-|\Delta \zeta|^2}{2}}
\end{align}
with $\Delta\zeta=\zeta_i-\zeta_j$ and $\phi=(\zeta^*_j\zeta_i-\zeta_j \zeta^*_i)/2$. All that's left is solving the integral to find
\begin{align}
\tr[\hat{D}(\zeta_i)\hat{\rho}_\mathrm{th}\hat{D}^\dagger(\zeta_j)]&=e^{\phi-|\Delta\zeta|^2(\frac{1}{2}+\bar{n})}.
\end{align}
As the overlap goes with $\exp(-\bar{n})$, it goes to zero very quickly for large occupation numbers.

\section{Analysis of Wave Functions Structure}\label{App:Analysis of Wave Functions Structure}
\begin{figure}
\centering

\subfloat[\label{fig:calpha}]{%
  \includegraphics[width=0.3\columnwidth]{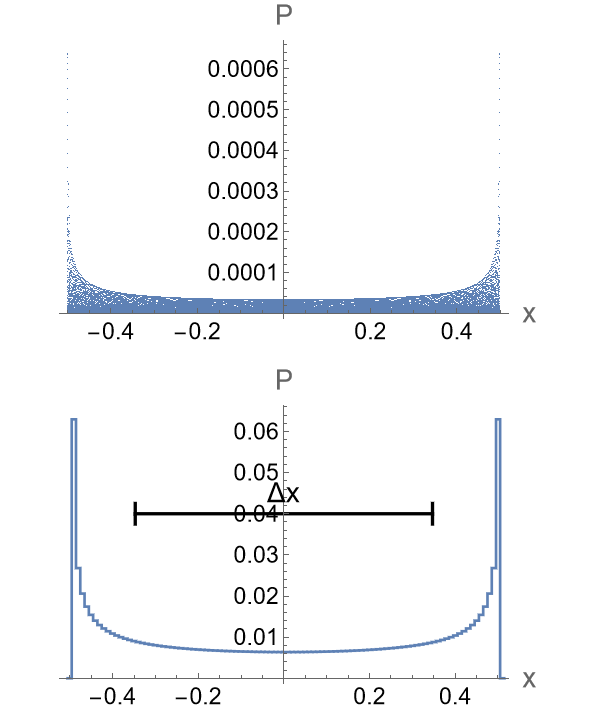}%
}%
\subfloat[\label{fig:ckappacalphal}]{%
\centering
  \includegraphics[width=0.3\columnwidth]{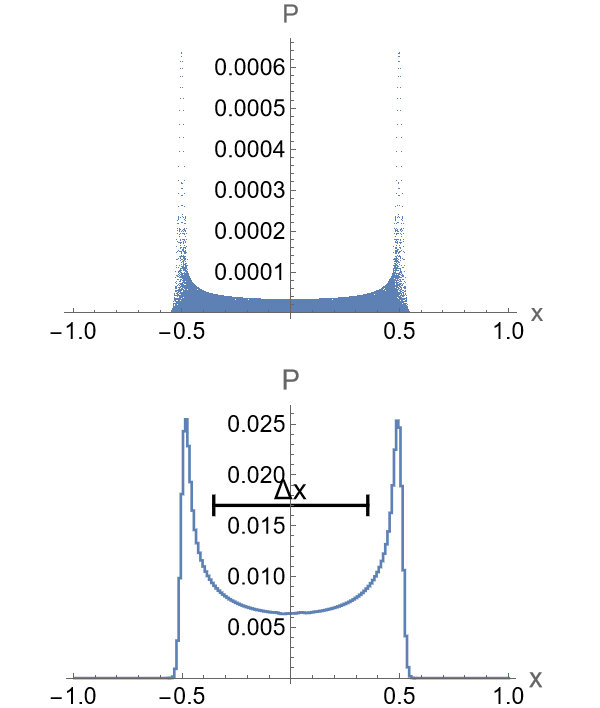}%
}
\subfloat[\label{fig:ckappa}]{%
  \includegraphics[width=0.3\columnwidth]{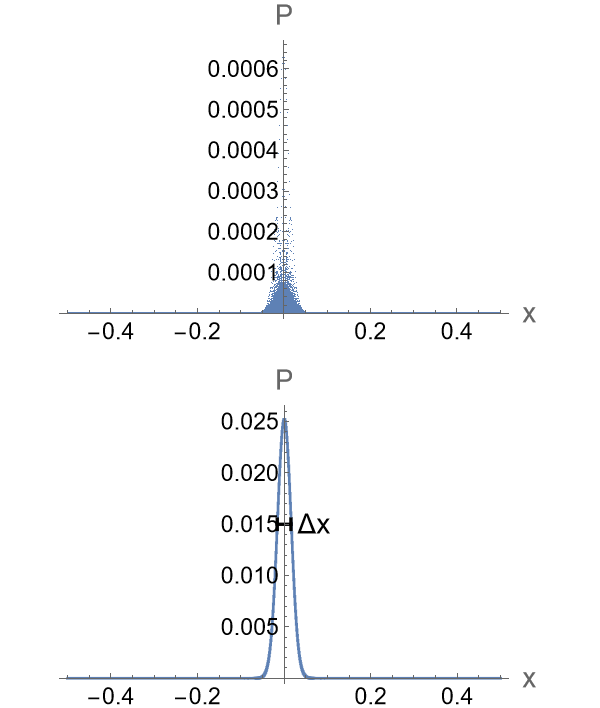}%
}
\caption{Here, the position distribution of the wave packet after the second $\pi/2$ pulse for some arbitrary numbers is given. In (a) one can see the  the wave packet for $c_\kappa=0$, which shown two distinct peaks at the very end of the wave function, in (b) for $c_\kappa=c_\alpha$ and in (c) for $c_\alpha=0$. For $c_\alpha \ll c_\kappa$, the peaks wander towards the middle and approach the Gaussian one expects for $c_\alpha=0$. In the top plots, each dot represents the position and probability of one trajectory $(\kappa,\alpha)$. To understand the probability distribution better, in the bottom plot, one sees the corresponding histogram, where all probabilities in a given position range have been added up. $\Delta x=\sqrt{\Delta x^2}$ is the width of the wave function as defined in \cref{eq:Deltax}.}
\label{fig:PosDistribution}
\end{figure}

\begin{figure}
\centering

\subfloat[\label{fig:Dxckappa}]{%
  \includegraphics[width=0.3\columnwidth]{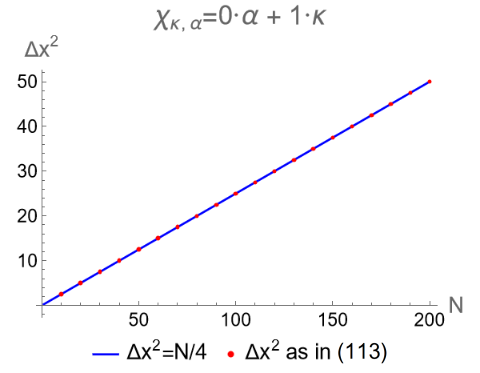}%
}%
\subfloat[\label{fig:Dxcalpha}]{%
\centering
  \includegraphics[width=0.3\columnwidth]{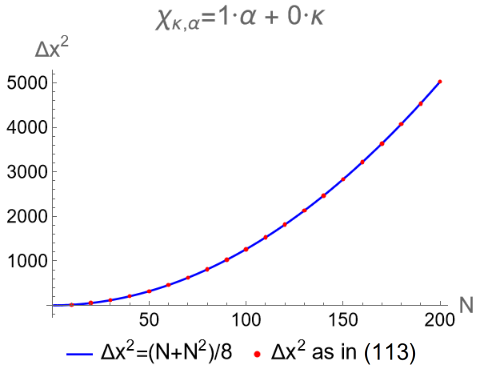}%
}
\subfloat[\label{fig:Dxckappacalpha}]{%
  \includegraphics[width=0.3\columnwidth]{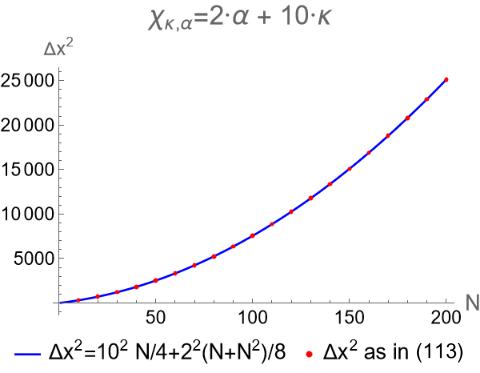}%
}
\caption{Dependence of the wave-function width on $N$. As can be seen in (a), the width grows linearly with $N$ as $\Delta x_\kappa = N/4$ when $c_\alpha = 0$ and $c_\kappa = 1$, while for $c_\kappa = 0$ and $c_\alpha = 1$ the width grows as $\Delta x_\alpha = (N + N^2)/8$, as shown in (b). For a general superposition $\chi_{\kappa,\alpha} = c_\kappa \kappa + c_\alpha \alpha$, the width behaves as $\Delta x^2 = c_\kappa^2 \Delta x_\kappa^2 + c_\alpha^2 \Delta x_\alpha^2$,
as shown in (c). Although we are not able to prove this analytically, it is consistent with all parameter sets we have tested.
}
\label{fig:Dx}
\end{figure}

Initially, the wave packet is prepared in a thermal state. After applying the first $\pi/2$ pulse, the wave packet will split up into $N+1$ gaussian wave packets, which we numerate with $\kappa\in{0,N}$, where $\kappa$ denotes how many spins point up. The position of the $\kappa^\mathbf{th}$ trajectory is proportional to $\kappa$ at any time and the probability for it to be realized is given by a binomial distribution $\binom{N}{k}$. When the second $\pi/2$ pulse is applied, the structure of the wave packet becomes significantly more complex. Each trajectory again splits into $N+1$ trajectories. Therefore, each trajectory is denoted by $(\kappa,\alpha)$, where $\alpha$ is in $[0,N]$. The probability for the $(\kappa,\alpha)^\mathrm{th}$ to be realized is now given by 
\begin{equation}
    P(\alpha,\kappa)=\binom{N}{\kappa} \left[\left(\frac{\pi}{2}\right)_{\kappa,\alpha}^{\otimes N}\right]^2.
\end{equation}
The position of the $(\kappa, \alpha)^\mathrm{th}$ trajectory it at all times of the form $c_\kappa \kappa+c_\alpha\alpha$, where $c_{\kappa,\alpha}$ depends on the time and the parameters of the experiment.
As we are limited in the number of spins, we analyze the behavior with the wave packets width for a limited number of spins and different $c_\kappa, c_\alpha$ in order to later extrapolate the wave functions width for larger $N$.
\\
For some arbitrary numbers, the position distribution of the wave function is shown in \cref{fig:PosDistribution}, as well as how the width $\Delta x$ grows with $N$ for $c_\kappa$ and $x_\alpha$.
We define the width as
\begin{equation}
    \Delta x^2(t)=\sum_{\kappa,\alpha}^N \chi^2_{\kappa,\alpha}(t) P(\kappa,\alpha) =\sum_{\kappa,\alpha}^N \left(c_\kappa(t) \kappa + c_\alpha (t)\alpha \right)^2 P(\kappa,\alpha)
    \label{eq:Deltax}
\end{equation}
where $\chi_{\kappa,\alpha}(t)$ is the position of the $(\kappa,\alpha)^\mathrm{th}$ trajectories at time $t$.
As expected, for $c_\alpha=0$, the wave function is distributed according to a binomial distribution and $\Delta x^2(t)$ grows with $N$. For $c_\kappa=0$ on the other hand, the wave function exhibits two distinct peaks and $\Delta x^2(t)$ grows like $N+N^2$. 
Therefore, we expect that the wave functions width $\Delta x(t)=\sqrt{\Delta x^2(t)}$ grows with $\sqrt{N}$  for small $N$ and for large $N$, the width eventually grows linearly in $N$. 
In order to find an expression of the variance in dependence of $c_\alpha, c_\kappa$ and $N$, we define
\begin{equation}
    \Delta x_\kappa^2=\sum_{\kappa,\alpha}^N \kappa^2  P(\kappa,\alpha)=\frac{N}{4},\quad \Delta x_\alpha^2=\sum_{\kappa,\alpha}^N \alpha^2  P(\kappa,\alpha) =\frac{N+N^2}{8}
\end{equation}
and then use the Ansatz $\Delta x^2(t)=c_\kappa^2(t)\Delta x_\kappa^2+c_\alpha^2(t)\Delta x_\alpha^2$, which fits perfectly as shown in \cref{fig:Dx}. While we were not able to prove this relation, it fits with all parameters we tested.

\section{Spin Flips}\label{App:Spin Flips}

The $T_1$ time gives a time scale, at which a nuclear spin statistically flips. In this subsection, we want to analyze how such a flip affects the time evolution of the levitated nanoparticle. For that we split the time evolution into the time evolution before and after the spin flip. 
If the spins are initially fully polarized, the state is given by 
\begin{equation}
    \hat{\rho}(0)=\frac{1}{2^N}\sum_{\kappa=0}^N \sqrt{\binom{N}{\kappa}} \ket{\kappa}\otimes\hat{\rho}_x.
\end{equation}
Until time $\tau$, the nanoparticle evolves undisturbed according to the given protocol. The time evolution of the $\kappa^\mathrm{th}$ branch is then given by
\begin{align}
    \hat{U}_\kappa(\tau)=e^{-i \phi_\kappa(\tau)}\hat{D}\left(\frac{\chi}{2 x_0}\zeta_\kappa(\tau)\right)e^{-i \tau \hat{a}^\dagger\hat{a}}.
\end{align}
If one spin flips, a Dicke state transforms like
\begin{align}
    \sqrt{\binom{N}{\kappa}}\ket{\kappa}_N \rightarrow\;& \sqrt{\binom{N-1}{\kappa-1}} \ket{0}_1\otimes\ket{\kappa-1}_{N-1} + \sqrt{\binom{N-1}{\kappa+1}} \ket{1}_1\otimes\ket{\kappa+1}_{N-1}.
\end{align}
Therefore, each branch in turn splits into an incoherent superposition of two branches. One with $\kappa-1$ spins up and one with $\kappa+1$ spins up. The time evolution for the full protocol along the $\kappa^\mathrm{th}$ trajectory is therefore given by
\begin{align}
     \hat{U}^\pm_\kappa(T_\mathrm{tot},\tau)=&  e^{-i \phi_{\kappa\pm 1}(T_\mathrm{tot}-\tau)}\hat{D}\left(\frac{\chi}{2 x_0}\zeta_{\kappa\pm 1}(T_\mathrm{tot}-\tau)\right) e^{-i (T_\mathrm{tot}-\tau) \hat{a}^\dagger\hat{a}} \nonumber \\
     \times&e^{-i \phi_\kappa(\tau)}\hat{D}\left(\frac{\chi}{2 x_0}\zeta_\kappa(\tau)\right)e^{-i \tau \hat{a}^\dagger\hat{a}}.
\end{align}
As $\zeta_\kappa$ is linear in $(2 \kappa-N)$, we can write it as $\zeta_{\kappa}(t)=(2 \kappa-N)\tilde{\zeta}(t)$. $\zeta_{\kappa\pm 1}$ can therefore be rewritten as $\zeta_{\kappa\pm 1}=\zeta_\kappa(t) \pm 2 \tilde{\zeta}$. The effect of the spin-flip can be separated from the undisturbed time evolution, by rewriting the displacement operator like
\begin{align}
\hat{D}\left(\frac{\chi}{2 x_0}\zeta_{\kappa\pm 1}(T_\mathrm{tot}-\tau)\right)&=\hat{D}\left(\pm\frac{\chi}{ x_0}\tilde{\zeta}(T_\mathrm{tot}-\tau)\right) \hat{D}\left(\frac{\chi}{2 x_0}\zeta_{\kappa}(T_\mathrm{tot}-\tau)\right).  
\end{align}
We therefore find
\begin{align}
     \hat{U}^{\pm}_\kappa(T_\mathrm{tot},\tau)=&  e^{-i \tilde{\phi}_{\kappa}(T_\mathrm{tot},\tau)}  \hat{D}\left(\pm\frac{\chi}{x_0}\tilde{\zeta}(T_\mathrm{tot}-\tau)\right) e^{-i \phi_{\kappa}(T_\mathrm{tot})} \hat{D}\left(\frac{\chi}{2 x_0}\zeta_\kappa(T_\mathrm{tot})\right)e^{-i T_\mathrm{tot} \hat{a}^\dagger\hat{a}}.
\end{align}
The second line describes the unperturbed evolution, while the first line describes the effect of the flip. The flip induces an additional phase $\tilde{\phi}_{\kappa}(T_\mathrm{tot},\tau)=\phi_{\kappa\pm 1}(T_\mathrm{tot}-\tau)-\phi_{\kappa}(T_\mathrm{tot}-\tau)$.  The protocol is designed such that, if the initial state is a thermal state, all trajectories recombine in the center. 
The displacement operator in the first line indicates that, when a flip happens, the trajectories either recombine in $\hat{D}(\frac{\chi}{x_0}\tilde{\zeta}(T_\mathrm{tot}-\tau))$ and or in $\hat{D}(-\frac{\chi}{x_0}\tilde{\zeta}(T_\mathrm{tot}-\tau))$, each with a probability of 50\,\%. The distance in phase space is given by $|\Delta\zeta(\tau)|=2\chi/x_0|\tilde{\zeta}(T_\mathrm{tot}-\tau)|$. For the first protocol described here, $|\tilde{\zeta}(T_\mathrm{tot}-\tau)|$ is bounded by two. This bound is realized if the flip happens while the wave function is maximally spread at $\tau=\pi/\Omega$. The maximal displacement is therefore given by $|\Delta\zeta(\tau)|< 4 \chi/x_0$.
\hfill

\end{document}